\documentclass[10pt,twocolumn,twoside]{IEEEtran}

\usepackage{amsmath}
\usepackage{amssymb}
\usepackage{latexsym}
\usepackage{verbatim}
\usepackage{subfigure}
\usepackage[final]{graphicx}
\usepackage{psfrag}
\usepackage{hyperref}

\usepackage{tcolorbox}
\definecolor{darkred}{RGB}{250,0,0}
\definecolor{darkgreen}{RGB}{0,150,0}
\definecolor{myblue}{RGB}{0,0,250}
\definecolor{darkblue}{RGB}{0,0,200}
\hypersetup{colorlinks=true, linkcolor=darkred, citecolor=myblue, urlcolor=darkblue}

\newtheorem{theorem}{Theorem}
\newtheorem{proposition}{Proposition}
\newtheorem{lemma}{Lemma}

\newtheorem{remark}{Remark}

{\begin{list}               
    {$\bullet$ \hfill}{
        \setlength{\leftmargin}{\parindent}
        \setlength{\parsep}{0.04\baselineskip}
        \setlength{\itemsep}{0.5\parsep}
        \setlength{\labelwidth}{\leftmargin}
        \setlength{\labelsep}{0em}}
    }
{\end{list}}

\providecommand{\cref}[1]{Chapter~\ref{chap:#1}}

\providecommand{\R}{\ensuremath{\mathbb{R}}}

\providecommand{\abs}[1]{\lvert#1\rvert}
\providecommand{\norm}[1]{\lVert#1\rVert}

\providecommand{\bydef}{\overset{\text{def}}{=}}

\renewcommand{\vec}[1]{\ensuremath{\boldsymbol{#1}}}
\providecommand{\mat}[1]{\ensuremath{\boldsymbol{#1}}}

\providecommand{\mA}{\mat{A}} \providecommand{\mB}{\mat{B}}
\providecommand{\mC}{\mat{C}} 
\providecommand{\mD}{\mat{D}}
\providecommand{\mH}{\mat{H}}
\providecommand{\mI}{\mat{I}}  
  
\providecommand{\mM}{\mat{M}} \providecommand{\mP}{\mat{P}} 
 
\providecommand{\mS}{\mat{S}} \providecommand{\mU}{\mat{U}} 
\providecommand{\mV}{\mat{V}}

\providecommand{\mT}{\mat{T}}

\providecommand{\mGm}{\mat{\Gamma}} \providecommand{\mG}{\mat{G}}

\providecommand{\va}{\vec{a}} \providecommand{\vb}{\vec{b}}
\providecommand{\vc}{\vec{c}} 
\providecommand{\vh}{\vec{h}} 
  
 \providecommand{\vp}{\vec{p}}
 \providecommand{\vs}{\vec{s}}
 \providecommand{\vr}{\vec{r}}
\providecommand{\vg}{\vec{g}}
\providecommand{\vu}{\vec{u}} \providecommand{\vw}{\vec{w}}
\providecommand{\vx}{\vec{x}} \providecommand{\vy}{\vec{y}}
\providecommand{\vz}{\vec{z}} 
 
 \providecommand{\vv}{\vec{v}}

\usepackage{lipsum}    
\usepackage{ifthen}
\usepackage{tcolorbox}
\newboolean{showcomments}
\setboolean{showcomments}{true}
\newcommand{\oussama}[1]{\ifthenelse{\boolean{showcomments}}
{ \textcolor{red}{(Oussama says:  #1)}}{}}
\newcommand{\christos}[1]{\ifthenelse{\boolean{showcomments}}
{ \textcolor{blue}{(Christos says: #1)} } {} }
\newcommand{\yue}[1]{\ifthenelse{\boolean{showcomments}}
{ \textcolor{magenta}{(Yue says:  #1)}}{}}

\usepackage{hyperref}

\usepackage{enumitem}

\providecommand{\abs}[1]{\lvert#1\rvert}
\providecommand{\norm}[1]{\lVert#1\rVert}

\providecommand{\bydef}{\overset{\text{def}}{=}}

\renewcommand{\vec}[1]{\ensuremath{\boldsymbol{#1}}}
\providecommand{\mat}[1]{\ensuremath{\boldsymbol{#1}}}

\providecommand{\mA}{\mat{A}} \providecommand{\mB}{\mat{B}}
\providecommand{\mC}{\mat{C}} 
\providecommand{\mD}{\mat{D}}
\providecommand{\mF}{\mat{F}}
\providecommand{\mI}{\mat{I}}  
  
\providecommand{\mM}{\mat{M}} \providecommand{\mP}{\mat{P}} 
 
\providecommand{\mS}{\mat{S}} \providecommand{\mU}{\mat{U}} 
\providecommand{\mV}{\mat{V}}

\providecommand{\mZ}{\mat{Z}}

\providecommand{\mGm}{\mat{\Gamma}} \providecommand{\mG}{\mat{G}}
\providecommand{\mOm}{\mat{\Omega}}

\usepackage{cases}

\DeclareMathOperator{\erf}{erf}
\DeclareMathOperator{\sign}{sign}

\providecommand{\va}{\vec{a}} \providecommand{\vb}{\vec{b}}
\providecommand{\vc}{\vec{c}} 
\providecommand{\vh}{\vec{h}} 
  
 \providecommand{\vp}{\vec{p}}
 \providecommand{\vs}{\vec{s}}
 \providecommand{\vr}{\vec{r}}
\providecommand{\vf}{\vec{f}} 
\providecommand{\vg}{\vec{g}}
\providecommand{\vu}{\vec{u}} \providecommand{\vw}{\vec{w}}
\providecommand{\vx}{\vec{x}} \providecommand{\vy}{\vec{y}}
\providecommand{\vz}{\vec{z}} 
 
 \providecommand{\vv}{\vec{v}}

\providecommand{\vepsilon}{\vec{\epsilon}}

\usepackage{algorithm}
\usepackage{algpseudocode}
\algnewcommand\algorithmicforeach{\textbf{Until :}}
\algnewcommand\algorithmicendif{\textbf{End}}
\algnewcommand\ForEach{\item[ \algorithmicforeach]}
\algnewcommand\EndiFF{\item[ \algorithmicendif]}

\usepackage{cite}

\usepackage{color}

\usepackage{enumitem}
\usepackage{commath}

\usepackage{hyperref}

\usepackage{esdiff}

\usepackage{pgf,tikz,psfrag}

\usepackage{dsfont}

\newcommand{\argmax}{\operatornamewithlimits{argmax}}
\newcommand{\argmin}{\operatornamewithlimits{argmin}}

\providecommand{\vxi}{\vec{\xi}}

\begin{document}
%
\title{A Precise Performance Analysis of Learning with Random Features}

\author{Oussama Dhifallah and Yue M. Lu
\thanks{O. Dhifallah and Y. M. Lu are with the John A. Paulson School of Engineering and Applied Sciences, Harvard University, Cambridge, MA 02138, USA (e-mails: oussama$\_$dhifallah@g.harvard.edu and yuelu@seas.harvard.edu).}
\thanks{This work was supported by the Harvard FAS Dean's Fund for Promising Scholarship, and by the US National Science Foundation under grants CCF-1718698 and CCF-1910410. A preliminary version of this work will be presented at the 2020 Asilomar Conference on Signals, Systems and Computers.}
}
\maketitle


\begin{abstract}
We study the problem of learning an unknown function using random feature models. Our main contribution is an exact asymptotic analysis of such learning problems with Gaussian data. Under mild regularity conditions for the feature matrix, we provide an exact characterization of the asymptotic training and generalization errors, valid in both the under-parameterized and over-parameterized regimes. The analysis presented in this paper holds for general families of feature matrices, activation functions, and convex loss functions. Numerical results validate our theoretical predictions, showing that our asymptotic findings are in excellent agreement with the actual performance of the considered learning problem, even in moderate dimensions. Moreover, they reveal an important role played by the regularization, the loss function and the activation function in the mitigation of the ``double descent phenomenon'' in learning.
\end{abstract}


\section{Introduction}\label{sec:intro}

Suppose we are given a collection of training data $\lbrace (y_i,\va_i) \rbrace_{i=1}^{m}$, where $\va_i\in\mathbb{R}^n$ and the labels $\lbrace y_i \rbrace_{i=1}^{m}$ are generated according to the following model
\begin{align}\label{cmodel}
y_i=\varphi(\va_i^\top \vxi)+\Delta\epsilon_i,\qquad 1 \le i \le m.
\end{align}
Here, $\vxi$ is an unknown and fixed vector with $\norm{\vxi}=\rho$, $\lbrace \epsilon_i \rbrace_{i=1}^{m}$ are independent and identically distributed standard Gaussian random variables, $\Delta>0$ is a fixed positive constant, and $\varphi(\cdot)$ is a scalar (deterministic or probabilistic) function. We consider the problem of fitting the available data $\lbrace (y_i,\va_i) \rbrace_{i=1}^{m}$ using the random feature model \cite{RR08}, which corresponds to a restricted family of functions in the form of 
\begin{align}
\mathcal{F}_{\text{RF}}=\Big\lbrace g_{\vw}(\va) = \vw^\top \sigma(\mF^\top \va),~\vw\in\mathbb{R}^k \Big\rbrace.
\end{align}
Here, $\mF\in\mathbb{R}^{n\times k}$ is a random feature matrix drawn from some matrix ensembles, and $\sigma(\cdot)$ is a scalar activation function applied to each element of $\mF^\top \va$. The weight vector $\vw$ is learned by solving an optimization problem
\begin{align}\label{formulation}
\widehat{\vw}=\argmin_{\vw\in\mathbb{R}^k} \frac{1}{m} \sum_{i=1}^{m} {\ell}\left(y_i,\vw^\top \sigma(\mF^\top \va_i) \right)+\tfrac{\lambda}{2} \norm{\vw}^2,
\end{align}
with some loss function $\ell(\cdot, \cdot)$ and positive regularization constant $\lambda$. Note that one can also view this model as a two-layer neural network, with $k$ hidden neurons and the first layer weights (i.e., the matrix $\mF$) fixed in the learning process.

In this paper, we assume that the loss function ${\ell}(\cdot, \cdot)$ in \eqref{formulation} takes one of the following two forms
\begin{subnumcases}~
{\ell}(y,z)=\widehat{\ell}(z-y) &\text{for regression tasks}\label{reg_form}\\
{\ell}(y,z)=\widehat{\ell}(yz) &\text{for classification tasks},\label{class_form}
\end{subnumcases}
where $\widehat{\ell}(\cdot)$ is a convex function. For example, $\widehat{\ell}(\cdot)$ can be the squared loss for regression problems and the logistic loss for classification problems.

Given a fresh data sample $\va_\text{new}\in\mathbb{R}^{n}$, the prediction of the corresponding label $\widehat{y}_{\text{new}}$ can be expressed as
\begin{align}
\widehat{y}_{\text{new}}=\widehat{\varphi} [\widehat{\vw}^\top \sigma(\mF^\top \va_\text{new})],
\end{align}
where $\widehat{\vw}\in\mathbb{R}^k$ denotes the optimal solution of \eqref{formulation} and $\widehat{\varphi}$ is some fixed function. We measure the performance of the learning process via the generalization error, defined as 
\begin{align}\label{testerr_def}
\mathcal{G}_{n,\text{test}}=\frac{1}{4^\upsilon} \mathbb{E}\left[ \varphi(\vxi^\top \va_{\text{new}})-\widehat{\varphi}(\widehat{\vw}^\top \sigma(\mF^\top \va_{\text{new}})) \right]^2.
\end{align}
The expectation is taken over the distribution of the new data vector $\va_{\text{new}}$ and the function $\varphi$. The constant $\upsilon$ in \eqref{testerr_def} is set to $0$ for linear regression (e.g. when $\varphi$ is the identity function) and to $1$ for binary classification problems (e.g., when $\varphi$ is the sign function). Moreover, we use the training error
\begin{align}\label{trainerr_def}
\mathcal{G}_{n,\text{train}}= \frac{1}{m}\sum_{i=1}^{m}  \ell\left(y_i,\widehat{\vw}^\top \sigma(\mF^\top \va_i)\right)+\tfrac{\lambda }{2} \norm{\widehat{\vw}}^2,
\end{align}
as a performance measure on the training process. It is exactly the optimal cost value of the problem given in \eqref{formulation}.
\subsection{Main Contribution}   
The main contribution of this paper is to precisely characterize the asymptotic performance of the generalization and training errors for a general family of feature matrices, activation functions, and convex loss functions. Our analysis is based on the so-called \emph{uniform Gaussian equivalence conjecture} (uGEC), which states that the performance of \eqref{formulation} can be fully characterized by analyzing the following asymptotically equivalent formulation 
\begin{equation}\label{GEC_for}
\min_{\vw\in\mathbb{R}^k} \frac{1}{m} \sum_{i=1}^{m}  \ell \left(y_i,\vw^\top (\mu_0 \vec{1}_k+\mu_1 \mF^\top \va_i+\mu_\star \vz_i) \right) + \tfrac{\lambda}{2} \norm{\vw}^2,
\end{equation}
where $\lbrace \vz_i \rbrace_{i=1}^{m}$ are independent standard Gaussian random vectors and independent of $\lbrace \va_i \rbrace_{i=1}^{m}$. Moreover, $\mu_0=\mathbb{E}[\sigma(z)]$, $\mu_1=\mathbb{E}[z\sigma(z)]$, and $\mu^2_\star=\mathbb{E}[\sigma(z)^2]-\mu_0^2-\mu_1^2$, where $z$ is a standard Gaussian random variable. In what follows, we shall refer to the original problem \eqref{formulation} as the \emph{feature formulation}, and refer to \eqref{GEC_for} as the \emph{Gaussian formulation}.

\begin{figure}[t]
    \centering
    \subfigure[]{\label{fig:fig_int1}
    \includegraphics[width=0.47\linewidth]{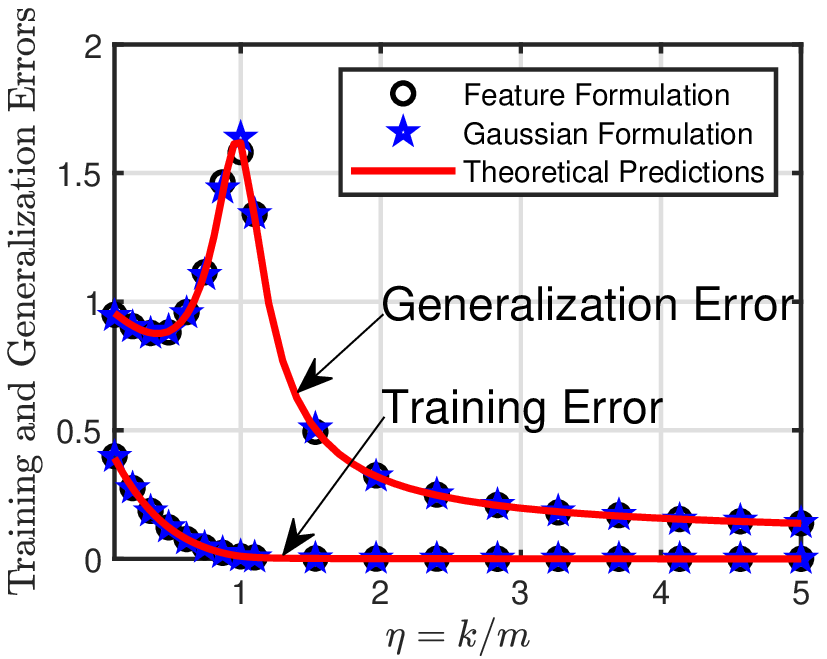}
    }
    \subfigure[]{\label{fig:fig_int2}
        \includegraphics[width=0.47\linewidth]{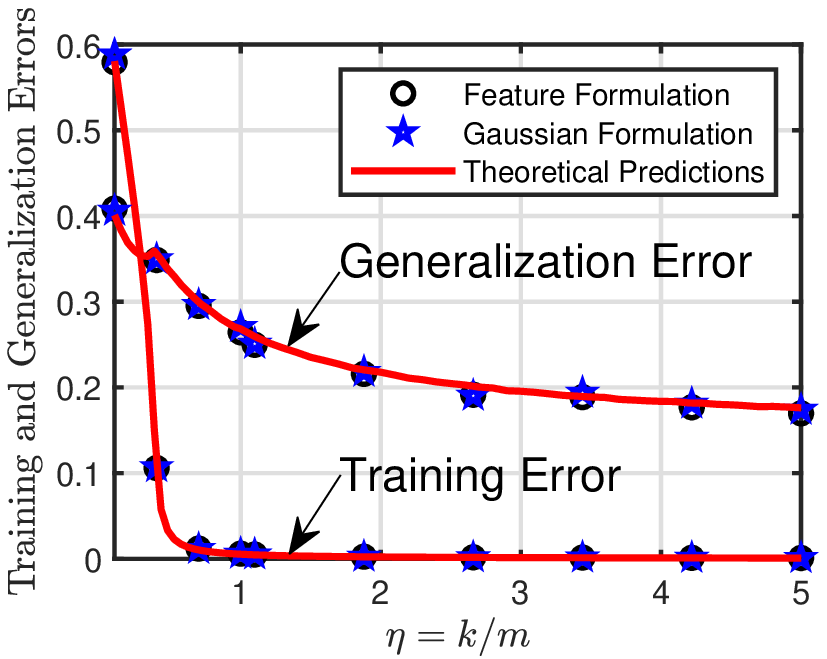}}

    \caption{Theoretical predictions v.s. numerical simulations. (a) We consider a deterministic model where $\varphi$ and $\widehat{\varphi}$ are both the identity function, $\widehat{\ell}$ is the squared loss given in \eqref{reg_loss}, $\sigma:x\to \max(x,0)$ is the ReLu activation function, $\lambda=10^{-3}$, $\alpha=m/n=2$ and $\Delta=0.1$. (b) We consider a binary classification problem where the labels $\lbrace y_i \rbrace_{i=1}^{m}$ are binary $\{\pm 1\}$ numbers. Both ${\varphi}$ and $\widehat{\varphi}$ are the sign function, $\widehat{\ell}$ is the logistic loss given in \eqref{loss_class}, $\sigma$ is the sign activation function, $\lambda=10^{-4}$, $\alpha=m/n=3$ and $\Delta=0$. The hidden vector $\vxi$ in \eqref{cmodel} has norm $\rho=1$. The feature matrix $\mF$ can be expressed as $\mF=\frac{1}{\sqrt{n}} \mV$, where $\mV\in\mathbb{R}^{n\times k}$ has independent standard Gaussian random components. The results shown in the figures are averaged over $50$ independent Monte Carlo trials and we set $n=400$. }
        \label{fig_int}
\end{figure}

The asymptotic equivalence of the feature and Gaussian formulations has been observed in several earlier papers in the literature (see, e.g., \cite{RF_monta_1,RF_monta_2, RF_rep_1, CS13, PW17}). It has also been validated by extensive numerical simulations. (See Figure~\ref{fig_int} for yet another demonstration.) In this work, we build our analysis on this conjecture, and study the Gaussian formulation \eqref{GEC_for} as a surrogate of the original feature formulation \eqref{formulation}. Under mild regularity assumptions on the functions $\varphi$ and $\widehat{\varphi}$ and the feature matrix $\mF$, we show that the training and generalization errors converge in probability to deterministic limit functions as the dimensions $m, n, k$ tend to infinity. These limit functions can be explicitly computed by solving a four-dimensional deterministic optimization problem.  Our analysis rigorously verifies the predictions given in \cite{RF_rep_1}, which were obtained by using the non-rigorous replica method \cite{Mezard:1986} from statistical physics.  

Figure \ref{fig_int} compares our theoretical predictions with empirical simulations. It considers a linear regression and a binary classification problem. Figure \ref{fig_int} shows that our theoretical results are in excellent agreement with the actual performance of the feature formulation in \eqref{formulation}, and this validates our predictions for \eqref{GEC_for} as well as the equivalence conjecture. 
Note that the generalization error follows a U-shaped curve for small model complexity $\eta \bydef k / m$. Specifically, the generalization error first decreases, then, it increases until it reaches a peak known as the interpolation threshold \cite{BHMM18}. After the peak, the generalization error decreases monotonically as a function of the model complexity $\eta$. This behaviour is known as the ``double descent'' phenomenon \cite{BMM18,BHMM18}. Moreover, note that the peak occurs when the training error converges to zero, i.e., $\eta=1$ for linear regression and $\eta \approx 0.4$ for binary classification. 

\subsection{Related Work}

Random feature models \cite{RR08} have attracted significant attention in the literature (see, e.g., \cite{Bach15,aless,ghorbani}). Closely related to our work are several recent papers \cite{RF_monta_1,RF_monta_2, RF_rep_1, Ba2020Generalization} on analyzing the high-dimensional performance of such models. In \cite{RF_monta_1}, the authors precisely characterized the generalization errors of ridge regression with Gaussian feature matrices. This corresponds to the case where the function $\varphi$ is the identity function and $\ell(y, z)$ is the squared loss. In a subsequent work, \cite{RF_monta_2} provides a precise asymptotic characterization of the maximum-margin linear classifier in the overparametrized regime using the convex Gaussian min-max theorem (CGMT) \cite{chris:151,chris:152}. The unregularized least squares regression problem with two-layer neural network, with the first or the second layer weights fixed, is analyzed in \cite{Ba2020Generalization}. That work provides a bias-variance decomposition of the generalization error and precisely characterizes the variance term for the feature model with  Gaussian feature matrix and a generic data model. Under mild restrictions on the feature matrix, \cite{RF_rep_1} uses the non-rigorous replica method \cite{Mezard:1986} from statistical physics to precisely analyze the feature model for a generic convex loss function. 

In this paper, we use the same technical tool, namely CGMT, as in \cite{RF_monta_2} to precisely characterize the performance of the equivalent Gaussian formulation, in both the under-parameterized and over-parameterized regimes. Our model is different from and generalizes the one considered in \cite{RF_monta_2} in that the latter assumes that the labels $\lbrace y_i \rbrace_{i=1}^m$ are generated from $\vb_i = \sigma(\mF^\top \va_i)$, instead of the standard Gaussian vectors $\lbrace \va_i \rbrace_{i=1}^m$ as in our case. In addition, the theoretical analysis in this paper is valid for a much more general family of convex loss functions and feature matrices. Moreover, this paper provides a precise characterization of the bias and variance terms of the generalization error for a general data model which extends the results in \cite{Ba2020Generalization} and rigorously verify the replica predictions in \cite{RF_rep_1}.


The rest of the paper is organized as follows. We summarize the main technical assumptions and theoretical predictions in Section \ref{pr_analysis}. To further illustrate these results, we present additional numerical examples in Section \ref{sim_res}. The derivations of our theoretical predictions are detailed in Section \ref{tech_deta}. Section \ref{concd} concludes the paper. Finally, the appendix collects the proofs of all the technical results introduced in previous sections.


\section{Precise Performance Analysis}\label{pr_analysis}

\subsection{Technical Assumptions}\label{ass_not}
The asymptotic predictions derived in this paper are based on the following technical assumptions.
\begin{enumerate}[label=\text{A.\arabic*}]
\item \label{itm:data_Gauss} The data vectors $\lbrace \va_i \rbrace_{i=1}^{m}$ are known and drawn independently from $\mathcal{N}(0, \mI_n)$. 
\item \label{itm:asy_lim} The number of samples and the number of hidden neurons satisfy $m=m(n)$ and $k=k(n)$ with $\alpha_n=m(n)/n \to \alpha>0$ and $\eta_n=k(n)/m(n) \to \eta>0$ as $n\to \infty$.
\item \label{itm:xi} The unknown signal $\vxi\in\mathbb{R}^n$ is independent of the feature matrix $\mF\in\mathbb{R}^{n\times k}$, where $\norm{\vxi}=\rho$ is known.
\item \label{itm:act_fun} The activation function $\sigma$ satisfies the conditions that $0 < \mathbb{E}[\sigma(z)^2]<+\infty$ and $ \mathbb{E}[z\sigma(z)]>0$, where $z\sim\mathcal{N}(0,1)$.
\item \label{itm:cost} The loss function $\widehat{\ell}$ defined in \eqref{reg_form} and \eqref{class_form} is a proper convex function in $\mathbb{R}$. Moreover, it satisfies the following three properties:
\begin{itemize}
\item[(1)] If the activation function $\sigma$ is not odd, the function $\widehat{\ell}$ is strongly convex in any compact set and strictly convex in $\mathbb{R}$. 
\item[(2)] If the activation function $\sigma$ is not odd, there exists a universal constant ${C}_{1}>0$ such that the loss function satisfies the following scaling condition
\begin{align}
\frac{1}{m}\abs{\mathcal{L}(\zeta \vec{1}_m+\vv)} \leq {C}_{1},\frac{\norm{\vv}^2}{m}\leq {C}_{1} \implies \abs{\zeta} \leq {C}_{1},\nonumber
\end{align}
with probability going to $1$ as $n$ goes to $+\infty$, where $\zeta\in\mathbb{R}$, $\vv\in\mathbb{R}^m$ and  $\mathcal{L}(\vx)=\sum_{i=1}^{m} \ell(y_i,x_i)$.
\item[(3)] There exists a universal constant ${C}_{2}>0$ such that the sub-differential set of the loss function satisfies the following scaling condition
\begin{align}
\norm{\vv} \leq {C}_{2} \sqrt{m} \implies \sup_{\vs\in \partial \mathcal{L}(\vv)} \norm{\vs} \leq {C}_{2} \sqrt{m},
\end{align}
with probability going to $1$ as $n$ goes to $+\infty$, where $\partial \mathcal{L}$ denotes the sub-differential set of the  function $\mathcal{L}$.
\end{itemize}
\item \label{itm:fun_fwf} The function $\varphi$ is independent of the data vectors $\lbrace \va_i \rbrace_{i=1}^{m}$ and generates independent and identically distributed labels $\lbrace y_i \rbrace_{i=1}^{m}$. Moreover, it satisfies the following property
\begin{align}\label{ass_pi0}
\mathbb{P}(y < 0) > 0,~\text{and}~\mathbb{P}(y > 0) > 0,
\end{align}
where $y$ is generated according to \eqref{cmodel}. We assume that the condition in \eqref{ass_pi0} is true only in the classification task. Furthermore, the functions $\varphi$ and $\widehat{\varphi}$ satisfy the following 
\begin{itemize}
\item[(1)] The function $\widehat{\varphi}$ is almost surely continuous in $\mathbb{R}$. Moreover,  $\varphi$ and $\widehat{\varphi}$ satisfy $0<\mathbb{E}({\varphi(z)}^2 )<+\infty$ and $0<\mathbb{E}({\widehat{\varphi}(z)}^2 )<+\infty$, where $z\sim\mathcal{N}(0,h)$ and $h>0$. 
\item[(2)] There exists a function $g$ such that $\abs{\widehat{\varphi}(\zeta+\chi x)}^2 \leq g(x)$ for any $x\in\mathbb{R}$, $\zeta$ and $\chi$ in compact sets. Furthermore, the function $g$ satisfies $\mathbb{E}({g(z)} )<+\infty$, where $z\sim\mathcal{N}(0,1)$.
\end{itemize} 
\item \label{itm:ass_F} Consider the following decomposition $\mF=\mU\mS\mV$, where $\mU\in\mathbb{R}^{n\times n}$ and $\mV\in\mathbb{R}^{k\times k}$ are orthogonal matrices and $\mS\in\mathbb{R}^{n\times k}$ is a diagonal matrix formed by the singular values of the feature matrix $\mF$. Then, the matrix $\mU$ is a Haar-distributed random unitary matrix. Define the matrix $\mT$ as follows 
\begin{align}
\mT=\begin{cases}
\mF\mF^\top &\text{if}~\delta > 1\\
\mF^\top\mF &\text{otherwise},
\end{cases}
\end{align}
where $\delta=k/n$. Define $\sigma_{\text{min}}(\mT)$ as the minimum eigenvalue of the matrix $\mT$ and $\sigma_{\text{max},1}(\mT)$ and $\sigma_{\text{max},2}(\mT)$ as the two largest eigenvalues of the matrix $\mT$ where $\sigma_{\text{max},2}(\mT) \leq \sigma_{\text{max},1}(\mT)$. Then, we have the following convergence in probability
\begin{align}
\begin{cases}
 \sigma_{\text{min}}(\mT) \xrightarrow{p} \kappa_{\text{min}},~\sigma_{\text{max},1}(\mT) \xrightarrow{p} \kappa_{\text{max}},\\
 \abs{\sigma_{\text{max},1}(\mT)-\sigma_{\text{max},2}(\mT)} \xrightarrow{p} 0.
 \end{cases}
\end{align}
Additionally, the empirical distribution of the eigenvalues of the matrix $\mT$ converges weakly to a probability distribution $\mathbb{P}_\kappa$ supported in $[\kappa_{\text{min}},\kappa_{\text{max}}]$, where $0< \kappa_{\text{min}}\leq \kappa_{\text{max}}< +\infty$.
\end{enumerate}
\begin{remark}
Assumption \ref{itm:asy_lim} also implies that $\delta_n=k(n)/n \to \delta>0$ as $n\to\infty$. Assumptions \ref{itm:cost} and \ref{itm:fun_fwf} are essential to proving our sharp asymptotic predictions. The first scaling property in Assumption \ref{itm:cost} corresponds to having $\lim_{\abs{x}\to +\infty} \widehat{\ell}(x)=+\infty$ for the loss functions of the regression task. Additionally, the first scaling property in Assumption \ref{itm:cost} combined with the condition in \eqref{ass_pi0} corresponds to having 
\begin{align}
\lim_{{x}\to +\infty} \widehat{\ell}(x)=+\infty,~\text{or}~\lim_{{x}\to -\infty} \widehat{\ell}(x)=+\infty,
\end{align} 
for the loss functions of the classification task. Assumption \ref{itm:fun_fwf} is also introduced to guarantee that the generalization error concentrates in the large system limit. Our theoretical analysis exploits the weak convergence of the empirical distribution of the eigenvalues of the matrix $\mT$ in \ref{itm:ass_F} to guarantee that the performance of the feature formulation given in \eqref{formulation} can be asymptotically characterized by a deterministic optimization problem. Our analysis shows that the deterministic optimization problem only depends on the asymptotic distribution of the eigenvalues of the matrix $\mT$ denoted by $\mathbb{P}_\kappa$.\\
Although our theoretical analysis is derived under the strong convexity property in Assumption \ref{itm:cost}, our simulation results show that our predictions are also valid for convex loss functions combined with not odd activation functions.

\end{remark}

\subsection{The Uniform Gaussian Equivalence Conjecture}\label{GEC}
The uniform Gaussian equivalence conjecture (uGEC) is a stronger version of an asymptotic equivalence theorem, referred to as the Gaussian equivalence theorem (GET), proved in \cite{RF_hmm_1}. Define a vector $\va$ with independent standard Gaussian entries, i.e. $\va\sim\mathcal{N}(0,\mI_n)$. Moreover, assume that the activation function $\sigma$ satisfies Assumption \ref{itm:act_fun}. Define the random variables $\nu_1$ and $\nu_2$ as follows
\begin{align}
\nu_1=\vxi^\top \va,~\nu_2=\vw^\top \sigma(\mF^\top \va),
\end{align}
where $\vxi\in\mathbb{R}^n$ and $\mF\in\mathbb{R}^{n\times k}$ are fixed and satisfy Assumption \ref{itm:xi} and Assumption \ref{itm:ass_F}, and where $\vw\in\mathbb{R}^k$ is a fixed vector. For fixed $\mF$, $\vxi$ and $\vw$, the Gaussian equivalence theorem (GET) shows that the random variables $\nu_1$ and $\nu_2$ are jointly Gaussian with mean vector $[0,\mu_0 \vw^\top \vec{1}_k]^\top$ and covariance matrix  
\begin{align}
\mGm_n=\begin{bmatrix}
\rho^2 & \mu_1 \vxi^\top \mF \vw \\
 \mu_1 \vxi^\top \mF \vw & \mu_1^2 \norm{\mF \vw}^2 + \mu_\star^2 \norm{\vw}^2
\end{bmatrix},
\end{align}
where $\mu_0=\mathbb{E}[\sigma(z)]$, $\mu_1=\mathbb{E}[z\sigma(z)]$ and $\mu_\star^2=\mathbb{E}[\sigma(z)^2]-\mu_0^2-\mu_1^2$, and where $z$ is a standard Gaussian random variable. This result is valid in the asymptotic regime, i.e. Assumption \ref{itm:asy_lim} is true and  $n\to \infty$. Note that the GET shows that the random variables $\nu_1$ and $\nu_2$ are statistically equivalent to the random variables $\nu_1$ and $\widehat{\nu}_2=\mu_0 \vw^\top \vec{1}_k+ \mu_1 \vw^\top\mF^\top\va+\mu_\star\vw^\top\vz$
in the large system limit, where $\vz \in\mathbb{R}^k$ is a standard Gaussian random vector independent of $\va$ and $\vec{1}_k$ is the all $1$ vector with size $k$.

This asymptotic result is valid for suitable choices of the feature matrix $\mF$. Reference \cite{RF_hmm_1} provides two balance conditions for  $\mF$ to ensure the Gaussian equivalence. This Gaussian equivalence property has also been mentioned and used in \cite{RF_monta_1,RF_monta_2}. 
Note that the GET is valid for fixed vectors $\vw\in\mathbb{R}^k$, $\vxi\in\mathbb{R}^n$ and matrix $\mF\in\mathbb{R}^{n\times k}$. In this paper, we require the validity of the GET uniformly in $\vw\in\mathbb{R}^k$. We conjecture that the GET can be extended to this stronger version which we refer to as the uniform Gaussian equivalence conjecture (uGEC). Using the uGEC, the asymptotic analysis of the feature formulation in \eqref{formulation} is equivalent to the analysis of the Gaussian formulation given in \eqref{GEC_for}. 
Similar conjecture is used in \cite{RF_rep_1}. Rigorously proving the uGEC is of interest and is left for future work. This conjecture is validated by extensive simulation examples. 
\subsection{Precise Analysis of the Feature Formulation}\label{pre_ana_fm}
In this section, we characterize the asymptotic behaviour of the generalization and training errors given in \eqref{testerr_def} and \eqref{trainerr_def} for general convex loss functions of the form given in \eqref{reg_form} and \eqref{class_form}. Before stating our asymptotic predictions, we introduce a few definitions. First, define the following min-max optimization problem
\begin{align}\label{scprob1}
\min_{\substack{  \beta \geq {\abs{q}}/{\sqrt{T_1}}\\ q\in\mathbb{R},\vartheta\in\mathbb{R}} } \sup_{t > -\theta }&~\frac{\lambda q^2}{2 T_1} (t+T_2-T_3(t)) -\frac{\lambda t \beta^2}{2}\nonumber\\
&+  \mathbb{E}\Big[ \mathcal{M}_{\widehat{\ell}} \Big( V({\vartheta,q,\beta}); \frac{ T_4(t)}{\lambda} Z \Big) \Big],
\end{align}
where $\widehat{\ell}$ is the loss function given in \eqref{reg_form} and  \eqref{class_form}, and the random variable $V({\vartheta,q,\beta})$ depends on the optimization variables $q$, $\beta$ and $\vartheta$ and is defined as follows for the regression task
\begin{align}
V({\vartheta,q,\beta})=&\beta H +\mu_0 \vartheta+\mu_1 q S- Y,\nonumber
\end{align}
and it can be expressed as follows for the classification task
\begin{align}
V({\vartheta,q,\beta})&=\beta Y H + \mu_0 Y \vartheta+\mu_1 q YS,\nonumber
\end{align}
where $S$ and $H$ are two independent standard Gaussian random variables and the random variable $Y$ depends on $S$ as follows
$Y=\varphi(\rho S)+\Delta \epsilon$,
where $\epsilon$ is a standard normal random variable. In \eqref{scprob1}, $Z=1$ for the regression task and $Z=Y^2$ for the classification task. Furthermore, the parameter $\theta$ satisfies $\theta=1/(\mu_1^2 \kappa_{\text{max}}+\mu_\star^2)$, where $\kappa_{\text{max}}$ is introduced in Assumption \ref{itm:ass_F}. The function $\mathcal{M}_{\widehat{\ell}}$ in the optimization problem \eqref{scprob1} denotes the Moreau envelope of the loss function $\widehat{\ell}$ given in \eqref{reg_form} and  \eqref{class_form} and is defined as follows
\begin{align}\label{m_env}
\mathcal{M}_{\widehat{\ell}}(a;x)=\min_{z\in\mathbb{R}}~\widehat{\ell}(z)+\frac{1}{2 x} \left( z-a \right)^2.
\end{align} 
The constant $T_1$ only depends on the asymptotic probability distribution $\mathbb{P}_\kappa$ introduced in Assumption \ref{itm:ass_F} and is given by
\begin{align}
T_1=\frac{e \mathbb{E}_\kappa\left[ {\kappa}/{(\mu_\star^2+\mu_1^2 \kappa)}\right] }{1-e+e \mu_\star^2 \mathbb{E}_\kappa\left[ {1}/{(\mu_\star^2+\mu_1^2 \kappa)}\right] },\nonumber
\end{align}
where $e=1$ if $\delta \geq 1$ and $e=\delta$ otherwise, and where the expectation is over the asymptotic probability distribution $\mathbb{P}_\kappa$. Based on Assumption \ref{itm:act_fun} and Assumption \ref{itm:ass_F}, note that $T_1>0$ which means that the optimization problem \eqref{scprob1} is well-defined. Also, $T_2$ can be expressed as follows
\begin{align}
T_2=\frac{ \mathbb{E}_\kappa\left[ {\kappa}/{(\mu_\star^2+\mu_1^2 \kappa)^2}\right]}{\left(1-e+e \mu_\star^2 \mathbb{E}_\kappa\left[ {1}/{(\mu_\star^2+\mu_1^2 \kappa)}\right] \right) \mathbb{E}_\kappa\left[ {\kappa}/{(\mu_\star^2+\mu_1^2 \kappa)}\right]}.\nonumber
\end{align}
Note that $T_2$ is a constant independent of the optimization variables in \eqref{scprob1}. For any feasible $t$, the function $T_3$ is defined as follows
\begin{align}
T_3(t)=T_2+t(1+\mu_1^2 T_1)-\frac{T_1}{e \mathbb{E}_\kappa\left[ {\kappa}/{(1+t\mu_\star^2+t\mu_1^2 \kappa)}\right]}.\nonumber
\end{align}
Moreover, $T_4$ depends on the optimization variable $t$ as follows
\begin{align}
T_4(t)=\frac{\eta}{d} \mathbb{E}_\kappa\left[ \frac{\mu_\star^2+\mu_1^2 \kappa}{1+t(\mu_\star^2+\mu_1^2 \kappa)}\right] + \eta\left(1-\frac{1}{d} \right) \frac{\mu_\star^2}{1+t \mu_\star^2},\nonumber
\end{align}
where $d=\delta$ if $\delta \geq 1$ and $d=1$ otherwise. Now, we are ready to state our main theoretical predictions.
\begin{theorem}\label{ther1}
Suppose that the assumptions in Section \ref{ass_not} are satisfied and the uGEC holds true. Then, the training error defined in \eqref{trainerr_def} converges in probability as follows 
\begin{align}\label{conv_trr_err}
\mathcal{G}_{n,\text{train}} \xrightarrow{~p~} C^\star_\ell(\lambda),
\end{align}
where $C^\star_\ell(\lambda)$ denotes the optimal cost value of the deterministic optimization problem \eqref{scprob1}.
Moreover, the generalization error given in \eqref{testerr_def} converges in probability as follows 
\begin{align}\label{conv_tst_err}
{\mathcal{G}}_{n,\text{test}} \xrightarrow{~p~} \frac{1}{4^\upsilon} \mathbb{E}\left[ \left( \varphi(\nu_1) -\widehat{\varphi}(\nu_2) \right)^2 \right],
\end{align}
where $\nu_1$ and $\nu_2$ have a bivariate Gaussian distribution with  mean vector $[0,\mu_0 \vartheta^\star]^\top$ and covariance matrix given by
\begin{align}
\mGm=\begin{bmatrix}
\rho^2 & \mu_1 \rho q^{\star} \\
 \mu_1 \rho q^\star &  \mu_1^2 (q^{\star})^2+(\beta^\star)^2
\end{bmatrix},\nonumber
\end{align}
where $q^\star$, $\beta^\star$ and $\vartheta^\star$ are the optimal solutions of \eqref{scprob1}.
\end{theorem}
The detailed proof of Theorem \ref{ther1} is provided in Section \ref{tech_deta}. Theorem \ref{ther1} accurately predicts the training and generalization errors of the feature formulation \eqref{formulation} in the high-dimensional limit. Note that our theoretical predictions require the strict and strong convexity properties only when the activation function is not an odd function, i.e. when $\mu_0\neq 0$. Moreover, the results presented in Theorem \ref{ther1} are valid for general activation function. To illustrate our theoretical results, we consider two applications: a non-linear regression model and a binary classification model.

\subsection{Application I: Regression Model}
\label{reg_model}
Consider a non-linear regression model where the labels $\lbrace y_i \rbrace_{i=1}^{m}$ are generated according to the following model 
\begin{align}
y_i=\max(\va_i^\top \vxi,0)+ \Delta \epsilon_i, \forall i\in\lbrace 1,\dots,m\rbrace,
\end{align}
where $\lbrace \epsilon_i \rbrace_{i=1}^{m}$ are independent and drawn from a standard Gaussian distribution. In this model, the function $\varphi$ introduced in \eqref{cmodel} is a deterministic function and it satisfies Assumption \ref{itm:fun_fwf}. Moreover, assume that the loss function $\widehat{\ell}$ is the squared loss, i.e.
\begin{align}\label{reg_loss}
{\ell}(y,z)=\frac{1}{2}(y-z)^2.
\end{align}
Note that the considered loss function satisfies Assumption \ref{itm:cost}. Furthermore, the formulation in \eqref{scprob1} can be simplified as follows
\begin{align}\label{scprob1_sim_reg}
&\min_{\substack{ q, \beta \geq {\abs{q}}/{\sqrt{T_1}} } } \sup_{t > -\theta}~\frac{\lambda q^2}{2 T_1} (t+T_2-T_3(t))  -\frac{\lambda t \beta^2}{2} \nonumber\\
&+\frac{\lambda}{2 T_4(t)+2\lambda} \Big( \gamma_1 + \beta^2+\mu_1^2 q^2-2\mu_1 q \gamma_2 - \gamma_3^2 \Big),
\end{align}
where the constants $\gamma_1$, $\gamma_2$ and $\gamma_3$ are defined as $\gamma_1=\mathbb{E}[Y^2]$, $\gamma_2=\mathbb{E}[Y S]$ and $\gamma_3=\mathbb{E}[Y]$, and therefore, are given by 
\begin{align}
\gamma_1=\rho^2\chi_2+\Delta^2,\gamma_2=\rho \chi_1,\gamma_3=\rho \chi_0,
\end{align}
where $\chi_2=\mathbb{E}[\max(z,0)^2]$, $\chi_1=\mathbb{E}[z\max(z,0)]$, and $\chi_0=\mathbb{E}[\max(z,0)]$, and $z$ is a standard Gaussian random variable.
The optimal solution $\vartheta^\star$ satisfies $\vartheta^\star=\gamma_3/\mu_0$ if $\mu_0\neq 0$ and $\vartheta^\star=0$ otherwise. 
Based on Lemma \ref{convx_prop}, the cost function of the deterministic optimization problem \eqref{scprob1_sim_reg} is jointly strongly convex in the variables $q$ and $\beta$, then, it can be efficiently solved. We assume that the function $\widehat{\varphi}$ is the identity function. According to Theorem \ref{ther1}, the generalization error given in \eqref{testerr_def} converges in probability as follows 
\begin{align}
\mathcal{G}_{n,\text{test}} \xrightarrow{~p~}& \rho^2 \chi_2 - 2 \mu_1 \chi_1 \rho q^\star+\mu_1^2 (q^{\star})^2+(\beta^\star)^2\nonumber\\
&-2 \rho \mu_0 \chi_0 \vartheta^\star+\mu_0^2 (\vartheta^\star)^2, \nonumber
\end{align}
where $q^\star$ and $\beta^\star$ are the optimal solutions of the  asymptotic optimization problem formulated in \eqref{scprob1_sim_reg}. 
\subsection{Application II: Binary Classification Model}\label{class_model}
In the second application, we consider a probabilistic model. Assume that the data $\lbrace y_i \rbrace_{i=1}^{m}$ is binary and generated according to the following probabilistic model 
\begin{align}\label{bn_mode}
y_i=\begin{cases}
\sign(\va_i^\top \vxi) & \text{with~probability}~1-p\\
-\sign(\va_i^\top \vxi) & \text{with~probability}~p,
\end{cases}
\end{align}
where $0 \leq p \leq 1/2$. In this model, the function $\varphi$ introduced in \eqref{cmodel} is a probabilistic function and it satisfies Assumption \ref{itm:fun_fwf}. 
We consider three convex loss functions, i.e. the hinge loss, the least absolute deviation (LAD) loss and the logistic loss, given by, respectively,
\begin{align}\label{loss_class}
\begin{cases}
{\ell}(y,x)=\max(1-yx,0)\\
{\ell}(y,x)=\abs{1-yx}\\
{\ell}(y,x)=\log(1+e^{-yx}).
\end{cases}
\end{align}
Note that the logistic loss satisfies Assumption \ref{itm:cost}. Although the statement in Theorem \ref{ther1} requires the strict and local strong convexity, we show empirically that our results are also valid for the hinge loss and LAD loss combined with a not odd activation function.
The Moreau envelope of the hinge loss and the LAD loss can be determined in closed-form. Moreover, the scalar optimization problem given in \eqref{scprob1} can be solved numerically. The objective is to predict the correct sign of any unseen sample $y_{\text{new}}$. Then, we fix the function $\widehat{\varphi}$ to be the sign function. If $\vartheta^\star=0$ or $\mu_0=0$, the generalization error given in \eqref{testerr_def} converges in probability as follows 
\begin{align}
\mathcal{G}_{n,\text{test}} \xrightarrow{~p~} p+ \frac{1-2p}{\pi} \cos^{-1}\hspace{-1mm}\Big( \frac{\mu_1 q^\star }{\sqrt{\mu_1^2 (q^{\star})^2+(\beta^\star)^2 }} \Big),
\end{align}
where $q^\star$, $r^\star$, and $\vartheta^\star$ are the optimal solutions of the scalar optimization problem given in \eqref{scprob1}. 

\section{Simulation Results}\label{sim_res}
In this section, we provide additional simulation examples to validate our theoretical predictions given in Theorem \ref{ther1}. We consider the following two general forms of the feature matrix $\mF$ that satisfy the regularity assumptions introduced in \ref{itm:ass_F}.
\begin{enumerate}[label=\text{G.\arabic*}]
\item \label{RGFM} The columns of the feature matrix $\mF\in\mathbb{R}^{n\times k}$ are independent and drawn from a Gaussian distribution with zero mean and covariance matrix $\frac{1}{n} \mI_n$. In this case, we refer to $\mF$ as the Gaussian feature matrix.
\item \label{ROFM} The feature matrix $\mF$ can be decomposed as follows $\mF= \mU \mD \mV$, where $\mU\in\mathbb{R}^{n\times n}$ and $\mV\in\mathbb{R}^{k\times k}$ are two random orthogonal matrices and where $\mD$ is a diagonal matrix with diagonal entries $d=\max(\sqrt{\delta},1)$. In this case, we refer to $\mF$ as the random orthogonal feature matrix.
\end{enumerate}
In \ref{ROFM}, the singular-values of the feature matrix $\mF$ are uniformly equal to $d=\max(\sqrt{\delta},1)$ to guarantee a fair comparison with  the Gaussian feature matrix. To illustrate our theoretical predictions, we consider two different models: the non-linear regression model discussed in Section \ref{reg_model} and the binary classification model presented in Section \ref{class_model}.
\subsection{Regression Model}
\begin{figure}[h!]
    \centering
    \subfigure[]{\label{fig:nmse_1}
    \includegraphics[width=0.47\linewidth]{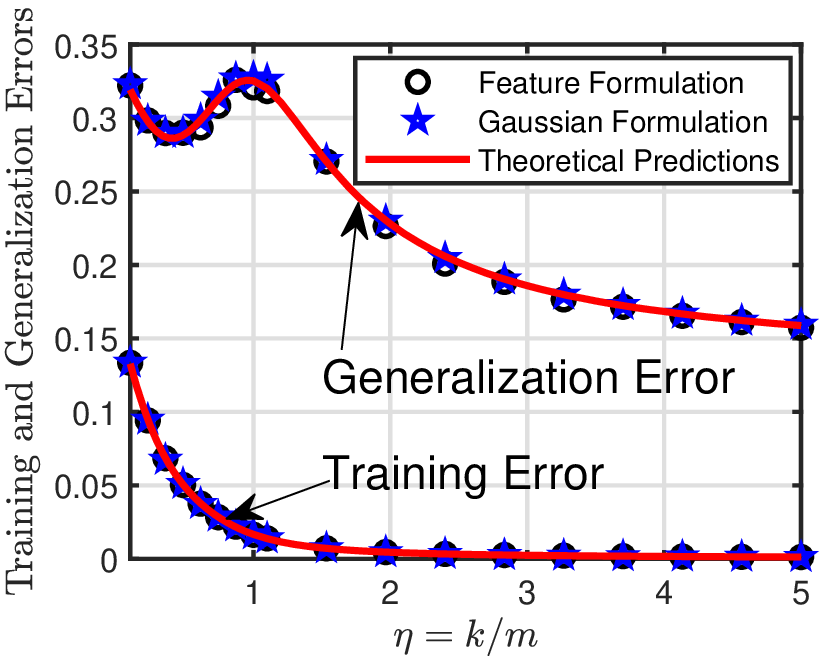}
    }
    \subfigure[]{\label{fig:nmse_2}
        \includegraphics[width=0.47\linewidth]{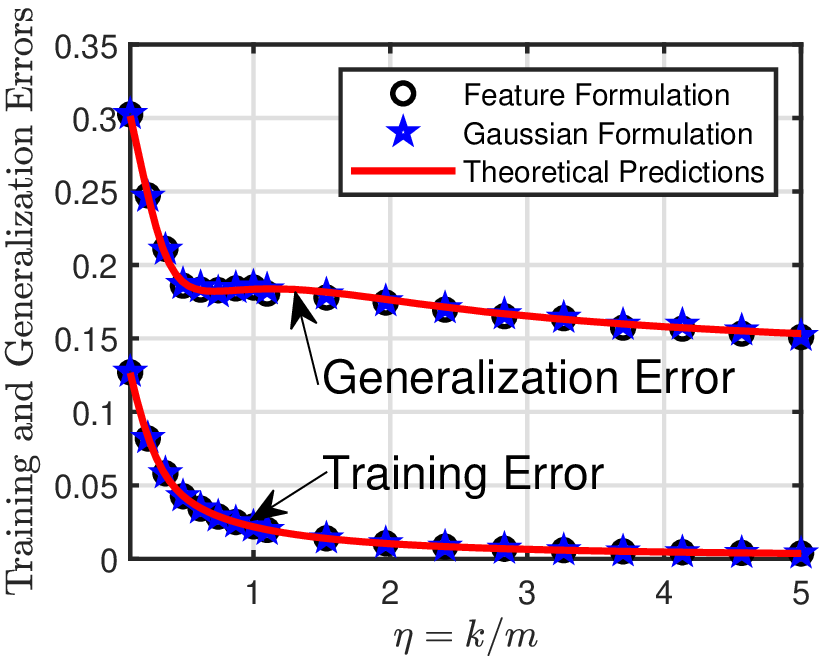}}

    \caption{Theoretical predictions v.s. numerical simulations. (a)  $\sigma$ is the ReLu activation function, i.e. $\sigma:x\to \max(x,0)$; (b) $\sigma$ is the SoftPlus activation function, i.e. $\sigma:x\to \log(1+\text{e}^{x})$. The functions $\varphi$ and $\widehat{\varphi}$ are as given in Section \ref{reg_model}, $\widehat{\ell}$ is the squared loss. Moreover, $\lambda=10^{-2}$, and $\Delta=0.05$. The sampling ratio $\alpha=m/n=3$ and the hidden signal $\vxi$ has norm $\rho=1$. The feature matrix $\mF$ is the Gaussian feature matrix. The results are averaged over $50$ independent Monte Carlo trials and we set $n=400$. }
        \label{fig1}
\end{figure}

In the first simulation example, we consider the regression model introduced in Section \ref{reg_model}. Figure \ref{fig1} compares the numerical predictions and our theoretical predictions given in Theorem \ref{ther1}.  
First, our theoretical predictions summarized in Section \ref{reg_model} are in excellent agreement with the actual performance of the learning problem \eqref{formulation} and its Gaussian formulation \eqref{GEC_for}. Moreover, observe that the SoftPlus activation function outperforms the ReLu activation function in the sense that it provides a lower generalization error. Figure \ref{fig1} also reveals the important role played by the activation function in reducing the generalization error and in the mitigation of the double descent phenomenon. Specifically, it suggests that an optimized activation function can significantly improve the generalization error and reduce the interpolation threshold peak. Furthermore, Figure \ref{fig1} shows that the performance of the Gaussian formulation matches the performance of the feature formulation, which validates the conjecture discussed in Section \ref{GEC}.
\subsection{Classification Model}
\begin{figure}[h!]
    \centering
    \subfigure[]{\label{fig:nmse_1}
    \includegraphics[width=0.47\linewidth]{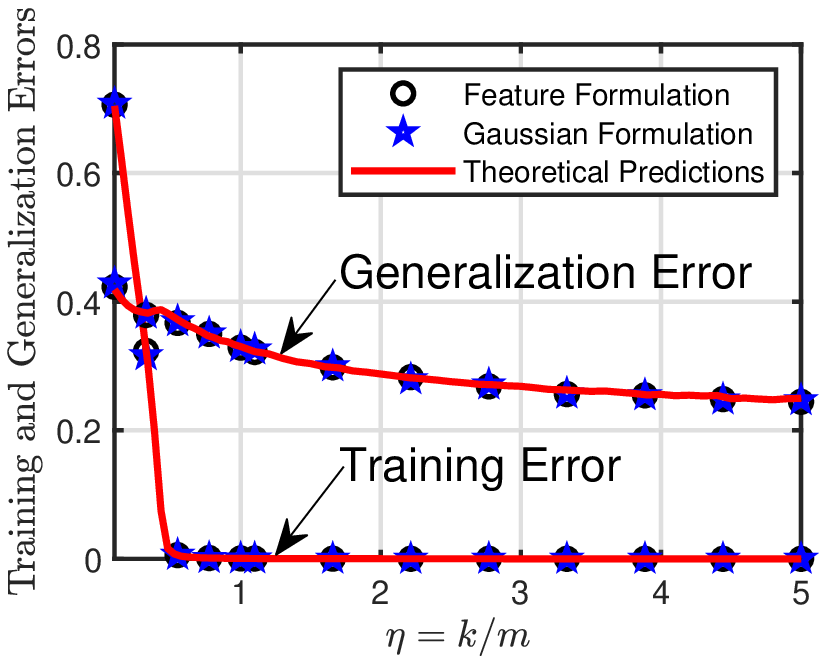}
    }
    \subfigure[]{\label{fig:nmse_2}
        \includegraphics[width=0.47\linewidth]{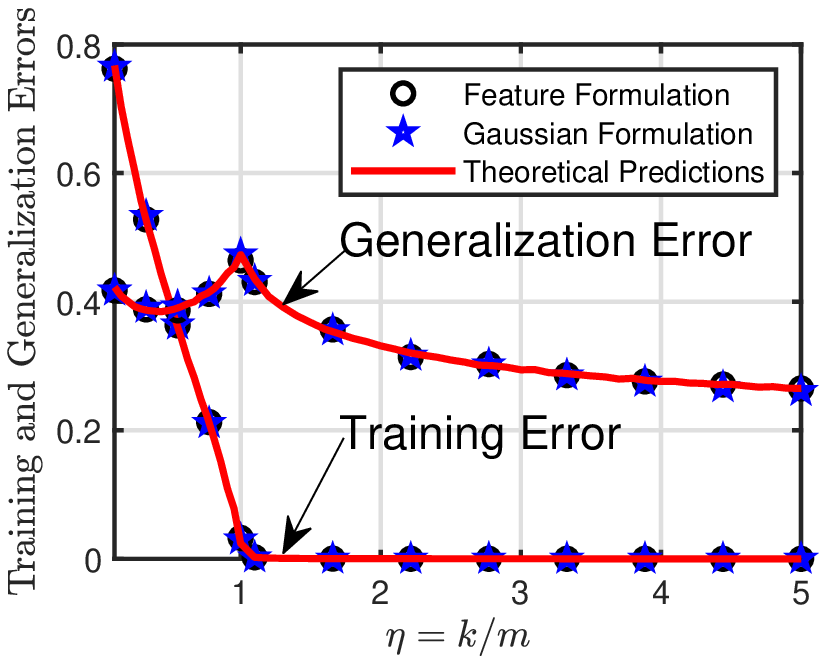}}

    \caption{Theoretical predictions v.s. numerical simulations. (a)  The loss function $\widehat{\ell}$ is the hinge loss; (b) The loss function $\widehat{\ell}$ is the LAD loss. The sampling ratio $\alpha=m/n=3$, $\Delta=0$ and $\lambda=10^{-4}$. The functions $\varphi$ and $\widehat{\varphi}$ are as given in Section \ref{class_model}, $\sigma$ is the binary step activation function, i.e. $\sigma(x)=1$, if $x\geq 0$ and $\sigma(x)=0$ otherwise, and the probability $p$ in \eqref{bn_mode} is set to $p=0.05$. The feature matrix $\mF$ is the Gaussian feature matrix. The hidden signal $\vxi$ has norm $\rho=1$. The results are averaged over $50$ independent Monte Carlo trials and we set $n=200$.}
        \label{fig2}
\end{figure}

Now, we focus on the binary classification problem discussed in Section \ref{class_model}. A comparison between the numerical simulation and the CGMT theoretical predictions is provided in Figure \ref{fig2}. 
Our simulation example shows again that the CGMT predictions match perfectly the actual performance of the feature formulation given in \eqref{formulation} and its Gaussian formulation. Moreover, observe that the hinge loss provides a lower generalization error as compared to the LAD loss. Figure \ref{fig2} also shows that the generalization error of the LAD loss follows a double descent curve with a higher interpolation threshold peak as compared to the hinge loss. This shows the important role played by the loss function in the mitigation of the double descent phenomenon. Additionally, Figure \ref{fig2} shows that the theoretical predictions in Theorem \ref{ther1} are valid even when we relax the strict and strong convexity properties considered in Assumption \ref{itm:cost}. Again, the conjecture discussed in Section \ref{GEC} is validated by observing that the performance of the Gaussian formulation is in excellent agreement with the performance of the feature formulation.
\subsection{Double Descent Phenomenon}
\begin{figure}[h!]
    \centering
    \subfigure[]{\label{fig:nmse_1}
    \includegraphics[width=0.47\linewidth]{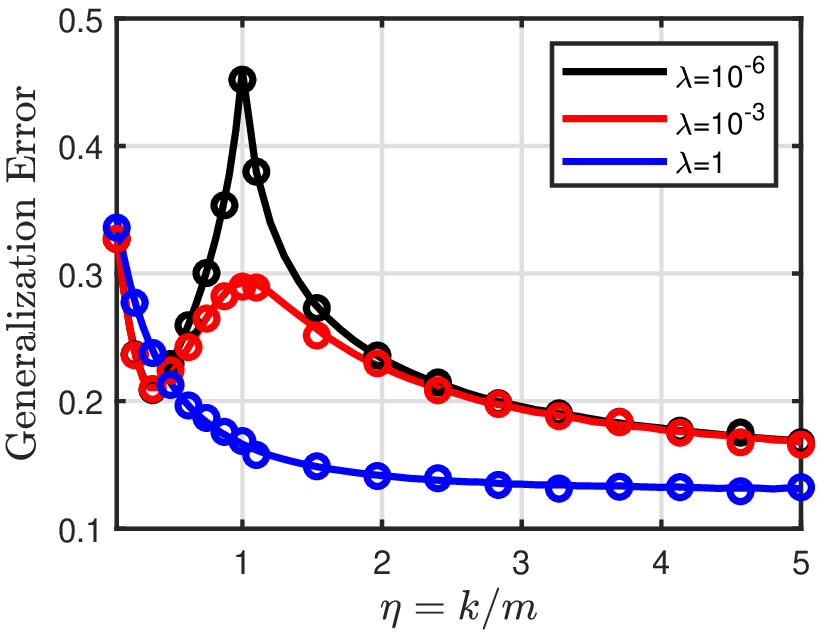}
    }
    \subfigure[]{\label{fig:nmse_2}
        \includegraphics[width=0.47\linewidth]{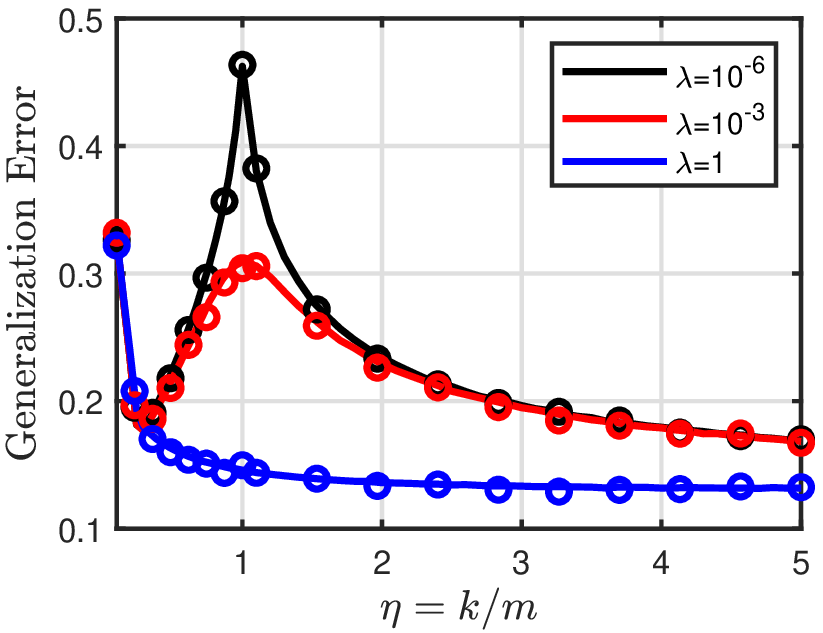}}

    \caption{Theoretical predictions v.s. numerical simulations. Continuous line: CGMT predictions, circles: numerical simulation (feature formulation). (a) $\mF$ is the Gaussian feature matrix and the activation function is $\sigma=\tanh$; (b) $\mF$ is the orthogonal feature matrix and the activation function is $\sigma=\erf$. The functions $\varphi$ and $\widehat{\varphi}$ are the sign function, $\widehat{\ell}$ is the squared loss, and $\Delta=0$. The sampling ratio $\alpha=m/n=4$. The hidden signal $\vxi$ has norm $\rho=1$. The results are averaged over $50$ independent Monte Carlo trials and we set $n=200$.}
        \label{fig_dd}
\end{figure}

In this part, we provide a simulation example to illustrate the double descent phenomenon in the binary classification problem. Figure \ref{fig_dd} considers the squared loss and the $\sign$ activation function.  
First, note that our theoretical predictions match the actual performance of the considered problem in \eqref{formulation}. Figure \ref{fig_dd} shows that the generalization error follows a U-shaped curve for small model complexity $\eta$. Then, after reaching a peak, the generalization error decreases monotonically as a function of the model complexity $\eta$. Moreover, note that the interpolation threshold decreases for larger values of $\lambda$. Figure \ref{fig_dd} also shows that $\lambda=1$ provides the best performance. In particular, it leads to a monotonically decreasing generalization error. This matches the results stated in \cite{opt_reg} where the authors show that optimal regularization can mitigate double descent.

\section{Technical Details: Analysis of the Random Feature Formulation}\label{tech_deta}
In this section, we use the CGMT framework \cite[Section~6]{chris:151} to precisely analyze the feature formulation given in \eqref{formulation} under the assumptions introduced in Section \ref{pr_analysis}.  In the rest of the paper, we suppose that the assumptions provided in Section \ref{pr_analysis} are all satisfied. Assumption \ref{itm:ass_F} also supposes that there exist two constants $c_F>0$ and $C_F>0$ such that the maximum and minimum eigenvalues of the matrix $\mT$, denoted by $\sigma_{\text{max}}(\mT)$ and $\sigma_{\text{min}}(\mT)$, satisfies 
\begin{align}
c_F \leq \sigma_{\text{min}}(\mT) \leq \sigma_{\text{max}}(\mT)\leq C_F,
\end{align}
on events with probability going to $1$ when $n$ goes to $+\infty$. Then, it suffices to prove our theoretical results conditioned on those events. 
\subsection{Technical Tool: Convex Gaussian Min-Max Theorem}\label{CGMT_fram}
The CGMT replaces the precise analysis of a generally hard primary optimization (PO) problem with a simplified auxiliary optimization (AO) problem. The CGMT considers primary problems of the following form
\begin{equation}\label{eq:PO}
\Phi(\mB)=\min\limits_{\vw \in \mathcal{S}_{\vw}} \max\limits_{\vu\in\mathcal{S}_{\vu}} \vu^\top \mB \vw + \psi(\vw,\vu),
\end{equation}
and formulates the corresponding AO problem as follows
\begin{equation}\label{eq:AO}
\hspace{-1mm}\phi(\vg,\vh)=\min\limits_{\vw \in \mathcal{S}_{\vw}} \max\limits_{\vu\in\mathcal{S}_{\vu}}  \norm{\vu} \vg^\top \vw +\norm{\vw} \vh^\top \vu  + \psi(\vw,\vu).\nonumber
\end{equation}
Before showing the equivalence between the PO and AO, the CGMT assumes that $\mB\in\mathbb{R}^{\widetilde{m}\times \widetilde{n}}$, $\vg \in \mathbb{R}^{\widetilde{n}}$ and $\vh\in\mathbb{R}^{\widetilde{m}}$, all have i.i.d standard normal entries, the feasibility sets $\mathcal{S}_{\vw}\subset\R^{\widetilde{n}}$ and $\mathcal{S}_{\vu}\subset\R^{\widetilde{m}}$ are convex and compact, and the function $\psi: \mathbb{R}^{\widetilde{n}} \times \mathbb{R}^{\widetilde{m}} \to \mathbb{R}$ is continuous \emph{convex-concave} on $\mathcal{S}_{\vw}\times \mathcal{S}_{\vu}$. Moreover, the function $\psi$ is independent of the matrix $\mB$. Under these assumptions, the CGMT \cite[Theorem 6.1]{chris:151} shows that for any $\mu \in \mathbb{R}$ and $t>0$, it holds
\begin{equation}\label{eq:cgmt}
\mathbb{P}\left( \abs{\Phi(\mB)-\mu} > t\right) \leq 2 \mathbb{P}\left(  \abs{\phi(\vg,\vh)-\mu} > t \right).
\end{equation}
The CGMT uses \eqref{eq:cgmt} and strict convexity conditions for the AO problem to prove that concentration of the set of optimal solutions of the AO implies concentration of the set of optimal solutions of the PO problem to the same set. 
Therefore, the CGMT allows us to analyze the generally easy AO to infer asymptotic properties of the generally hard PO. Next, we use the CGMT \cite{chris:151,chris:152} to rigorously prove the technical results presented in Theorem \ref{ther1}. 
\subsection{Random Feature Model Analysis}
In this part, we prove the asymptotic predictions stated in Theorem \ref{ther1}. Specifically, the objective is to precisely analyze the  following feature formulation denoted by $\mathfrak{{V}}_{n,1}$ using the CGMT framework
\begin{align}\label{main_prob}
\mathfrak{{V}}_{n,1}:\min_{\vw\in\mathbb{R}^k} \frac{1}{m} \sum_{i=1}^{m} \ell\left(y_i;\vw^\top \sigma(\mF^\top \va_i) \right)+\frac{\lambda}{2} \norm{\vw}^2.
\end{align}
Given that the assumptions in Section \ref{pr_analysis} are all satisfied, it suffices to analyze the Gaussian formulation  \eqref{GEC_for} to fully characterize the training and generalization errors of the feature formulation. Our approach is to formulate and simplify the auxiliary problem corresponding to the formulation in  \eqref{GEC_for}.
\subsubsection{Gaussian Equivalent Problem}
The asymptotic analysis of the feature formulation \eqref{main_prob} is equivalent to the asymptotic analysis of the following Gaussian formulation
\begin{align}\label{gmain_prob}
\hspace{-2mm}\mathfrak{{V}}^g_{n,1}:\min_{\vw\in\mathbb{R}^k}& \frac{1}{m} \sum_{i=1}^{m} \ell\left(y_i; \mu_0 \vw^\top\vec{1}_k+\vw^\top\vp_i  \right)+\frac{\lambda}{2} \norm{\vw}^2,
\end{align}
where $\vp_i=\mu_1 \mF^\top \va_i + \mu_{\star} \vz_i$, for any $i\in\lbrace 1,\dots,m \rbrace$ and where $\mu_0=\mathbb{E}[\sigma(z)]$, $\mu_1=\mathbb{E}[z\sigma(z)]$, $\mu^2_{\star}=\mathbb{E}[\sigma(z)^2]-\mu_0^2-\mu_1^2$, the vectors $\lbrace \vz_i \rbrace_{i=1}^{m}$ are drawn independently from a standard Gaussian distribution and where the data vectors $\lbrace \va_i \rbrace_{i=1}^{m}$ are independent from the vectors  $\lbrace \vz_i \rbrace_{i=1}^{m}$. Based on Assumption \ref{itm:cost}, the formulation in \eqref{gmain_prob} is strongly convex where $\lambda$ is a strong convexity parameter. Furthermore, the cost function of the optimization problem \eqref{gmain_prob} is a proper and continuous  function. This means that  \eqref{gmain_prob} attains its minimum in the interior of the feasibility set. An essential assumption in the CGMT framework \cite[Theorem 6.1]{chris:151} is the compactness of the feasibility sets. The following lemma shows that the unique optimal solution of the unconstrained formulation in $\mathfrak{{V}}^g_{n,1}$ belongs to a compact set.
\begin{lemma}[Primal Compactness]
\label{prl_comp}
Assume that $\widehat{\vw}_n \in\mathbb{R}^k$ is the unique optimal solution of the optimization problem given in \eqref{gmain_prob}. Then, there exist large positive constants $\widehat{C}_w>0$, and $\widehat{C}_\vartheta>0$ independent of $n$ such that
\begin{align}
\mathbb{P}\Big( \norm{ \widehat{\vw}_n } \leq \widehat{C}_w \Big) \xrightarrow{n \to \infty} 1,~\mathbb{P}\Big( \abs{\vec{1}_k^\top \widehat{\vw}_n} \leq \widehat{C}_\vartheta \Big) \xrightarrow{n \to \infty} 1,\nonumber
\end{align}
where the second asymptotic result is valid only when $\mu_0 \neq 0$.
\end{lemma}
The proof of Lemma \ref{prl_comp} is deferred to Appendix \ref{pf_prl_comp}. The asymptotic result stated in Lemma \ref{prl_comp} shows that the analysis of the formulation in \eqref{gmain_prob} is equivalent to studying the properties of the following constrained optimization problem
\begin{align}\label{gmain_prob2}
\hspace{-2mm}\mathfrak{{V}}^g_{n,2}:\hspace{-1mm}\min_{ \vw \in \mathcal{F}_{\vw} }& \frac{1}{m} \sum_{i=1}^{m} \ell\left(y_i;  \mu_0 \vw^\top\vec{1}_k+\vw^\top\vp_i \right) +\frac{\lambda}{2} \norm{\vw}^2,
\end{align}
in the large system limit, where the feasibility set $\mathcal{F}_{\vw}$ is defined as follows
\begin{align}
\mathcal{F}_{\vw}=\lbrace \vw\in\mathbb{R}^k:~\norm{ {\vw} } \leq \widehat{C}_w,~\abs{\vec{1}_k^\top {\vw} } \leq \widehat{C}_\vartheta  \rbrace,
\end{align}
and $\widehat{C}_w>0$, and $\widehat{C}_\vartheta>0$ are any two large positive constants independent of $n$ and guarantee the result in Lemma \ref{prl_comp}. 
The optimization problem formulated in \eqref{gmain_prob2} can be expressed in terms of two independent optimization variables. Before presenting this theoretical result, define the following optimization problem
\begin{align}\label{gmain_prob3}
\mathfrak{{V}}^g_{n,3}:\min_{\substack{~\norm{ {\vw} } \leq \widehat{C}_w \\ \abs{\vartheta} \leq \widehat{C}_\vartheta  }}& \frac{1}{m} \sum_{i=1}^{m} \ell\left(y_i;\mu_0 \vartheta + \vw^\top (\mu_1 \mF^\top \va_i + \mu_{\star} \vz_i)  \right) \nonumber\\
&+{\lambda}/{2} \norm{\vw}^2.
\end{align}
Note that the formulation in \eqref{gmain_prob3} replaces the term $\vw^\top \vec{1}_k$ corresponding to the mean $\mu_0$ in the formulation \eqref{gmain_prob2} by an additional optimization variable $\vartheta$, independent of the vector $\vw$. 
Clearly, the formulations in \eqref{gmain_prob2} and \eqref{gmain_prob3} are equivalent when the activation function is odd, i.e. $\mu_0=0$. The following proposition rigorously proves that they are asymptotically equivalent for general activation function.
\begin{proposition}[High-dimensional Equivalence I]\label{prop_in_v}
Define $\mathcal{S}^{g}_{n,2}$ and ${\mathcal{S}}^{g}_{n,3}$ as the sets of optimal solutions of the minimization problems in $\mathfrak{{V}}^g_{n,2}$ and $\mathfrak{{V}}^g_{n,3}$, as follows
\begin{align}
&\mathcal{S}^{g}_{n,2}=\lbrace (\widehat{\vw},\widehat{\vartheta}): \widehat{\vartheta}=\widehat{\vw}^\top \vec{1}_k, \widehat{\vw}~\text{is optimal for}~\mathfrak{{V}}^g_{n,2} \rbrace\nonumber\\
&\mathcal{S}^{g}_{n,3}=\lbrace (\widetilde{\vw},\widetilde{\vartheta}): \widetilde{\vartheta}~\text{and} ~\widetilde{\vw}~\text{are optimal for}~\mathfrak{{V}}^g_{n,3} \rbrace.\nonumber
\end{align}
Moreover, let $O^{g}_{n,2}$ and $O^{g}_{n,3}$ be the optimal objective values of the optimization problems $\mathfrak{{V}}^g_{n,2}$ and $\mathfrak{{V}}^g_{n,3}$, respectively. Then, the following convergence in probability holds
\begin{equation}
\abs{ O^{g}_{n,3}-O^{g}_{n,2} } \overset{p}{\longrightarrow} 0,~\text{and}~\mathbb{D}( {\mathcal{S}}^{g}_{n,3},{\mathcal{S}}^{g}_{n,2} )  \overset{p}{\longrightarrow} 0,
\end{equation}
where $\mathbb{D}( \mathcal{A},\mathcal{B} )$ denotes the deviation between the sets $\mathcal{A}$ and $\mathcal{B}$ and is defined as $\mathbb{D}( \mathcal{A},\mathcal{B} )=\sup_{\vx_1\in\mathcal{A}} \inf_{\vx_2\in\mathcal{B}} \norm{\vx_1-\vx_2}$.
\end{proposition}
The proof of Proposition \ref{prop_in_v} is provided in Appendix \ref{pf_prop_in_v}. Based on Proposition \ref{prop_in_v}, it suffices to analyze the optimization problem $\mathfrak{{V}}^g_{n,3}$ using the CGMT framework.
\subsubsection{Formulating the Primary and Auxiliary Optimization Problems}
Based on the asymptotic results stated in Lemma \ref{prl_comp} and Proposition \ref{prop_in_v}, there exist three sufficiently large constants $C_w>0$, $C_q>0$ and $C_\vartheta>0$ such that the asymptotic analysis of the following formulation
\begin{align}\label{cons_po}
\mathfrak{{V}}^g_{n,4}:\min_{ \substack{\vw\in\mathcal{P}_{\vw}\\ \abs{\vartheta}\leq C_\vartheta } }&~ \frac{1}{m} \sum_{i=1}^{m} \ell \Big(y_i;\mu_0 \vartheta + \vw^\top (\mu_1 \mF^\top \va_i + \mu_{\star} \vz_i)  \Big) \nonumber\\
&+{\lambda}/{2} \norm{\vw}^2,
\end{align}
is equivalent to the asymptotic analysis of the feature formulation given in \eqref{main_prob},
where the feasibility set $\mathcal{P}_{\vw}$ is defined as follows
\begin{align}
\mathcal{P}_{\vw}=\Big\lbrace \vw \in\mathbb{R}^k :~ \norm{ \mM^{\frac{1}{2}} {\vw} } \leq C_w, \abs{ \vc^\top \vw} \leq C_q \Big\rbrace,
\end{align}
where $\mM=\mu_1^2 \mF^\top \mP^{\perp}_{\vxi} \mF+\mu^2_\star \mI_k$, $\vc= \mF^\top  \bar{\vxi}$, $\mP^{\perp}_{\vxi}=\mI_n-\bar{\vxi} {\bar{\vxi}}^\top$ denotes the projection matrix onto the orthogonal complement of the space spanned by the vector $\vxi$ and where $\bar{\vxi}=\vxi/\norm{\vxi}$. Based on \cite[Corollary 1.10]{var_ana}, the formulation in \eqref{cons_po} has a unique optimal $\vw$ for any fixed feasible $\vartheta$.
To simplify the analysis, we show in the following proposition that one can analyze the optimization problem given in \eqref{cons_po} for any fixed feasible $\vartheta$, then, minimize its asymptotic limit over the scalar $\vartheta$ to infer the asymptotic properties of the feature formulation.
\begin{proposition}[Fixed Scalar Variable]\label{in_vartheta}
Assume that $\vartheta$ is in the feasibility set $\lbrace \vartheta: \abs{\vartheta} \leq C_\vartheta \rbrace$ and $\mu_0\neq 0$. Define the set $\mathcal{S}_{n,\vartheta,\epsilon}$ as follows
\begin{align}
\mathcal{S}_{n,\vartheta,\epsilon}=&\Big\lbrace \vw\in\mathbb{R}^k: \abs{q_n-q^\star_\vartheta } \leq \epsilon, \abs{\beta_n-\beta^\star_\vartheta} \leq \epsilon;\nonumber\\
&~~~q_n=\bar{\vxi}^\top\mF \vw, {\beta}_n=\sqrt{{\vw}^\top \mM {\vw}} \Big\rbrace,
\end{align}
for a fixed $\epsilon>0$, where $q^\star_\vartheta$ and $\beta^\star_\vartheta$ are two deterministic constants.
Moreover, assume that $\Phi_n(\vartheta)$ and $\widehat{\vw}_{n,\vartheta}$ are the optimal cost and the optimal solution of the formulation in $\mathfrak{{V}}^g_{n,4}$ for fixed feasible $\vartheta$. Assume that the following properties are all satisfied
\begin{itemize}
\item[(1)] There exists a constant $\phi(\vartheta)$ such that the optimal cost $\Phi_n(\vartheta)$ converges in probability to $\phi(\vartheta)$ as $n$ goes to $+\infty$, for any feasible $\vartheta$.
\item[(2)] The event $\lbrace \widehat{\vw}_{n,\vartheta} \in \mathcal{S}_{n,\vartheta,\epsilon} \rbrace$ has probability going to $1$ as $n$ goes to $+\infty$, for any $\epsilon>0$ and any feasible $\vartheta$.
\item[(3)] The function $\vartheta \to \phi(\vartheta)$ is continuous, convex in $\vartheta$ and has a unique minimizer $\vartheta^\star$.
\end{itemize}
Then, the following convergence in probability holds
\begin{align}
\abs{ \Phi_n -\phi(\vartheta^\star) } &\overset{p}{\longrightarrow} 0,~\mathbb{P}( \widehat{\vw}_n \in \mathcal{S}_{n,\vartheta^\star,\epsilon}  )  \overset{n\to \infty}{\longrightarrow} 0,\nonumber\\
&~~~\text{and}~\widehat{\vartheta}_n^\star \overset{p}{\longrightarrow} \vartheta^\star,
\end{align}
for any $\epsilon>0$, where $\Phi_n$ and $(\widehat{\vw}_n,\widehat{\vartheta}_n^\star)$ are the optimal cost and any optimal solution of the optimization problem \eqref{cons_po}.
\end{proposition}
The detailed proof of Proposition \ref{in_vartheta} is provided in Appendix \ref{pf_in_vartheta}. 
Note that if the activation function is odd, i.e. $\mu_0=0$, the formulation in \eqref{cons_po} is independent of the variable $\vartheta$. Now, when $\mu_0\neq0$, Proposition \ref{in_vartheta} allows us to apply the same analysis for odd activation functions where the only difference is that the loss function is shifted with the term $\mu_0 \vartheta$.
We continue our analysis by assuming that $\vartheta$ is fixed in the feasibility set $\lbrace \vartheta :~\abs{{\vartheta}} \leq C_\vartheta \rbrace$ and we show later that the assumptions in Proposition \ref{in_vartheta} are all satisfied. Note that the feasibility set is now convex and compact based on \cite[ Theorem 1.6]{var_ana}. The next step is to rewrite the optimization problem $\mathfrak{{V}}^g_{n,4}$ in the form of the PO formulation given in \eqref{eq:PO}. To this end, we introduce additional optimization variables. Given that the loss function is proper, continuous, and convex, the optimization problem $\mathfrak{{V}}^g_{n,4}$ can be equivalently formulated as follows
\begin{align}\label{cons_po2}
\mathfrak{{V}}^g_{n,4}&:\min_{ \vw\in\mathcal{P}_{\vw} } \max_{ \vu\in\mathbb{R}^m } -\frac{1}{m} \sum_{i=1}^{m} \ell^\star \left(y_i;u_i\right)+\frac{\lambda}{2} \norm{\vw}^2 \nonumber\\
&+ \frac{1}{m} \sum_{i=1}^{m} u_i \Big(\mu_0 \vartheta+\vw^\top (\mu_1 \mF^\top \va_i + \mu_{\star} \vz_i) \Big),
\end{align}
where $\ell^\star$ is the convex conjugate function \cite{var_ana}  of the convex loss function $\ell$. The CGMT framework further assumes that the feasibility set of the optimization vector $\vu\in\mathbb{R}^m$ is convex and compact. The following lemma shows that this assumption is also satisfied in our case.
\begin{lemma}[Dual Compactness]
\label{du_comp}
Assume that $\widehat{\vu}_n$ is the optimal solution of the optimization problem $\mathfrak{{V}}^g_{n,4}$ given in \eqref{cons_po2}. Then, there exists a positive constant $C_u>0$ independent of $n$ such that
\begin{align}
\mathbb{P}\left( \norm{\widehat{\vu}_n}/\sqrt{m} \leq C_u \right) \xrightarrow{n \to \infty} 1.
\end{align}
\end{lemma}
The detailed proof of Lemma \ref{du_comp} is provided in Appendix \ref{pf_du_comp}. Based on this result, the asymptotic analysis of the formulation given in \eqref{cons_po2} can be replaced by the asymptotic analysis of the following formulation
\begin{align}
\mathfrak{{V}}^g_{n,5}:&\min_{ \vw \in\mathcal{P}_{\vw} }\max_{\vu\in\mathcal{D}_{\vu}} -\frac{1}{m} \sum_{i=1}^{m} \ell^\star \left(y_i;u_i\right)+\frac{\lambda}{2} \norm{\vw}^2 \nonumber\\
&+ \frac{1}{m}  \sum_{i=1}^{m} u_i \Big(\mu_0 \vartheta + \vw^\top (\mu_1 \mF^\top \va_i + \mu_{\star} \vz_i) \Big),
\end{align}
where the dual feasibility set $\mathcal{D}_{\vu}$ is given by $\mathcal{D}_{\vu}=\lbrace \vu\in\mathbb{R}^m: {\norm{{\vu}}}/\sqrt{m} \leq C_u \rbrace$, and $C_u>0$ is any fixed constant independent of $n$ satisfying the result in Lemma \ref{du_comp}.
Note that the feasibility sets of the optimization problem $\mathfrak{{V}}^g_{n,5}$ are now convex and compact. Furthermore, the formulation given in $\mathfrak{{V}}^g_{n,5}$ can be rewritten as follows
\begin{align}\label{PO1}
\mathfrak{{V}}^g_{n,5}&:\min_{ \vw \in\mathcal{P}_{\vw} }\max_{\vu\in\mathcal{D}_{\vu}}-\frac{1}{m} \sum_{i=1}^{m} \ell^\star \left(y_i;u_i\right)+\frac{\lambda}{2} \norm{\vw}^2 \nonumber\\
& + \frac{\mu_1}{m}\vw^\top\mF^\top \mA^\top \vu+\frac{\mu_\star}{m}\vw^\top\mZ^\top \vu+ \frac{\mu_0}{m} \vartheta \vu^\top \vec{1}_m,
\end{align}
where the data matrix $\mA=[\va_1,\dots,\va_m]^\top \in\mathbb{R}^{m\times n}$, the matrix $\mZ=[\vz_1,\dots,\vz_m]^\top \in\mathbb{R}^{m\times k}$. Note that the labels $\lbrace y_i \rbrace_{i=1}^{m}$  depend on the data matrix $\mA$ as follows $y_i=\varphi(\va_i^\top \vxi)+\Delta \epsilon_i$, where $\epsilon_i$ is a standard Gaussian random variable. Then, we decompose the matrix $\mA$ as follows
\begin{align}
\mA=\mA \mP_{\vxi}+\mA \mP^{\perp}_{\vxi}=\mA \bar{\vxi} \bar{\vxi}^\top +\mA \mP^{\perp}_{\vxi},
\end{align}
where $\mP_{\vxi}\in\mathbb{R}^{n\times n}$ denotes the projection matrix onto the space spanned by the vector $\vxi\in\mathbb{R}^n$, $\mP^{\perp}_{\vxi}=\mI_n-\bar{\vxi} {\bar{\vxi}}^\top$ denotes the projection matrix onto the orthogonal complement of the space spanned by the vector $\vxi$ and where $\bar{\vxi}=\vxi/\norm{\vxi}$. Note that the matrix $\mA \bar{\vxi} \bar{\vxi}^\top$ is independent of the matrix $\mA \mP^{\perp}_{\vxi}$. Then, $\mA$ can be expressed as follows without changing its statistics
\begin{align}
\mA=\vs \bar{\vxi}^\top + \mH \mP^{\perp}_{\vxi},
\end{align}
where $\vs \sim \mathcal{N}(0,\mI_m)$, the components of the matrix $\mH\in\mathbb{R}^{m\times n}$ are drawn independently from a standard Gaussian distribution and where $\vs$ and $\mH$ are independent. This means that the high-dimensional analysis of the optimization problem \eqref{PO1} can be replaced by the high-dimensional analysis of the following formulation
\begin{align}\label{PO3}
&\mathfrak{{V}}^g_{n,6}:\min_{ \vw \in\mathcal{P}_{\vw} }\max_{\vu\in\mathcal{D}_{\vu}} -\frac{1}{m} \sum_{i=1}^{m} \ell^\star \left(y_i;u_i\right)+\frac{\lambda}{2} \norm{\vw}^2 + \frac{\mu_0}{m} \vartheta \vu^\top \vec{1}_m\nonumber\\
& + \frac{\mu_1}{m} \vu^\top \vs \bar{\vxi}^\top \mF \vw+ \frac{\mu_1}{m} \vu^\top \mH \mP^{\perp}_{\vxi} \mF \vw+ \frac{\mu_{\star}}{m} \vu^\top \mZ \vw.
\end{align}
Note that the matrix $\mu_1 \mH \mP^{\perp}_{\vxi} \mF+ \mu_\star\mZ$ can be expressed as follows without changing its statistics
\begin{align}
\mu_1 \mH \mP^{\perp}_{\vxi} \mF+ \mu_\star \mZ=\mB \mM^{1/2},
\end{align}
where the components of $\mB \in\mathbb{R}^{m\times k}$ are drawn independently from a standard Gaussian distribution and the matrix $\mM$ is given by $\mM=\mu_1^2 \mF^\top \mP^{\perp}_{\vxi} \mF+\mu^2_\star \mI_k$. Given that $\mM$ is a positive definite matrix, the analysis of the optimization problem \eqref{PO3} is equivalent to the analysis of the following formulation
\begin{align}\label{PO4}
&\mathfrak{{V}}^g_{n,7}:\min_{ \vw \in \widehat{\mathcal{P}}_{\vw} }\max_{\vu\in\mathcal{D}_{\vu}} -\frac{1}{m} \sum_{i=1}^{m} \ell^\star \left(y_i;u_i\right) + \frac{\mu_0}{m} \vartheta \vu^\top \vec{1}_m\nonumber\\
&+ \frac{\lambda}{2} \vw^\top \mM^{-1} \vw + \frac{\mu_1}{m} \vu^\top \vs \bar{\vxi}^\top \mF \mM^{-\frac{1}{2}} \vw+ \frac{1}{m} \vu^\top \mB \vw,
\end{align}
where we perform the change of variable $\vw_{\text{new}}=\mM^{\frac{1}{2}} \vw$, then, we replace $\vw_{\text{new}}$ by $\vw$. Additionally, the primal feasibility set $\widehat{\mathcal{P}}_{\vw}$ is defined as follows 
\begin{align}
\widehat{\mathcal{P}}_{\vw}=\lbrace \vw \in\mathbb{R}^k :~ \norm{ {\vw} } \leq C_w, \abs{\vv^\top \vw} \leq C_q \rbrace,
\end{align}
where the vector $\vv\in\mathbb{R}^k$ is defined as follows 
\begin{align}\label{v_vect}
\vv= \mM^{-\frac{1}{2}} \mF^\top  \bar{\vxi}.
\end{align}
Now, we are ready to formulate the optimization problem $\mathfrak{{V}}^g_{n,7}$ in the form of the PO problem given in \eqref{eq:PO}. Specifically, the problem $\mathfrak{{V}}^g_{n,7}$ can be expressed as follows
\begin{align}
{\mathfrak{V}}^g_{n,7}&:\min_{\vw \in \widehat{\mathcal{P}}_{\vw}}\max_{\vu\in\mathcal{D}_{\vu}}~\frac{1}{m} \vu^\top \mB \vw+\psi(\vw,\vu),\nonumber
\end{align}
where the function $\psi$ is convex in the argument $\vw\in\mathbb{R}^k$ and concave in the argument $\vu\in\mathbb{R}^m$ and can be expressed as follows
\begin{align}
\psi(\vw,\vu)&=-\frac{1}{m} \sum_{i=1}^{m} \ell^\star \left(y_i;u_i\right) + \frac{\mu_0}{m} \vartheta \vu^\top \vec{1}_m\nonumber\\
&+ \frac{\lambda}{2} \vw^\top \mM^{-1} \vw + \frac{\mu_1}{m} \vu^\top \vs \bar{\vxi}^\top \mF \mM^{-\frac{1}{2}} \vw.
\end{align}
Note that the function $\psi$ is continuous and convex-concave, and the feasibility sets are convex and compact. Then, the corresponding AO problem can be formulated as follows
\begin{align}
&\widehat{\mathfrak{V}}_{n,1}:\min_{\vw \in \widehat{\mathcal{P}}_{\vw}}\max_{\vu\in\mathcal{D}_{\vu}}\frac{\norm{\vu}}{m}  \vg^\top \vw + \frac{\norm{\vw}}{m}  \vh^\top \vu +\frac{\mu_0}{m} \vartheta \vu^\top \vec{1}_m  \nonumber\\
&-\frac{1}{m} \sum_{i=1}^{m} \ell^\star \left(y_i;u_i\right)+ \frac{\lambda}{2} \vw^\top \mM^{-1} \vw + \frac{\mu_1}{m} \vu^\top \vs \bar{\vxi}^\top \mF \mM^{-\frac{1}{2}} \vw,\nonumber
\end{align}
where $\vg\sim\mathcal{N}(0,\mI_k)$, $\vh \sim \mathcal{N}(0,\mI_m)$, and where $\vg$ and $\vh$ are independent. Following the CGMT framework, we focus on analyzing the AO formulation $\widehat{\mathfrak{V}}_{n,1}$. Specifically, the objective is to simplify the optimization problem $\widehat{\mathfrak{V}}_{n,1}$ and study its asymptotic properties.
\subsubsection{Simplifying the AO Problem}\label{ao_form_sim}
Assume that $\mB_{\vv}^{\perp} \in\mathbb{R}^{k\times (k-1)}$ is formed by an orthonormal basis orthogonal to the vector $\vv\in\mathbb{R}^k$. Then, any vector $\vw\in\mathbb{R}^k$ can be decomposed as follows
\begin{align}\label{w_decomp}
\vw= (\bar{\vv}^\top \vw) \bar{\vv} + \mB_{\vv}^{\perp}\vr,
\end{align}
where $\bar{\vv}=\vv/\norm{\vv}$ and $\vr \in\mathbb{R}^{k-1}$.
Moreover, define the scalar $q\in\mathbb{R}$ as follows $q={\vv}^\top \vw$. Therefore,  the AO formulation given in $\widehat{\mathfrak{V}}_{n,1}$ corresponding to our primary formulation in ${\mathfrak{V}}^g_{n,7}$ can be expressed as follows 
\begin{align}\label{v1_v2}
&\widehat{\mathfrak{V}}_{n,1}: \min_{(q,\vr) \in {\mathcal{P}}_{q,\vr}}\max_{\vu\in\mathcal{D}_{\vu}}~~\frac{\norm{\vu} q }{m \sqrt{T_{n,1}}}   \vg^\top \bar{\vv} + \sqrt{\frac{q^2}{T_{n,1}} + \norm{\vr}^2} \frac{\vh^\top \vu}{m}   \nonumber\\
&-\frac{1}{m} \sum_{i=1}^{m} \ell^\star \left(y_i;u_i\right) + \frac{\mu_1 }{m } q \vu^\top \vs + \frac{\lambda q^2}{2 T_{n,1}} T_{n,2} + \frac{ \lambda q}{\sqrt{T_{n,1}}} \vf^\top \vr \nonumber\\
&+ \frac{\lambda}{2} \vr^\top \mG \vr+\frac{\norm{\vu}}{m}   \vg^\top \mB_{\vv}^{\perp} {\vr}+\frac{\mu_0 \vartheta}{m}  \vu^\top \vec{1}_m,
\end{align}
where the vector $\vf \in\mathbb{R}^{k-1}$ and the matrix $\mG\in\mathbb{R}^{(k-1)\times(k-1)}$, $T_{n,1}$ and $T_{n,2}$ are defined as follows
\begin{align}\label{Tn12}
&\vf= (\mB_{\vv}^{\perp})^\top \mM^{-1} \bar{\vv},~\mG=(\mB_{\vv}^{\perp})^\top \mM^{-1} \mB_{\vv}^{\perp}\nonumber\\
&T_{n,1}=\norm{\vv}^2,~T_{n,2}=\bar{\vv}^\top \mM^{-1} \bar{\vv}.
\end{align}
Based on Assumption \ref{itm:ass_F}, $T_{n,1}>0$ which means that the optimization problem \eqref{v1_v2} is well-defined.
Moreover, the feasibility set ${\mathcal{P}}_{q,\vr}$ of the optimization variables $q$ and $\vr$ is defined as follows
\begin{align}
{\mathcal{P}}_{q,\vr}=\Big\lbrace q\in\mathbb{R},\vr\in\mathbb{R}^{k-1}: \abs{q} \leq C_q,\frac{q^2}{T_{n,1}}+\norm{\vr}^2\leq C_w^2 \Big\rbrace.\nonumber
\end{align}
Next, the main objective is to formulate the optimization problem given in $\widehat{\mathfrak{V}}_{n,1}$ in terms of scalar optimization variables. Our approach is to find the closed-form solution over the direction of the vector $\vr\in\mathbb{R}^{k-1}$, then, simplify the obtained formulation over the dual vector $\vu\in\mathbb{R}^m$. To this end, define the following optimization problem
\begin{align}
&\widehat{\mathfrak{V}}_{n,2}:\hspace{-2mm}\min_{(q,r) \in {\mathcal{P}}_{q,r}}\hspace{-0.5mm} \max_{\vu\in\mathcal{D}_{\vu}}\min_{ \norm{\vr}=r }  \frac{\norm{\vu} q \vg^\top \bar{\vv} }{m \sqrt{T_{n,1}}}    + \sqrt{\frac{q^2}{T_{n,1}} + \norm{\vr}^2}  \frac{\vh^\top \vu}{m}    \nonumber\\
&-\frac{1}{m} \sum_{i=1}^{m} \ell^\star \left(y_i;u_i\right) + \frac{\mu_1 }{m } q \vu^\top \vs + \frac{\lambda q^2}{2 T_{n,1}} T_{n,2} + \frac{ \lambda q}{\sqrt{T_{n,1}}} \vf^\top \vr \nonumber\\
&+ \frac{\lambda}{2} \vr^\top \mG \vr+\frac{\norm{\vu}}{m}   \vg^\top \mB_{\vv}^{\perp} {\vr}+\frac{\mu_0 \vartheta}{m}  \vu^\top \vec{1}_m,
\end{align}
where the feasibility set $\mathcal{P}_{q,r}$ is defined as follows
\begin{align}
{\mathcal{P}}_{q,r}=\Big\lbrace q\in\mathbb{R},r\in\mathbb{R}^{+}: \abs{q} \leq C_q,\frac{q^2}{T_{n,1}}+r^2\leq C_w^2 \Big\rbrace.
\end{align}
Note that the formulation in $\widehat{\mathfrak{V}}_{n,2}$ is obtained by switching the minimization over a non-convex feasibility set of the vector $\vr$ and the maximization over the variable $\vu$ in the formulation $\widehat{\mathfrak{V}}_{n,1}$. Hence, the optimization problems $\widehat{\mathfrak{V}}_{n,1}$ and $\widehat{\mathfrak{V}}_{n,2}$ are not necessarily equivalent. The following proposition shows that it suffices to precisely analyze the formulation $\widehat{\mathfrak{V}}_{n,2}$ to infer the asymptotic properties of our primary optimization problem.
\begin{proposition}[High-dimensional Equivalence II]\label{swi_norm}
Assume that $\vartheta$ is fixed in the feasibility set $\lbrace \vartheta: \abs{\vartheta} \leq C_\vartheta \rbrace$ and define the set $\mathcal{S}_{n,\vartheta,\epsilon}$ as follows
\begin{align}
\mathcal{S}_{n,\vartheta,\epsilon}=&\Big\lbrace \vw\in\mathbb{R}^k: \abs{q_n-q^\star_\vartheta} < \epsilon, \abs{\beta_n-\beta^\star_\vartheta} < \epsilon;\nonumber\\
&~~~q_n=\vv^\top \vw, {\beta}_n=\norm{\vw}  \Big\rbrace,
\end{align}
for a fixed $\epsilon>0$, where $\vv$ is given in \eqref{v_vect} and $q^\star_\vartheta$ and $\beta^\star_\vartheta$ are two deterministic constants.
Moreover, define the set $\mathcal{S}^c_{n,\vartheta,\epsilon}={\mathcal{P}}_{q,r} \setminus \mathcal{S}_{n,\vartheta,\epsilon}$. Let $\phi_n(\vartheta)$ and $\phi^c_{n}(\vartheta)$ be the optimal cost values of the formulation in $\widehat{\mathfrak{V}}_{n,2}$ with feasibility sets ${\mathcal{P}}_{q,r}$ and $\mathcal{S}^c_{n,\vartheta,\epsilon}$, respectively. Assume that the following properties are all satisfied
\begin{itemize}
\item[(1)] There exists a constant $\phi(\vartheta)$ such that the optimal cost $\phi_n(\vartheta)$ converges in probability to $\phi(\vartheta)$ as $n$ goes to $+\infty$.
\item[(2)] There exists a constant $\phi^c(\vartheta)$ such that the optimal cost $\phi^c_{n}(\vartheta)$ converges in probability to $\phi^c(\vartheta)$ as $n$ goes to $+\infty$, for any fixed $\epsilon>0$.
\item[(3)] There exists a positive constant $\zeta>0$ such that $\phi^c(\vartheta) \geq \phi(\vartheta)+\zeta$, for any fixed $\epsilon>0$.
\end{itemize}
Then, the following convergence in probability holds
\begin{equation}
\abs{ \Phi_n(\vartheta) -\phi_n(\vartheta) } \overset{p}{\longrightarrow} 0,~\text{and}~\mathbb{P}( \widehat{\vw}_{n,\vartheta} \in \mathcal{S}_{n,\vartheta,\epsilon} )  \overset{n\to\infty}{\longrightarrow} 1,\nonumber
\end{equation}
for any fixed $\epsilon>0$, where $\Phi_n(\vartheta)$ and $\widehat{\vw}_{n,\vartheta}$ are the optimal cost and the optimal solution of the primary formulation \eqref{PO4}.
\end{proposition}
The proof of Proposition \ref{swi_norm} is omitted since it follows similar steps of \cite[Lemma A.3]{chris:151}. Next, we focus on asymptotically analyzing the optimization problem $\widehat{\mathfrak{V}}_{n,2}$ and we show later that all the assumptions in Proposition \ref{swi_norm} are satisfied. Note that the optimization over the direction of the vector $\vr$ can be formulated as follows
\begin{align}\label{opt_r}
\mathfrak{R}_n: &\min_{\vr\in\mathbb{R}^{k-1}} \vb_n^\top \vr + \frac{1}{2} \vr^\top \mG \vr \nonumber\\
&~~\text{s.t.}~~\norm{\vr}=r,
\end{align}
where we ignore constant terms independent of $\vr$, and where the vector $\vb_n\in\mathbb{R}^n$ is given by
\begin{align}\label{bz1z2}
\vb_n= \frac{q}{\sqrt{T_{n,1}}} \vf + \frac{\norm{\vu}}{\lambda m} (\mB_{\vv}^{\perp})^\top \vg.
\end{align}
Note that the variables $r$ and $q$ are fixed in the feasibility set ${\mathcal{P}}_{q,r}$ and the vector $\vu$ is fixed in the feasibility set $\mathcal{D}_{\vu}$.
The optimization problem \eqref{opt_r} is well studied in the literature and it is known as the trust region subproblem \cite{TRS}.  The formulation in $\mathfrak{R}_n$ is not convex due to the norm equality constraint. Assuming that $\vb_n=0$, the optimal cost of the optimization problem \eqref{opt_r} denoted by $C^\star_n$ can be expressed as
\begin{align}
C^\star_n={r^2} \sigma_{\text{min}}(\mG)/{2},
\end{align}
where $\sigma_{\text{min}}(\mG)$ denotes the minimum eigenvalue of the matrix $\mG$.
Next, assume that $\vb_n\neq0$.  Note that the Karush-Kuhn-Tucker (KKT) conditions \cite{byd_conv} corresponding to the non-convex optimization problem \eqref{opt_r} can be expressed as follows
\begin{enumerate}[label=\text{C.\arabic*}]
\item \label{Cd1} $\vb_n+ \mG \vr + t \vr=0$
\item \label{Cd3} $\norm{\vr}^2=r^2,$
\end{enumerate}
where $t\in\mathbb{R}$ represents the KKT multiplier. 
Based on \cite[Theorem 3.2]{TRS}, the optimal solution $\vr_{n}^{\star}$ of the non-convex optimization problem \eqref{opt_r} can be expressed as follows
\begin{align}\label{vr_opt}
\vr_{n}^{\star}=- \left[ \mG+t_n^\star \mI_{k-1} \right]^{-1} \vb_n,
\end{align}
where the optimal KKT multiplier $t_n^\star$ is the unique solution of the equality constraint $\norm{\vr_{n}^{\star}}^2=r^2$ and it satisfies the following inequality constraint $t_n^\star > - \sigma_{\text{min}}(\mG)$. This means that the optimal cost of the optimization problem \eqref{opt_r} denoted by $C^\star_n$ can be expressed as follows
\begin{align}
C^\star_n=-\frac{1}{2} \vb_n^\top \left[ \mG+t_n^\star \mI_{k-1} \right]^{-1} \vb_n - \frac{t_n^\star}{2} \vb_n^\top \left[ \mG+t_n^\star \mI_{k-1} \right]^{-2} \vb_n, \nonumber
\end{align}
where the optimal KKT multiplier $t_n^\star$ guarantees that the optimal solution $\vr_{n}^{\star}$ is feasible. Specifically, $t_n^\star$ satisfies the following equality constraint
\begin{align}\label{eq_norm}
\vb_n^\top \left[ \mG+t_n^\star \mI_{k-1} \right]^{-2} \vb_n = r^2.
\end{align}
Note that the optimal cost $C^\star_n$ can be expressed in terms of a one dimensional optimization problem as follows
\begin{align}\label{opt_cost}
C^\star_n=\sup_{t > - \theta_n } \left\lbrace -\frac{1}{2} \vb_n^\top \left[ \mG+ t \mI_{k-1} \right]^{-1} \vb_n - \frac{t}{2} r^2 \right\rbrace, 
\end{align}
where $\theta_n=\sigma_{\text{min}}(\mG)$. The expression in \eqref{opt_cost} is valid for any $\vb_n\in\mathbb{R}^{k-1}$ since the first derivative of the cost function of the maximization problem formulated in \eqref{opt_cost} with respect to $t$ leads to the constraint in \eqref{eq_norm}. The above analysis shows that the optimization problem given in $\widehat{\mathfrak{V}}_{n,2}$ can be equivalently formulated as follows
\begin{align}\label{V2_form}
&\hspace{-1mm} \widehat{\mathfrak{V}}_{n,2}: \hspace{-2mm} \min_{(q,\beta) \in {\mathcal{P}}_{q,\beta}} \hspace{-1mm} \max_{ \substack{\vu\in\mathcal{D}_{\vu} \\ t > - \theta_n} } \frac{\norm{\vu} }{\sqrt{m}} \frac{q}{\sqrt{T_{n,1}}}   \frac{\vg^\top \bar{\vv}}{\sqrt{m}}  + \beta \frac{\vh^\top \vu }{m} +\frac{\mu_0}{m} \vartheta \vu^\top \vec{1}_m  \nonumber\\
&\hspace{-1mm} -\frac{1}{m} \sum_{i=1}^{m} \ell^\star \left(y_i;u_i\right) + \frac{\mu_1}{m } q \vu^\top \vs + \frac{\lambda q^2}{2 T_{n,1}}  T_{n,2} -\frac{\lambda q^2}{2 T_{n,1}} T_{n,3}(t) \nonumber\\
& \hspace{-1mm}-\frac{q \norm{\vu}}{\sqrt{m} \sqrt{T_{n,1}}} T_{n,5}(t) -\frac{\norm{\vu}^2}{2 \lambda m} T_{n,4}(t) -\frac{t \lambda}{2} \beta^2+\frac{\lambda t q^2}{2T_{n,1}},
\end{align}
where we perform the change of variable $\beta\bydef\sqrt{r^2+q^2/T_{n,1}}$, and where the feasibility set $\mathcal{P}_{q,\beta}$ is given by 
\begin{align}
{\mathcal{P}}_{q,\beta}=\Big\lbrace q\in\mathbb{R},\beta\in\mathbb{R}^{+}: \abs{q} \leq C_q, \beta \leq C_w, \beta \geq \abs{q}/\sqrt{T_{n,1}} \Big\rbrace.\nonumber
\end{align} 
Note that we replace the supremum in \eqref{opt_cost} by a maximization for simplicity of notation, and where the functions $T_{n,3}$, $T_{n,4}$ and $T_{n,5}$ depend on the optimization variable $t$ and can be expressed as follows
\begin{align}\label{Ts}
\begin{cases}
T_{n,3}(t)=\vf^\top \left[ \mG+ t \mI_{k-1} \right]^{-1} \vf \\
T_{n,4}(t)= \frac{1}{m} \vg^\top \mB_{\vv}^{\perp} \left[ \mG+ t \mI_{k-1} \right]^{-1} (\mB_{\vv}^{\perp})^\top \vg\\
T_{n,5}(t)= \frac{1}{\sqrt{m}}\vf^\top \left[ \mG+ t \mI_{k-1} \right]^{-1}  (\mB_{\vv}^{\perp})^\top \vg.
\end{cases}
\end{align}
Next, the main objective is to reformulate the optimization problem $\widehat{\mathfrak{V}}_{n,2}$ given in \eqref{V2_form} in terms of scalar optimization variables. To this end, we show that the formulation in \eqref{V2_form} is asymptotically equivalent to the following formulation 
\begin{align}\label{V3_form}
&\widehat{\mathfrak{V}}_{n,3}:\min_{(q,\beta) \in {\mathcal{P}}_{q,\beta}} \max_{ \substack{\vu\in\mathcal{D}_{\vu} \\ t > - \theta_n} } \beta \frac{\vh^\top \vu}{m}  +\frac{\mu_0}{m} \vartheta \vu^\top \vec{1}_m  -\frac{t \lambda}{2} \beta^2+\frac{\lambda t q^2}{2T_{n,1}} \nonumber\\
&-\frac{1}{m} \sum_{i=1}^{m} \ell^\star \left(y_i;u_i\right) + \frac{\mu_1}{m } q \vu^\top \vs + \frac{\lambda q^2 }{2 T_{n,1}}  T_{n,2} -\frac{\lambda q^2}{2 T_{n,1}}  T_{n,3}(t) \nonumber\\
& -\frac{1}{2 \lambda} \frac{\norm{\vu}^2}{m} T_{n,4}(t).
\end{align}
Note that \eqref{V3_form} only drops the terms that converge in probability to zero in the cost function of the formulation $\widehat{\mathfrak{V}}_{n,2}$ given in \eqref{V2_form}. 
The following proposition studies the asymptotic properties of the cost functions of the optimization problems $\widehat{\mathfrak{V}}_{n,2}$ and $\widehat{\mathfrak{V}}_{n,3}$.
\begin{lemma}[Partial Uniform Convergence]
\label{par_unf_con}
Define $\widehat{f}_{n,2}$ and $\widehat{f}_{n,3}$ as the cost functions of the optimization problems $\widehat{\mathfrak{V}}_{n,2}$   in \eqref{V2_form} and $\widehat{\mathfrak{V}}_{n,3}$  in \eqref{V3_form}, respectively. Then, the following convergence in probability holds
\begin{align}
\sup_{\vu\in\mathcal{D}_{\vu}}\abs{ \sup_{t > - \theta_n }\widehat{f}_{n,2}(q,\beta,t,\vu)-\sup_{t > - \theta_n }\widehat{f}_{n,3}(q,\beta,t,\vu) } \overset{p}{\longrightarrow} 0,\nonumber
\end{align}
for any fixed feasible $(q,\beta) \in {\mathcal{P}}_{q,\beta}$.
\end{lemma}
The detailed proof of Lemma \ref{par_unf_con} is deferred to Appendix \ref{pf_par_unf_con}. The asymptotic result in Lemma \ref{par_unf_con} shows that the cost functions of the optimization problems $\widehat{\mathfrak{V}}_{n,2}$ and $\widehat{\mathfrak{V}}_{n,3}$ converge uniformly in probability in the optimization vector $\vu\in\mathcal{D}_{\vu}$. We will show later that this property is sufficient to conclude that the formulations $\widehat{\mathfrak{V}}_{n,2}$ and $\widehat{\mathfrak{V}}_{n,3}$ are asymptotically equivalent. We continue our analysis by focusing on studying the asymptotic properties of the formulation in $\widehat{\mathfrak{V}}_{n,3}$. Our approach is to express the optimization over the vector $\vu$ in \eqref{V3_form} in terms of a separable function as shown in the following Lemma.
\begin{lemma}[Moreau Envelope Representation]
\label{sep_u_vn3}
Assume that $(q,\beta) \in {\mathcal{P}}_{q,\beta}$ and $t > - \theta_n$ and define the maximization problem over the vector $\vu$ in the formulation given in \eqref{V3_form} as follows
\begin{align}\label{max_mor_rep}
I_{n,q,\beta,t}&\hspace{-0.5mm}=\max_{ \substack{\vu\in\mathcal{D}_{\vu} } } \beta \frac{\vh^\top \vu}{m}  +\frac{\mu_0}{m} \vartheta \vu^\top \vec{1}_m \hspace{-0.5mm}  + \frac{\mu_1}{m } q \vu^\top \vs \nonumber\\
& -\frac{\norm{\vu}^2}{2 m \lambda}  T_{n,4}(t) -\frac{1}{m} \sum_{i=1}^{m} \ell^\star \left(y_i;u_i\right).
\end{align}
Then, the maximization problem given in $I_{n,q,\beta,t}$ can be expressed in terms of a separable function as follows
\begin{align}
I_{n,q,\beta,t}&=\frac{1}{m} \sum_{i=1}^{m} \mathcal{M}_{\ell(y_i;.)}\Big( \beta h_i + \mu_1 q s_i+\mu_0 \vartheta  ; {T_{n,4}(t)}/{\lambda} \Big),\nonumber
\end{align} 
on events with probability going to one as $n$ goes to $+\infty$ and uniformly over $(q,\beta) \in {\mathcal{P}}_{q,\beta}$ and $t > - \theta_n$, where $\lbrace h_i \rbrace_{i=1}^{m}$ and $\lbrace s_i \rbrace_{i=1}^{m}$ represent the components of the vectors $\vh$ and $\vs$, respectively, and where the function $\mathcal{M}_{\ell(y_i;.)}$ denotes the Moreau envelope of the loss function ${\ell(y_i;.)}$.
\end{lemma}
The detailed proof of Lemma \ref{sep_u_vn3} is deferred to Appendix \ref{pf_sep_u_vn3}. The result in Lemma \ref{sep_u_vn3} shows that the formulation in \eqref{V3_form} can be equivalently formulated as follows
\begin{align}\label{SOP_v1}
&\widehat{\mathfrak{V}}_{n,3}:\min_{(q,\beta) \in {\mathcal{P}}_{q,\beta}} \sup_{ t > - \theta_n } \frac{\lambda q^2}{2 T_{n,1}}  ( t+ T_{n,2}-T_{n,3}(t) ) -\frac{t \lambda}{2} \beta^2 \nonumber\\
&+ \frac{1}{m} \sum_{i=1}^{m} \mathcal{M}_{\ell(y_i;.)}\Big( \beta h_i + \mu_1 q s_i+\mu_0 \vartheta  ; {T_{n,4}(t)}/{\lambda} \Big).
\end{align}
Note that the loss function $\ell$ has two different forms as given in \eqref{reg_form} and \eqref{class_form}. For the regression task, the Moreau envelope in the cost function of the optimization problem \eqref{SOP_v1} can be expressed as follows
\begin{align}
\mathcal{M}_{\ell(y_i;.)}\Big( V_{i}(\vartheta,q,\beta); \frac{T_{n,4}(t)}{\lambda} \Big)=\mathcal{M}_{\widehat{\ell}}\Big( \widehat{V}_{i}(\vartheta,q,\beta); \frac{T_{n,4}(t)}{\lambda} \Big),\nonumber
\end{align}
where the functions $V_{i}$ and $\widehat{V}_{i}$ can be expressed as follows
\begin{align}
\begin{cases}
V_{i}(\vartheta,q,\beta)= \beta h_i + \mu_1 q s_i+\mu_0 \vartheta\\
\widehat{V}_{i}(\vartheta,q,\beta)= \beta h_i + \mu_1 q s_i+\mu_0 \vartheta-y_i.
\end{cases}
\end{align}
For the regression task, the Moreau envelope satisfies the following
\begin{align}
&\mathcal{M}_{\ell(y_i;.)}\Big( V_{i}(\vartheta,q,\beta); \frac{T_{n,4}(t)}{\lambda} \Big)=\mathcal{M}_{\widehat{\ell}}\Big( \widetilde{V}_{i}(\vartheta,q,\beta); \frac{T_{n,4}(t) y_i^2}{\lambda} \Big),\nonumber
\end{align}
where the functions $\widetilde{V}_{i}$ can be expressed as follows
\begin{align}
\widetilde{V}_{i}(\vartheta,q,\beta)= \beta y_i h_i + \mu_1 q y_i s_i+\mu_0 \vartheta y_i.
\end{align}
Next, we focus on studying the asymptotic properties of the scalar optimization problem \eqref{SOP_v1}. We refer to this problem as the scalar optimization problem (SOP).
\subsubsection{Asymptotic Analysis of the SOP} \label{asy_ana_ao}
In this part, we analyze the scalar optimization problem $\widehat{\mathfrak{V}}_{n,3}$ given in \eqref{SOP_v1}. To state our first asymptotic result, we define the following deterministic function
\begin{align}\label{dsf}
f_\vartheta(q,\beta,t)&=\frac{\lambda q^2}{2 T_1}  (t+ T_{2}-T_{3}(t)) -\frac{t \lambda}{2} \beta^2 \nonumber\\
&+\mathbb{E} \Big[ \mathcal{M}_{\widehat{\ell}}\Big( V(\vartheta,q,\beta);\frac{T_{4}(t) Z}{\lambda} \Big) \Big],
\end{align}
in the set $\mathcal{F}$ defined as follows
\begin{align}
\mathcal{F}=\Big\lbrace q,\beta,t: \abs{q}\leq C_q,\beta\leq C_w,\beta \geq \abs{q}/\sqrt{T_{1}}, t >-\theta \Big\rbrace,\nonumber
\end{align}
where $\theta$ is defined as $\theta=1/(\mu_\star^2+\mu_1^2 \kappa_{\text{max}})$ and the functions $T_1$, $T_2$, $T_3$ and $T_4$  and are given in Section \ref{pre_ana_fm}. Moreover, the random variable $Z$ satisfies $Z=1$ for the regression task and $Z=Y^2$ for the classification task, and the function $V$ can be expressed as follows
\begin{subnumcases}~
V(\vartheta,q,\beta)=\beta H + \mu_1 q S+\mu_0 \vartheta-Y~~\text{for regression}\nonumber\\
V(\vartheta,q,\beta)=\beta Y H + \mu_1 q Y S+\mu_0 \vartheta Y ~\text{for classification}\nonumber,
\end{subnumcases}
where $H$ and $S$ are two independent standard Gaussian random variables and $Y=\varphi(\rho S)+\Delta \epsilon$, and where $\epsilon$ is a standard Gaussian random variable independent of $H$ and $S$. We start our asymptotic analysis by studying the convergence behaviour of the cost function of the scalar optimization problem $\widehat{\mathfrak{V}}_{n,3}$ as stated in the following lemma.
\begin{lemma}[SOP Pointwise Convergence]
\label{sop_pt_conv}
Define $\widehat{f}_{n,3}$ as the cost function of the scalar optimization problem given in \eqref{SOP_v1}. Then, the function $\widehat{f}_{n,3}$ defined in the set 
\begin{align}
\mathcal{F}_{n,3}=\Big\lbrace q,\beta,t: \abs{q}\leq C_q,\beta\leq C_w,\beta \geq \abs{q}/\sqrt{T_{n,1}}, t >-\theta_n \Big\rbrace,\nonumber
\end{align}
converges pointwisely in probability to the function $f_\vartheta$
defined in the feasibility set ${\mathcal{F}}$.
\end{lemma}
The detailed proof of Lemma \ref{sop_pt_conv} is deferred to Appendix \ref{pf_sop_pt_conv}. 
Note that the function $f_\vartheta$ is not necessarily a convex function given the negative quadratic term in \eqref{dsf}. The following Lemma provides convexity properties of the deterministic function $f_\vartheta$.
\begin{lemma}[Convexity Property]
\label{convx_prop}
The deterministic function $f_\vartheta$ is strictly concave in $t$ for fixed feasible $(q,\beta)$. Moreover, define the function $\widehat{f}_\vartheta$ as follows
\begin{align}
\widehat{f}_\vartheta(q,\beta)=\sup_{t>-\theta}f_\vartheta(q,\beta,t),
\end{align}
in the set $\mathcal{F}_{q,\beta}$ defined as follows
\begin{align}\label{last_feas}
\mathcal{F}_{q,\beta}=\Big\lbrace q,\beta: \abs{q}\leq C_q,\beta\leq C_w,\beta \geq \abs{q}/\sqrt{T_{1}} \Big\rbrace.
\end{align}
Then, the function $\widehat{f}_\vartheta$ is jointly strongly convex in $q$ and $\beta$ and convex in $\vartheta$, where a strong convexity parameter is $\theta \lambda\min(\mu_1^2,1)$. Moreover, the function defined as follows
\begin{align}\label{phi_fun_def}
\phi(\vartheta)=\min_{(q,\beta) \in \mathcal{F}_{q,\beta}} \sup_{ t > - \theta } f_\vartheta (q,\beta,t),
\end{align}
is convex in its argument and it has a unique minimizer in the set $\lbrace \vartheta:\abs{\vartheta} \leq C_\vartheta \rbrace$. 
\end{lemma}
The detailed proof of Lemma \ref{convx_prop} is deferred to Appendix \ref{pf_convx_prop}. Next, we use the convexity properties summarized in Lemma \ref{convx_prop} and the asymptotic results in Lemma \ref{par_unf_con} and Lemma \ref{sop_pt_conv} to show that the following deterministic problem  
\begin{align}\label{detprob}
&{\mathfrak{V}}_\vartheta=\min_{(q,\beta) \in \mathcal{F}_{q,\beta}} \sup_{ t > - \theta } f_\vartheta (q,\beta,t),
\end{align}
is the converging limit of the formulation $\widehat{\mathfrak{V}}_{n,2}$ given in \eqref{V2_form} as stated in the following proposition.
\begin{proposition}[Consistency of the SOP]\label{cons_sop}
Define $\widehat{\mathcal{S}}^{\star}_{n,2}$ and $\widehat{O}^{\star}_{n,2}$ as the set of optimal $(q,\beta)$ and the optimal objective value of the formulation $\widehat{\mathfrak{V}}_{n,2}$ given in \eqref{V2_form}. Moreover, let $\widehat{\mathcal{S}}^{\star}$ and $\widehat{O}^{\star}$ be the set of optimal $(q,\beta)$ and the optimal objective value of the {deterministic}  problem formulated in \eqref{detprob}. Then, the following convergence in probability holds
\begin{equation}
\widehat{O}^{\star}_{n,2}  \xrightarrow{p} \widehat{O}^{\star}~\mathrm{and}~\mathbb{D}( \widehat{\mathcal{S}}^{\star}_{n,2},\widehat{\mathcal{S}}^{\star} )  \xrightarrow{p} 0,
\end{equation}
where $\mathbb{D}( \mathcal{A},\mathcal{B} )$ denotes the deviation between the sets $\mathcal{A}$ and $\mathcal{B}$ and is defined as $\mathbb{D}( \mathcal{A},\mathcal{B} )=\sup_{\vx_1\in\mathcal{A}} \inf_{\vx_2\in\mathcal{B}} \norm{\vx_1-\vx_2}$.
\end{proposition}
The detailed proof of Proposition \ref{cons_sop} is provided in Appendix \ref{pf_cons_sop}. Now that we obatined the asymptotic problem, it remains to study the convergence properties of the training and generalization errors. Specifically, our approach is to show that all the assumptions in Proposition \ref{in_vartheta} and Proposition \ref{swi_norm} are satisfied. It is also important to mention that the extreme values $C_\vartheta$, $C_q$ and $C_w$ can be any finite strictly positive constants as long as they satisfy the asymptotic results in Lemma \ref{prl_comp}. 

\subsubsection{Asymptotic Analysis of the Training and Generalization Errors}
First, the generalization error is given by
\begin{align}\label{testerr_ana}
\mathcal{G}_{n,\text{test}}&=\frac{1}{4^\upsilon} \mathbb{E}\left[ \left( \varphi(\va_{\text{new}}^\top\vxi) -\widehat{\varphi}(\widehat{\vw}^\top \sigma(\mF^\top \va_{\text{new}})) \right)^2 \right],
\end{align}
where $\va_{\text{new}}$ is an unseen data sample and $\widehat{\vw}$ is the unique optimal solution of the feature formulation given in \eqref{formulation}.
Based on Section \ref{GEC} and Proposition \ref{prop_in_v}, the asymptotic properties of the generalization error given in \eqref{testerr_ana} is equivalent to the asymptotic properties of $\widehat{\mathcal{G}}_{n,\text{test}}$ defined as follows
\begin{align}\label{genhat}
\widehat{\mathcal{G}}_{n,\text{test}}=\frac{1}{4^\upsilon} \mathbb{E}\Big[ &\Big( \varphi(\va_{\text{new}}^\top\vxi) -\widehat{\varphi}(\mu_0 \widehat{\vartheta}^\star_n + \mu_1 \widehat{\vw}^\top \mF^\top \va_{\text{new}}\nonumber\\
&+\mu_\star  \widehat{\vw}^\top \vz) \Big)^2 \Big], 
\end{align}
where $\vz$ is independent of $\va_{\text{new}}$ and drawn from a standard Gaussian distribution and $\widehat{\vw}$ and $\widehat{\vartheta}^\star_n$ are the optimal solutions of the formulation given in \eqref{gmain_prob3}. The expectation in \eqref{genhat} is over the distribution of the random vectors $\va_{\text{new}}$ and $\vz$. Now, consider the following two random variables
\begin{align}
\nu_1=  \va_{\text{new}}^\top\vxi,~\text{and}~\nu_2= \mu_0 \widehat{\vartheta}^\star_n+\mu_1 \widehat{\vw}^\top \mF^\top \va_{\text{new}}+\mu_\star  \widehat{\vw}^\top \vz.\nonumber
\end{align}
Given $\widehat{\vw}$ and $\widehat{\vartheta}^\star_n$, note that $\nu_1$ and $\nu_2$ have a bivaraite Gaussian distribution with mean vector $[0,\mu_0 \widehat{\vartheta}^\star_n]^\top$ and covariance matrix given by
\begin{align}\label{cov_mtx}
\mGm_n=\begin{bmatrix}
\norm{\vxi}^2 & \mu_1 \vxi^\top \mF \widehat{\vw} \\
 \mu_1 \vxi^\top \mF \widehat{\vw} &  \mu_1^2 (\bar{\vxi}^\top \mF \widehat{\vw} )^2 + \widehat{\vw}^\top \mM \widehat{\vw}
\end{bmatrix}.
\end{align}
To precisely analyze the asymptotic behaviour of the generalization error, it suffices to analyze the properties of the mean vector and the covariance matrix. Define the random variables $\widehat{q}_n^\star$ and $\widehat{\beta}_n^\star$ as follows
\begin{align}\label{opt_po}
\widehat{q}_n^\star=\overline{\vxi} ^\top \mF \widehat{\vw},~\text{and}~\widehat{\beta}_n^\star=\sqrt{\widehat{\vw}^\top \mM \widehat{\vw}}.
\end{align}
Then, the covariance matrix $\mGm_n$ given in \eqref{cov_mtx} can be expressed as follows
\begin{align}\label{cov_mat_ps}
\mGm_n=\begin{bmatrix}
\rho^2 & \mu_1 \rho \widehat{q}^{\star}_{n} \\
 \mu_1 \rho \widehat{q}^\star_{n} &  \mu_1^2 (\widehat{q}^{\star}_{n})^2+ (\widehat{\beta}_n^\star)^2
\end{bmatrix}.
\end{align}
Hence, to study the asymptotic properties of the generalization error, it suffices to study the asymptotic properties of $\widehat{\vartheta}^\star_n$, $\widehat{q}^\star_n$ and $\widehat{\beta}_n^\star$. The asymptotic result in Proposition \ref{cons_sop} shows that the following convergence in probability holds
\begin{align}
&q_n^\star \xrightarrow{p} q^\star,~\text{and}~\beta_n^\star \xrightarrow{p} \beta^\star,\nonumber
\end{align}
for any fixed $\vartheta$ in the set $\lbrace \vartheta: \abs{\vartheta} \leq C_\vartheta \rbrace$,
where $q_n^\star$ and $\beta_n^\star$ are any optimal solutions of the formulation given in \eqref{V2_form} and where $q^\star$ and $\beta^\star$ are the unique optimal solutions of the optimization problem in \eqref{detprob}. 
Next, we show that the assumptions in Proposition \ref{in_vartheta} and Proposition \ref{swi_norm} are valid to prove that $\widehat{\vartheta}_n^\star$, $\widehat{q}_n^\star$ and $\widehat{\beta}_n^\star$ concentrate around the optimal solution of the following minimization problem 
\begin{align}\label{detprob2}
&{\mathfrak{V}}=\min_{ \substack{ (q,\beta) \in \mathcal{F}_{q,\beta}\\ \abs{\vartheta} \leq C_\vartheta } } \sup_{ t > - \theta } f_\vartheta (q,\beta,t).
\end{align}
The following proposition summarizes these asymptotic properties.
\begin{proposition}[Feature Formulation Performance]
\label{ff_perf}
The optimal values $\widehat{\vartheta}_n^\star$, $\widehat{q}^\star_n$  and $\widehat{\beta}^\star_n$  converge in probability as follows
\begin{align}
\widehat{\vartheta}_n^\star \xrightarrow{p} \vartheta^\star,~\widehat{q}_n^\star \xrightarrow{p} q^\star,~\text{and}~\widehat{\beta}_n^\star \xrightarrow{p} \beta^\star,
\end{align}
where $\vartheta^\star$ is the unique minimizer of the function
\begin{align}\label{last_cost}
\vartheta \to \min_{(q,\beta) \in {\mathcal{F}}_{q,\beta}} \sup_{ t > - \theta } f_\vartheta (q,\beta,t),
\end{align}
in the set $\lbrace \vartheta:\abs{\vartheta} \leq C_\vartheta \rbrace$. Additionally, the optimal cost value $\Phi_n$ of the feature formulation satisfies the following asymptotic result
\begin{align}\label{ff_cost_conv}
\abs{ \Phi_n -\phi(\vartheta^\star) } \overset{p}{\longrightarrow} 0.
\end{align}
\end{proposition}
The detailed proof of Proposition \ref{ff_perf} is provided in Appendix \ref{pf_ff_perf}. 
Now, to show the convergence in \eqref{conv_tst_err} in Theorem \ref{ther1}, it suffices to show that $\widehat{\mathcal{G}}_{n,\text{test}}$ is a continuous function in $\widehat{\vartheta}_n^\star$, $\widehat{q}_n^\star$, and $\widehat{\beta}_n^\star$. Observe that the optimal solutions $\widehat{\vartheta}_n^\star$, $\widehat{q}_n^\star$, and $\widehat{\beta}_n^\star$ are bounded. Based on Assumption \ref{itm:fun_fwf} and continuity under integral sign property \cite{schilling_2005}, the continuity of $\widehat{\mathcal{G}}_{n,\text{test}}$ follows. Then, the convergence result given in \eqref{conv_tst_err} in Theorem \ref{ther1} is valid. Based on the result in \eqref{ff_cost_conv}, the optimal cost value of the feature formulation converges to the optimal cost value of the minimization problem given in \eqref{detprob2}. Then, the convergence of the training error $\mathcal{G}_{n,\text{train}}$ given in \eqref{conv_trr_err} is valid since $\mathcal{G}_{n,\text{train}}$ is the optimal cost value of the feature formulation. \\
The convergence result in Proposition \ref{ff_perf} holds for any $C_\vartheta>0$, $C_q>0$, and $C_w>0$ that guarantee the asymptotic results in Lemma \ref{prl_comp}. Furthermore, note that the optimal solutions, $\widehat{\vartheta}_n^\star$, $\widehat{q}_n^\star$ and $\widehat{\beta}_n^\star$, of the feature formulation \eqref{formulation} are independent of the extreme  values $C_\vartheta>0$, $C_q>0$, and $C_w>0$. This means that the optimal solution of the minimization problem in \eqref{detprob2} is in the interior of the domain. Combining this with the convexity properties in Lemma \ref{convx_prop}, the minimization problem in \eqref{detprob2} is equivalent to the formulation in \eqref{scprob1}.

\section{Conclusion}\label{concd}
In this paper, we presented a precise characterization of the asymptotic properties of a general convex formulation of the learning problem with random feature matrices. Our predictions are based on the uGEC and the CGMT framework. The analysis presented in this paper is valid for a general family of feature matrices, generic activation function and generic convex loss function. Moreover, our theoretical results rigorously verify previous analysis derived using the non-rigorous replica method from statistical physics. Simulation results validate our theoretical analysis and show that the generalization error follows a double descent curve.

\section{Appendix: Technical details}
\label{app}
\subsection{Proof of Lemma \ref{prl_comp}: Primal Compactness}\label{pf_prl_comp}
Assume that $\widehat{\vw}_n \in\mathbb{R}^k$ is the unique optimal solution of the optimization problem given in \eqref{gmain_prob}. Based on Assumption \ref{itm:cost}, the loss function is a proper function which means that there exists $C_1>0$ such that 
\begin{align}
\ell(y,z) \geq -C_1,~\forall z \in \mathbb{R}.
\end{align}
Define $V_{n,1}^\star$ as the optimal cost of the optimization problem formulated in \eqref{gmain_prob}. Then, there exists $C_1>0$ such that  the following inequality is valid
\begin{align}
\frac{\lambda}{2} \norm{\widehat{\vw}_n}^2 \leq V_{n,1}^\star+C_1.
\end{align}
Given that $\vec{0}_k$, the all zero vector, is a feasible solution in the formulation given in \eqref{gmain_prob}, we obtain the following inequality
\begin{align}
\frac{\lambda}{2} \norm{\widehat{\vw}_n}^2 \leq \frac{1}{m} \sum_{i=1}^{m} \ell(y_i;0)+C_1.
\end{align}
Note that the loss function satisfies the generic form given in \eqref{reg_form} and \eqref{class_form}. For the classification task, $\ell(y_i;0)=\widehat{\ell}(0)$ which is bounded by a constant given the continuity of the loss function in $\mathbb{R}$. Now, consider the loss function of the regression task, i.e. $\ell(y_i;0)=\widehat{\ell}(-y_i)$. Given that the loss function is convex in $\mathbb{R}$ and using the subgradient mean value Theorem \cite{conv_ana_opt}, there exists $\gamma_i\in(0,-y_i)$ such that 
\begin{align}
\widehat{\ell}(-y_i)=\widehat{\ell}(0)-s_g(\gamma_i) y_i,
\end{align}
where $s_g(\gamma_i)$ is a subgradient of the function $\widehat{\ell}$ evaluated at $\gamma_i$, and we assume without loss of generality that $y_i\leq0$. This means that the following inequality holds true
\begin{align}
\frac{1}{m} \abs{\sum_{i=1}^{m} \ell(y_i;0)} \leq \abs{\widehat{\ell}(0)} + \frac{1}{m} \norm{\vs_g} \norm{\vy},
\end{align}
where the components of both vectors $\vs_g$ and $\vy$ are $\lbrace s_g(\gamma_i) \rbrace_{i=1}^{m}$ and $\lbrace y_i \rbrace_{i=1}^{m}$, respectively.
Based on the weak law of large numbers (WLLN) \cite[Theorem 5.14]{prob_cour}, there exists $C_2>0$ such that
\begin{align}
\frac{1}{m}\sum_{i=1}^{m} \abs{\gamma_i}^2 \leq \frac{1}{m}\sum_{i=1}^{m} \abs{y_i}^2 \leq {C}_2,
\end{align}
with probability going to $1$ as $n$ goes to $+\infty$. Then, based on Assumption \ref{itm:cost} and the WLLN, there exist two constants ${C}_3>0$ and ${C}_4>0$ such that $\norm{\vs_g} \leq {C}_3 \sqrt{m}$ and $\norm{\vy} \leq {C}_4 \sqrt{m}$. 
Combining this with the continuity of the loss function in $\mathbb{R}$, we conclude that there exists $C_5>0$ such that the following holds
\begin{align}\label{p_b_w}
\norm{\widehat{\vw}_n}^2 \leq C_5,
\end{align}
with probability going to one as $n \to \infty$. Next, assume that $\mu_0\neq 0$. The above analysis also shows that there exists $C_6>0$ such that the following inequality holds
$\abs{ V_{n,1}^\star } \leq C_6
$
with probability going to one as $n \to \infty$. Combining this with the result in \eqref{p_b_w}, there exists $C_7>0$ such that
\begin{align}\label{L_bnd}
 \frac{1}{m} \abs{\mathcal{L}\left(\vv\right) } \leq C_7,
\end{align}
with probability going to one as $n \to \infty$, where $\mathcal{L}(\vv)=\sum_{i=1}^{m} \ell \left(y_i;v_i\right)$ and $\vv=\mu_0 (\vec{1}_k^\top \widehat{\vw}_n)  \vec{1}_m +\mu_1 \mA \mF \widehat{\vw}_n + \mu_{\star} \mZ \widehat{\vw}_n$, where $\mA^\top=[\va_1,\dots,\va_m] \in\mathbb{R}^{n\times m}$ and $\mZ^\top=[\vz_1,\dots,\vz_m]\in\mathbb{R}^{k\times m}$.  Assume that $\mB=\mu_1 \mA \mF + \mu_\star \mZ$, then, note that the following inequality always holds true
\begin{align}\label{B_boun}
{\norm{\mB}}/{\sqrt{m}}&={\norm{\mu_1 \mA \mF +\mu_\star \mZ}}/{\sqrt{m}} \nonumber\\
&\leq \frac{\abs{\mu_1}}{\sqrt{m}} \norm{\mF} \norm{\mA} +\frac{\abs{\mu_\star}}{\sqrt{m}} \norm{\mZ}.
\end{align}
Based on \cite[Theorem 2.1]{eigen_cons}, the following convergence in probability holds
\begin{align}\label{un_bond2}
\begin{cases}
\frac{\norm{\mA}}{\sqrt{m}} \xrightarrow{p} \frac{\sqrt{\alpha}+1}{\sqrt{\alpha}}\\
\frac{\norm{\mZ}}{\sqrt{m}} \xrightarrow{p} 1+\sqrt{\eta}.
\end{cases}
\end{align}
Combining this with Assumption \ref{itm:ass_F} shows that there exists a positive constant $C_8>0$ such that ${\norm{\mB}/\sqrt{m}} \leq  C_8$ with probability going to $1$ as $n$ goes to $\infty$. Together with the result in \eqref{p_b_w} shows that the following inequality
\begin{align}\label{bw_bnd}
\frac{1}{\sqrt{m}} \norm{\mB \widehat{\vw}_n} \leq C_8 \sqrt{C_5},
\end{align}
holds with probability going to $1$ as $n$ goes to $\infty$.
Based on Assumption \ref{itm:cost}, the asymptotic result in \eqref{L_bnd} and the asymptotic result in \eqref{bw_bnd}, we conclude that there exists $C_9>0$ such that
\begin{align}
(\vec{1}_k^\top \widehat{\vw}_n)^2 \leq C_9,
\end{align} 
with probability going to one as $n \to \infty$. This completes the proof of Lemma \ref{prl_comp}.
\subsection{Proof of Proposition \ref{prop_in_v}:High-dimensional Equivalence I}\label{pf_prop_in_v}
Let $O^{g}_{n,2}$ and $O^{g}_{n,3}$ be the optimal objective values of the optimization problems $\mathfrak{{V}}^g_{n,2}$ and $\mathfrak{{V}}^g_{n,3}$  given in \eqref{gmain_prob2} and \eqref{gmain_prob3}, respectively. 
The optimization problem $\mathfrak{{V}}^g_{n,2}$ can be expressed as follows
\begin{align}
\mathfrak{{V}}^g_{n,2}:\min_{\substack{~\norm{ {\vw} } \leq \widehat{C}_w \\ \abs{\vartheta} \leq \widehat{C}_\vartheta, \vec{1}_k^\top \vw=\vartheta  }}& \frac{1}{m} \sum_{i=1}^{m} \ell\left(y_i;\mu_0 \vartheta + \vw^\top \vp_i \right)+\frac{\lambda}{2} \norm{\vw}^2,\nonumber
\end{align}
where $\vp_i=(\mu_1 \mF^\top \va_i + \mu_{\star} \vz_i)$, for all $i\in\lbrace 1,\dots,m \rbrace$.
Moreover, the optimization problem $\mathfrak{{V}}^g_{n,3}$ can be expressed as follows
\begin{align}
\mathfrak{{V}}^g_{n,3}:\min_{\substack{~\norm{ {\vw} } \leq \widehat{C}_w \\ \abs{\vartheta} \leq \widehat{C}_\vartheta  }}& \frac{1}{m} \sum_{i=1}^{m} \ell\left(y_i;\mu_0 \vartheta + \vw^\top \vp_i \right)+\frac{\lambda}{2} \norm{\vw}^2.\nonumber
\end{align}
 Now, consider the following formulation referred to as $\widetilde{\mathfrak{{V}}}^g_{n,3}$
\begin{align}
\widetilde{\mathfrak{{V}}}^g_{n,3}:\hspace{-2mm}\min_{\substack{~\norm{ {\vw} } \leq \widehat{C}_w \\ \abs{\vartheta} \leq \widehat{C}_\vartheta  }}& \frac{1}{m} \sum_{i=1}^{m} \ell\left(y_i;\mu_0 \bar{\vartheta}+ \vw^\top \vp_i  \right) +\frac{\lambda}{2}  \norm{\vw}^2,\nonumber
\end{align}
where $\bar{\vartheta}={\vartheta}/{2}+{\vec{1}_k^\top \vw}/{2}$. Next, we show that the optimization problems ${\mathfrak{{V}}}^g_{n,3}$ and $\widetilde{\mathfrak{{V}}}^g_{n,3}$ have the same optimal cost asymptotically.  Based on the analysis in Appendix \ref{pf_prl_comp}, one can show that the optimal solution ${\vw}^\star$ of the optimization problem $\widetilde{\mathfrak{{V}}}^g_{n,3}$ satisfies
\begin{align}
\abs{\vec{1}_k^\top {\vw}^\star} \leq 3\widehat{C}_\vartheta,
\end{align} 
with probability going to one as $n \to \infty$. This means that the optimal solution $(\vartheta^\star,{\vw}^\star)$ of the optimization problem $\widetilde{\mathfrak{{V}}}^g_{n,3}$ satisfies 
\begin{align}
\abs{\vartheta^\star} \leq \widehat{C}_\vartheta,~\text{and}~\abs{\vec{1}_k^\top {\vw}^\star}\leq 3\widehat{C}_\vartheta,
\end{align}
asymptotically which means that $\bar{\vartheta}^\star={\vartheta}^\star/{2}+{\vec{1}_k^\top \vw^\star}/{2}$ satisfies $\abs{\bar{\vartheta}^\star} \leq 2\widehat{C}_\vartheta$ in the large system limit. Then, taking a sufficiently large $\widehat{C}_\vartheta$ implies that the optimal cost of the optimization problems $\mathfrak{{V}}^g_{n,3}$ and $\widetilde{\mathfrak{{V}}}^g_{n,3}$ are asymptotically equivalent. Moreover, the above properties show that
the optimal cost of the optimization problem $\widetilde{\mathfrak{{V}}}^g_{n,3}$ is asymptotically equivalent to the optimal cost of the following formulation
\begin{align}
\widehat{\mathfrak{{V}}}^g_{n,3}:\hspace{-2mm}\min_{\substack{~\norm{ {\vw} } \leq \widehat{C}_w \\ \abs{\vec{1}_k^\top {\vw}} \leq 3\widehat{C}_\vartheta}}\min_{\abs{\vartheta} \leq \widehat{C}_\vartheta  }& \frac{1}{m} \sum_{i=1}^{m} \ell\left(y_i;\mu_0 \bar{\vartheta}+ \vw^\top \vp_i  \right) +\frac{\lambda}{2}  \norm{\vw}^2.\nonumber
\end{align}
Now, performing the change of variable $\bar{\vartheta}={\vartheta}/{2}+{\vec{1}_k^\top \vw}/{2}$, we obtain the following  equivalent formulation
\begin{align}
\widehat{\mathfrak{{V}}}^g_{n,3}:\hspace{-2mm}\min_{\substack{~\norm{ {\vw} } \leq \widehat{C}_w \\ \abs{\vec{1}_k^\top {\vw}} \leq 3\widehat{C}_\vartheta}}\min_{\abs{{\vartheta}} \leq 2\widehat{C}_\vartheta  }& \frac{1}{m} \sum_{i=1}^{m} \ell\left(y_i;\mu_0 {\vartheta}+ \vw^\top \vp_i  \right) +\frac{\lambda}{2}  \norm{\vw}^2.\nonumber
\end{align}
Based on the analysis in Appendix \ref{pf_prl_comp}, the optimization problem $\widehat{\mathfrak{{V}}}^g_{n,3}$ is asymptotically equivalent to the following formulation
\begin{align}
\widehat{\mathfrak{{V}}}^g_{n,3}:\hspace{-2mm}\min_{\substack{~\norm{ {\vw} } \leq \widehat{C}_w \\ \abs{\vec{1}_k^\top {\vw}} \leq 3\widehat{C}_\vartheta}}\min_{\abs{{\vartheta}} \leq \widehat{C}_\vartheta  }& \frac{1}{m} \sum_{i=1}^{m} \ell\left(y_i;\mu_0 {\vartheta}+ \vw^\top \vp_i  \right) +\frac{\lambda}{2}  \norm{\vw}^2.\nonumber
\end{align}
This means that the following property is valid.\\
{\bf Property 1}: The optimal cost of the problem $\mathfrak{{V}}^g_{n,3}$ is equivalent to the optimal cost of the problem $\widehat{\mathfrak{{V}}}^g_{n,3}$ with probability going to one as $n \to \infty$.\\
Now, assume that $f$ is the cost value of the optimization problems $\mathfrak{{V}}^g_{n,2}$ and $\mathfrak{{V}}^g_{n,3}$  and $(\widehat{\vw},\widehat{\vartheta})$ is an optimal solution of $\mathfrak{{V}}^g_{n,2}$ and $({\vw}^\prime,{\vartheta}^\prime)$ is an optimal solution of $\widehat{\mathfrak{{V}}}^g_{n,3}$ and assume that $\bar{\vw}={\vw}^\prime+({\vartheta}^\prime-\vec{1}_k^\top {\vw}^\prime) \frac{\vec{1}_k}{k}$. Based on the analysis in Appendix \ref{pf_prl_comp}, we have the following 
\begin{align}\label{ineff}
0\leq f(\widehat{\vw},\widehat{\vartheta})-f({\vw}^\prime,{\vartheta}^\prime)\leq f(\bar{\vw},{\vartheta}^\prime)-f({\vw}^\prime,{\vartheta}^\prime).
\end{align}
Note that the right hand side of \eqref{ineff} satisfies the following properties
\begin{align}
&f(\bar{\vw},{\vartheta}^\prime)-f({\vw}^\prime,{\vartheta}^\prime)=\lambda ({\vartheta}^\prime-\vec{1}_k^\top {\vw}^\prime) \frac{\vec{1}_k^\top {\vw}^\prime}{k}+\frac{\lambda}{2} \frac{({\vartheta}^\prime-\vec{1}_k^\top {\vw}^\prime)^2}{k}\nonumber\\
&+\frac{1}{m} \sum_{i=1}^{m} \ell\left(y_i;\mu_0 {\vartheta}^\prime+ \bar{\vw}^\top \vp_i  \right) - \ell\left(y_i;\mu_0 {\vartheta}^\prime+ ({\vw}^\prime)^\top \vp_i  \right).\nonumber
\end{align}
The subgradient mean value Theorem \cite{conv_ana_opt} implies that there exists $\gamma_i\in(\mu_0 {\vartheta}^\prime + ({\vw}^\prime)^\top \vp_i,\mu_0 {\vartheta}^\prime+ \bar{\vw}^\top \vp_i)$ such that 
\begin{align}
&\ell\left(y_i;\mu_0 {\vartheta}^\prime + \bar{\vw}^\top \vp_i  \right) - \ell\left(y_i;\mu_0 {\vartheta}^\prime+ ({\vw}^\prime)^\top \vp_i  \right)=\nonumber\\
&s_g(\gamma_i) ({\vartheta}^\prime-\vec{1}_k^\top {\vw}^\prime) \frac{\vec{1}_k^\top\vp_i}{k},
\end{align}
where $s_g(\gamma_i)$ is a subgradient of the function ${\ell}(y_i;.)$ evaluated at $\gamma_i$. Note that $({\vw}^\prime,{\vartheta}^\prime)$ is a solution of the formulation in $\widehat{\mathfrak{{V}}}^g_{n,3}$. This means that there exists $C_1>0$ such that
\begin{align}
\lambda ({\vartheta}^\prime-\vec{1}_k^\top {\vw}^\prime) \frac{\vec{1}_k^\top {\vw}^\prime}{k}+\frac{\lambda}{2} \frac{({\vartheta}^\prime-\vec{1}_k^\top {\vw}^\prime)^2}{k} \leq \frac{C_1}{k}.
\end{align}
Based on the analysis in Appendix \ref{pf_prl_comp}, there exist two constants ${C}_2>0$ and ${C}_3>0$ such that $\norm{\vs_g} \leq {C}_2 \sqrt{m}$ and $\norm{\mB {\vec{1}_k}/{\sqrt{k}}} \leq {C}_3 \sqrt{m}$ with probability going to $1$ as $n$ goes to $+\infty$, where the components of the vector $\vs_g$ are $\lbrace s_g(\gamma_i) \rbrace_{i=1}^{m}$ and where $\mB=\mu_1 \mA \mF + \mu_\star \mZ$. This implies that there exist $C_1>0$, $C_2>0$ and $C_3>0$ such that
\begin{align}
\abs{f(\bar{\vw},{\vartheta}^\prime)-f({\vw}^\prime,{\vartheta}^\prime)} \leq \frac{C_1}{k}+\frac{\abs{{\vartheta}^\prime-\vec{1}_k^\top {\vw}^\prime}}{\sqrt{k}} C_2 C_3,
\end{align}
with probability going to $1$ as $n$ goes to $+\infty$. Combining this with Property 1, we obtain the following Property.\\
{\bf Property 2}:
The optimal objective values of the optimization problems $\mathfrak{{V}}^g_{n,2}$ and $\mathfrak{{V}}^g_{n,3}$ given in \eqref{gmain_prob2} and \eqref{gmain_prob3} satisfy the following asymptotic property
\begin{align}
\abs{O^{g}_{n,2}-O^{g}_{n,3}} \overset{p}{\longrightarrow} 0.
\end{align}
Next, we focus without loss of generality on the loss function of the regression task. Based on Assumption \ref{itm:cost}, the loss function $\widehat{\ell}$ is strongly convex in compact sets and the regularizer is strongly convex with strong convexity parameter $\lambda$. Since the feasibility sets are compact and based on \eqref{B_boun} and \eqref{un_bond2}, the loss function $\widehat{\ell}$ is strongly convex with a strong convexity parameter $\varsigma>0$, on events with probability going to $1$ as $n$ goes to $+\infty$. This means that the objective function $f$ satisfies the following
\begin{align}
&f(\gamma \vartheta_1+(1-\gamma) \vartheta_2,\gamma \vw_1+(1-\gamma) \vw_2) \leq \gamma f(\vartheta_1,\vw_1)\nonumber\\
&+(1-\gamma) f(\vartheta_2,\vw_2)-\frac{\lambda}{2} \gamma (1-\gamma) \norm{\vw_1-\vw_2}^2\nonumber\\
&-\frac{\varsigma \gamma (1-\gamma)}{m} \sum_{i=1}^{m}(\mu_0 {\vartheta}_1+ {\vw}_1^\top \vp_i -\mu_0 {\vartheta}_2- {\vw}^\top_2 \vp_i )^2,
\end{align}
valid for any $\vartheta_1$, $\vartheta_2$, $\vw_1$ and $\vw_2$ in the feasibility set and $\gamma\in[0,1]$. Assume that $(\widetilde{\vw},\widetilde{\vartheta})$ is an optimal solution of the problem $\mathfrak{{V}}^g_{n,3}$  given in \eqref{gmain_prob3}. This means that the following inequality is valid
\begin{align}\label{stcocons}
\abs{O^{g}_{n,3}- O^{g}_{n,2}} &\geq \frac{\varsigma \gamma }{m} \norm{ \mu_0 (\widetilde{\vartheta}-\widehat{\vartheta}) \vec{1}_m + \mB(\widetilde{\vw}- \widehat{\vw}) }^2\nonumber\\
&~~~+\frac{\lambda \gamma}{2}  \norm{\widehat{\vw} - \widetilde{\vw} }^2.
\end{align}
Based on the asymptotic result in Property 2, we have the following convergence in probability $\norm{\widehat{\vw} - \widetilde{\vw} }^2 \overset{p}{\longrightarrow} 0$. Moreover, the inequality in \eqref{stcocons} implies the following
\begin{align}
\sqrt{\varsigma \gamma} \abs{\mu_0} \abs{\widetilde{\vartheta}-\widehat{\vartheta}} &\leq \sqrt{ \abs{O^{g}_{n,3}- O^{g}_{n,2}}-\frac{\lambda \gamma}{2}  \norm{\widehat{\vw} - \widetilde{\vw} }^2}\nonumber\\
&+\frac{\sqrt{\varsigma \gamma}}{\sqrt{m}} \norm{\mB} \norm{\widehat{\vw} - \widetilde{\vw}},
\end{align}
valid for any $\gamma\in[0,1]$.
Combining this with \eqref{B_boun} and \eqref{un_bond2} implies that the following convergence in probability $(\widetilde{\vartheta}-\widehat{\vartheta})^2 \overset{p}{\longrightarrow} 0$. Then, we obtain the following Property. \\
{\bf Property 3}:
Given the strong convexity property, the solutions $(\widehat{\vw},\widehat{\vartheta})\in\mathcal{S}^{g}_{n,2}$ and $(\widetilde{\vw},\widetilde{\vartheta})\in\mathcal{S}^{g}_{n,3}$ are unique. Moreover, we have the following asymptotic result
\begin{align}
\norm{\widehat{\vw} - \widetilde{\vw} }^2 \overset{p}{\longrightarrow} 0,~(\widetilde{\vartheta}-\widehat{\vartheta})^2 \overset{p}{\longrightarrow} 0.
\end{align}
Property 2 and 3 complete the proof of Proposition \ref{prop_in_v}.
\subsection{Proof of Proposition \ref{in_vartheta}: Fixed Scalar Variable}\label{pf_in_vartheta}
Note that there exists a constant $\phi(\vartheta)$ such that the optimal cost $\Phi_n(\vartheta)$ converges in probability to $\phi(\vartheta)$ as $n$ goes to $\infty$. The function $\vartheta \to \phi(\vartheta)$ is defined in a convex and compact set and it is convex in $\vartheta$ and has a unique minimizer $\vartheta^\star$. Moreover, the function $\phi$ is continuous. Then, based on \cite[Theorem 2.1]{NEWEY19942111}, we obtain the following asymptotic results
\begin{align}\label{conv_dp}
\abs{ \Phi_n -\phi(\vartheta^\star) } \overset{p}{\longrightarrow} 0, ~\widehat{\vartheta}_n^\star \overset{p}{\longrightarrow} \vartheta^\star,
\end{align} 
where $\Phi_n$ and  and $\widehat{\vartheta}_n^\star$ are the optimal cost and the optimal solution of the minimization problem of the function $\Phi_n(\vartheta)$. Note that the cost function of the optimization problem \eqref{cons_po} is jointly convex in $\vartheta$ and strongly convex in $\vw$ with a strong convexity parameter $\lambda$. This means that the following inequality 
\begin{align}\label{fn3_sconv}
&f_{n,4}(\gamma \vartheta_1+(1-\gamma) \vartheta_2,\gamma \vw_1+(1-\gamma) \vw_2) \leq \gamma f_{n,4}(\vartheta_1,\vw_1)\nonumber\\
&+(1-\gamma) f_{n,4}(\vartheta_2,\vw_2)-\frac{\lambda}{2} \gamma (1-\gamma) \norm{\vw_1-\vw_2}^2,
\end{align}
is valid for any feasible $\vartheta_1$, $\vartheta_2$, $\vw_1$, $\vw_2$ and $\gamma\in[0,1]$, where $f_{n,4}$ is the cost function of the optimization problem \eqref{cons_po}. Now, assume that $\widehat{\vartheta}_n^\star$ and $\widehat{\vw}_n$ are the optimal solution of the optimization problem \eqref{cons_po} and $\widehat{\vw}_{n,\vartheta^\star}$ is the optimal solution of the formulation $\mathfrak{{V}}^g_{n,4}$ given in \eqref{cons_po}, for a fixed $\vartheta=\vartheta^\star$. Then, the inequality in \eqref{fn3_sconv} can be expressed as follows
\begin{align}
&\Phi_n \leq \gamma f_{n,4}(\widehat{\vartheta}_n^\star,\widehat{\vw}_n)+(1-\gamma) f_{n,4}(\vartheta^\star,\widehat{\vw}_{n,\vartheta^\star})\nonumber\\
&-\frac{\lambda}{2} \gamma (1-\gamma) \norm{\widehat{\vw}_n-\widehat{\vw}_{n,\vartheta^\star}}^2.
\end{align}
This means that the following inequality 
\begin{align}
&\abs{\Phi_n -\Phi_n(\vartheta^\star)} \geq  \frac{\lambda}{2} \gamma  \norm{\widehat{\vw}_n-\widehat{\vw}_{n,\vartheta^\star}}^2,
\end{align}
is valid for any $\gamma\in[0,1]$. Based on the asymptotic result in \eqref{conv_dp}, we obtain the following convergence in probability
\begin{align}
\norm{\widehat{\vw}_n-\widehat{\vw}_{n,\vartheta^\star}}^2 \overset{p}{\longrightarrow} 0.
\end{align}
Note that the event $\lbrace \widehat{\vw}_{n,\vartheta^\star} \in \mathcal{S}_{n,\vartheta^\star,\epsilon} \rbrace$ has probability going to $1$ as $n$ goes to $\infty$, for any $\epsilon>0$. This implies that 
\begin{equation}
\mathbb{P}( \widehat{\vw}_n \in cl( \mathcal{S}_{n,\vartheta^\star,\epsilon} ) )  \overset{n\to \infty}{\longrightarrow} 0,
\end{equation}
for any $\epsilon>0$, where $cl( \mathcal{S}_{n,\vartheta^\star,\epsilon} )$ denotes the closure of the set $\mathcal{S}_{n,\vartheta^\star,\epsilon}$. Based on \cite[Theorem 1.6]{stoch_opt}, we have the following property $cl(\mathcal{S}_{n,\vartheta^\star,\epsilon})=\mathcal{S}_{n,\vartheta^\star,\epsilon}$. This completes the proof of Proposition \ref{in_vartheta}.
\subsection{Proof of Lemma \ref{du_comp}: Dual Compactness}\label{pf_du_comp}
Assume that $\widehat{\vu}_n$ is the optimal solution of the optimization problem $\mathfrak{{V}}^g_{n,4}$. Define the loss function $\mathcal{L}^\star(\vu)=\sum_{i=1}^{m} \ell^\star \left(y_i;u_i\right)$. Note that the optimal $\widehat{\vu}_n$  satisfies the following
\begin{align}
\widehat{\vu}_n=\argmax_{ \vu\in\mathbb{R}^m }~  \vu^\top \Big(\mu_0 \vartheta \vec{1}_m +\mu_1 \mA \mF \vw + \mu_{\star} \mZ \vw \Big)  - \mathcal{L}^\star(\vu),\nonumber
\end{align}
where the data matrix $\mA=[\va_1,\dots,\va_m]^\top \in\mathbb{R}^{m\times n}$, the matrix $\mZ=[\vz_1,\dots,\vz_m]^\top \in\mathbb{R}^{m\times k}$. Define $\partial \mathcal{L}^\star(\vu)$ as the sub-differential set of the loss function $\mathcal{L}^\star$ evaluated at $\vu$. Then, we have the following optimality condition
\begin{align}\label{opt_dual_u}
\mu_0 \vartheta \vec{1}_m +\mu_1 \mA \mF \vw + \mu_{\star} \mZ \vw \in \partial \mathcal{L}^\star(\widehat{\vu}_n).
\end{align}
Based on \cite[Proposition 11.3]{var_ana}, the optimality condition in \eqref{opt_dual_u} is equivalent the following optimality condition 
\begin{align}
\widehat{\vu}_n \in \partial \mathcal{L} \left( \mu_0 \vartheta \vec{1}_m +\mu_1 \mA \mF \vw + \mu_{\star} \mZ \vw \right),
\end{align}
where the loss function $\mathcal{L}(\vw)=\sum_{i=1}^{m} \ell \left(y_i;w_i\right)$ based on \cite[Proposition 11.22]{var_ana}. Based on the analysis in Appendix \ref{pf_prl_comp}, there exists $C_1 >0$ such that the following inequality holds
\begin{align}
\norm{ \mu_0 \vartheta \vec{1}_m +\mu_1 \mA \mF \vw + \mu_{\star} \mZ \vw }^2 \leq C_1 m,
\end{align}
with probability going to one as $n$ goes to $+\infty$. Combining this result with Assumption \ref{itm:cost}, we conclude that there exists $C_2>0$ such that the following inequality holds
\begin{align}
\norm{ \widehat{\vu}_n }^2 \leq C_2 m,
\end{align}
with probability going to one as $n$ goes to $+\infty$. This completes the proof of Lemma \ref{du_comp}.
\subsection{Proof of Lemma \ref{par_unf_con}: Partial Uniform Convergence}\label{pf_par_unf_con}
Following Appendix \ref{pf_sop_pt_conv}, we start our proof by making a change of variable $\bar{q}_n={q}/\sqrt{T_{n,1}}$ and $\bar{t}_n=t-\theta_n$. Then, the functions $\widehat{f}_{n,2}$ and $\widehat{f}_{n,3}$ are now defined in the set
\begin{align}
{\mathcal{S}}=\lbrace q,\beta,t,\vu: \abs{q} \leq C_q, \beta \leq C_w, \beta \geq \abs{q}, t > 0, \vu\in\mathcal{D}_u \rbrace.\nonumber
\end{align}
Now, fixed $q$ and $\beta$ in the above feasibility set. Note that the result follows if $\beta^2=q^2$, then, we assume that $\beta^2 \neq q^2$. First, we show that the function defined as follows
\begin{align}
V_n:(u,t) \to  -\frac{1}{2} \vb_n^\top \left[ \widehat{\mG}+ t \mI_{k-1} \right]^{-1} \vb_n - \frac{t-\theta_n}{2} (\beta^2-q^2),\nonumber
\end{align}
is jointly concave in its arguments in the feasibility set $\lbrace (u,t): 0 \leq u \leq C_u, t > 0 \rbrace$, where $\widehat{\mG}=\mG-\theta_n \mI_{k-1}$ and where the vector $\vb_n$ is given by
\begin{align}
\vb_n= q \vf + \frac{u}{\lambda} \frac{\mB_{\vv}^{\perp} \vg }{\sqrt{m}}.
\end{align}
First, observe that it suffices to show that the following function
\begin{align}
\widetilde{V}_n:(u,t) \to  -\frac{u^2}{2} \vc_n^\top \left[ \widehat{\mG}+ t \mI_{k-1} \right]^{-1} \vc_n,
\end{align}
is jointly concave in its arguments in the feasibility set, for any vector $\vc_n$. The function $\widetilde{V}$ is twice differentiable where its Hessian matrix $\mH_{\widetilde{V}}$ can be expressed as follows
\begin{align}
\begin{bmatrix}
-\vc_n^\top \left[ \widehat{\mG}+ t \mI_{k-1} \right]^{-1} \vc_n  & u \vc_n^\top \left[ \widehat{\mG}+ t \mI_{k-1} \right]^{-2} \vc_n  \\
u \vc_n^\top \left[ \widehat{\mG}+ t \mI_{k-1} \right]^{-2} \vc_n  & - u^2 \vc_n^\top \left[ \widehat{\mG}+ t \mI_{k-1} \right]^{-3} \vc_n 
\end{bmatrix}.\nonumber
\end{align}
Clearly, the trace of the Hessian matrix $\mH_{\widetilde{V}}$ is negative and its determinant is given by
\begin{align}
D_n&=u^2 \vc_n^\top \left[ \widehat{\mG}+ t \mI_{k-1} \right]^{-3} \vc_n \vc_n^\top \left[ \widehat{\mG}+ t \mI_{k-1} \right]^{-1} \vc_n \nonumber\\
&-u^2 \left( \vc_n^\top \left[ \widehat{\mG}+ t \mI_{k-1} \right]^{-2} \vc_n \right)^2.
\end{align}
Using the Cauchy--Schwarz inequality, the determinant $D_n$ is positive which implies that the Hessian matrix is negative semidefinite.  Therefore, the function $\widetilde{V}_n$ is jointly concave in its arguments which implies that the function $V_n$ is jointly concave in its arguments in the feasibility set. Based on the analysis in Appendix \ref{pf_sop_pt_conv}, the function $V_n$ defined in the set $\lbrace (u,t): 0 \leq {u} \leq C_u, t > 0 \rbrace$ converges in probability to the function
\begin{align}
V(u,t)= -\frac{q^2}{2 } T_{3}(t-\theta)  -\frac{u^2}{2 \lambda^2} T_{4}(t-\theta) -\frac{t-\theta}{2} (\beta^2-q^2),\nonumber
\end{align}
defined in the same set $\lbrace (u,t): 0\leq {u} \leq C_u, t > 0 \rbrace$
where the functions $T_3$ and $T_4$ are given in Section \ref{pre_ana_fm}. Moreover, observe that for any fixed feasible $u$, we have the following asymptotic result
\begin{align}
\lim_{t\to+\infty} V(u,t)=-\infty.
\end{align}
Then, using the result in \cite[Lemma B.1]{chris:151}, we obtain the following convergence result
\begin{align}\label{poin_conv_fn}
\sup_{t>0} V_n(u,t) \xrightarrow{p} \sup_{t > 0} V(u,t),
\end{align}
for any fixed feasible $u$. Given the joint concavity property of the function $V_n$ and based on \cite[Section 3.2]{convex_opt}, the convergence in \eqref{poin_conv_fn} is uniform \cite[Theorem II.1]{andersen1982}, i.e.
\begin{align}
\sup_{0 \leq {u} \leq C_u}\abs{ \sup_{t>0} V_n(u,t) -\sup_{t > 0} V(u,t) } \xrightarrow{p} 0.
\end{align}
Note that the same asymptotic results hold true when we ignore the cross term in the function $V_n$. Specifically, the same asymptotic properties are true for the function defined as follows
\begin{align}
&\widehat{V}_n(u,t)= -\frac{q^2}{2} \vf^\top \left[ \widehat{\mG}+ t \mI_{k-1} \right]^{-1} \vf\nonumber\\
&-\frac{u^2}{2\lambda} \frac{(\mB_{\vv}^{\perp} \vg)^\top}{\sqrt{m}}  \left[  \widehat{\mG}+ t \mI_{k-1} \right]^{-1} \frac{\mB_{\vv}^{\perp} \vg }{\sqrt{m}} - \frac{t-\theta_n}{2} (\beta^2-q^2),\nonumber
\end{align}
which means that the following convergence in probability holds
\begin{align}
\sup_{0 \leq {u}\leq C_u}\abs{ \sup_{t>0} \widehat{V}_n(u,t) -\sup_{t > 0} V(u,t) } \xrightarrow{p} 0.
\end{align}
Therefore, we conclude that the functions $V_n$ and $\widehat{V}_n$ satisfy the following asymptotic result
\begin{align}
\sup_{\vu\in\mathcal{D}_{\vu}  }\abs{ \sup_{t>0} {V}_n\Big(\frac{\norm{\vu}}{\sqrt{m}},t\Big) -\sup_{t>0} \widehat{V}_n\Big(\frac{\norm{\vu}}{\sqrt{m}},t\Big)  } \xrightarrow{p} 0.
\end{align}
Using the WLLN, the following convergence in probability holds
\begin{align}
\frac{\vg^\top \bar{\vv}}{\sqrt{m}} \xrightarrow{p} 0.
\end{align}
Given that $\widehat{f}_{n,2}$ and $\widehat{f}_{n,3}$ are the cost functions of the optimization problems $\widehat{\mathfrak{V}}_{n,2}$ and $\widehat{\mathfrak{V}}_{n,3}$, we conclude that the following asymptotic result holds
\begin{align}
\sup_{\vu\in\mathcal{D}_{\vu}}\abs{ \sup_{t > - \theta_n }\widehat{f}_{n,2}(q,\beta,t,\vu)-\sup_{t > - \theta_n }\widehat{f}_{n,3}(q,\beta,t,\vu) } \overset{p}{\longrightarrow} 0,\nonumber
\end{align}
which completes the proof of Lemma \ref{par_unf_con}.
\subsection{Proof of Lemma \ref{sep_u_vn3}: Moreau Envelope Representation}\label{pf_sep_u_vn3}
Assume that $(q,\beta) \in {\mathcal{P}}_{q,\beta}$ and $t > - \theta_n$ and define the unconstrained version of the maximization problem in \eqref{max_mor_rep} as follows
\begin{align}\label{ihat_b}
\widehat{I}_{n,q,\beta,t}&\hspace{-0.5mm}=\max_{ \substack{\vu\in\mathbb{R}^m } } \va^\top \vu - \frac{\tau_n}{2} \norm{\vu}^2 - \mathcal{L}^\star(\vu),
\end{align}
where $\va=\beta \vh+\mu_0 \vartheta \vec{1}_m+ \mu_1 q \vs$, $\tau_n=T_{n,4}(t)/\lambda$ and where $\mathcal{L}^\star(\vu)=\sum_{i=1}^{m} \ell^\star \left(y_i;u_i\right)$. First, the problem in \eqref{ihat_b} can be viewed as a sum of two concave functions. Moreover, $\widehat{I}_{n,q,\beta,t}$ can be upper-bounded as follows
\begin{align}\label{max_mor_bn}
\widehat{I}_{n,q,r,t} \leq \max_{ \substack{\vu\in\mathbb{R}^m } } \left[ \va^\top \vu  - \mathcal{L}^\star(\vu) \right] + \max_{ \substack{\vu\in\mathbb{R}^m }  } - \frac{\tau_n}{2} \norm{\vu}^2.
\end{align}
Based on the analysis in Appendix \ref{itm:cost}, the optimization problems, in the right hand side of \eqref{max_mor_bn}, attain their solutions in the interior of the feasibility set. Assume that $\vu_1^\star$ and $\vu_2^\star$ are the optimal solutions of the optimization problems in the right hand side of the inequality in \eqref{max_mor_bn}. Furthermore, assume that $\vu^\star$ is an optimal solution of the optimization problem in the left hand side of the inequality in \eqref{max_mor_bn}. Given that the optimization problems in \eqref{max_mor_bn} are all concave and seperable, then, there exists $\gamma_i \in [0,1]$ such that
\begin{align}
u^\star_i=\gamma_i u_{1i}^\star + (1-\gamma_i) u_{2i}^\star,
\end{align}
valid for any $i\in\lbrace 1,\dots,m\rbrace$, where $u_{i}^\star$, $u_{1i}^\star$ and $u_{2i}^\star$ denote the ith components of the vectors $\vu^\star$, $\vu_1^\star$ and $\vu_2^\star$, respectively. This means that the following inequality
\begin{align}
(u^\star_i)^2\leq ( \abs{u_{1i}^\star} + \abs{u_{2i}^\star} )^2,
\end{align}
is valid for any $i\in\lbrace 1,\dots,m\rbrace$ which means that
\begin{align}
\norm{\vu^\star}^2 \leq \norm{\vu_1^\star}^2+2\abs{\vu_1^\star}^\top \abs{\vu_2^\star}+\norm{\vu_2^\star}^2.
\end{align}
Given that $\tau_n \geq 0$, the optimal vector $\vu_2^\star$ is the all zero vector. Therefore, the following inequality
\begin{align}
\norm{\vu^\star}^2 \leq \norm{\vu_1^\star}^2,
\end{align}
always holds, where $\vu_1^\star$ satisfies the following 
\begin{align}
\vu_1^\star \in \partial \mathcal{L} \left( \va \right).
\end{align}
Based on the WLLN, there exists a constant $C_1>0$ such that $\norm{\va} \leq C_1 \sqrt{m}$ with probability going to one as $n$ goes to $+\infty$. Combining this with Assumption \ref{itm:cost}, there exists a constant $C_2>0$ such that $\norm{\vu_1^\star} \leq C_2 \sqrt{m}$ with probability going to one as $n$ goes to $+\infty$. Therefore, we conclude that there exists a constant $C_2>0$ such that $\norm{\vu^\star} \leq C_2 \sqrt{m}$ with probability going to one as $n$ goes to $+\infty$ and uniformly over $q$, $r$ and $t$. This means that the optimal solution of the unconstrained version of the maximization problem in \eqref{max_mor_rep} is bounded asymptotically. Combining this with the analysis in \cite[Example 11.26]{var_ana} completes the proof of Lemma \ref{sep_u_vn3}.
\subsection{Proof of Lemma \ref{sop_pt_conv}: SOP Pointwise Convergence}\label{pf_sop_pt_conv} 
In this Appendix, we assume that $\delta>1$ and $\widehat{\ell}$ is the loss function corresponding to the regression task. The proof for $\delta\leq 1$ and the loss function corresponding to the classification task is similar. First, we study the asymptotic properties of $T_{n,1}$ and $T_{n,2}$ introduced in \eqref{Tn12}. Note that 
\begin{align}
T_{n,1}=\norm{\vv}^2,~T_{n,2}=\bar{\vv}^\top \mM^{-1} \bar{\vv},
\end{align}
where $\vv= \mM^{-\frac{1}{2}} \mF^\top  \bar{\vxi}$ and $\mM=\mu_1^2 \mF^\top \mP^{\perp}_{\vxi} \mF+\mu^2_\star \mI_k$. First, note that $T_{n,1}$ can be expressed as follows
\begin{align}
T_{n,1}&=\bar{\vxi}^\top \mF (\mu_1^2 \mF^\top \mP^{\perp}_{\vxi} \mF+\mu^2_\star \mI_k)^{-1} \mF^\top  \bar{\vxi}\nonumber\\
&=\bar{\vxi}^\top \mF (\mOm-\mu_1^2 \mF^\top  \bar{\vxi} \bar{\vxi}^\top \mF)^{-1} \mF^\top  \bar{\vxi},
\end{align}
where $\mOm=\mu_1^2 \mF^\top  \mF+\mu^2_\star \mI_k$. Define the matrix $\mC$ as follows $\mC=(\mOm-\mu_1^2 \mF^\top  \bar{\vxi} \bar{\vxi}^\top \mF)^{-1}$. Using the matrix inversion lemma, the matrix $\mC$ can be expressed as follows
\begin{align}\label{mC}
\hspace{-2mm}\mC&=\mOm^{-1}+\frac{\mu_1^2}{1-\mu_1^2 \bar{\vxi}^\top \mF \mOm^{-1} \mF^\top  \bar{\vxi} } \mOm^{-1} \mF^\top  \bar{\vxi}  \bar{\vxi}^\top \mF \mOm^{-1}.
\end{align}
The matrix $\mC$ is well defined since all the eigenvalues of the matrix $\mu_1^2  \mF \mOm^{-1} \mF^\top$  are non-negative and strictly smaller than $1$. Based on Assumption \ref{itm:ass_F}, the left singular vectors matrix of the feature matrix $\mF\in\mathbb{R}^{n\times k}$ is a Haar-distributed random matrix. Then, using \cite[Proposition 3]{Debbah}, the following convergence in probability holds true
\begin{align}\label{conv_haar}
\overline{\vxi}^\top \mF \mOm^{-1} \mF^\top \overline{\vxi}-\frac{1}{n}\text{Tr}[\mF \mOm^{-1} \mF^\top]\xrightarrow{p} 0,
\end{align}
where $\text{Tr}[.]$ denotes the trace.
Assumption \ref{itm:ass_F} states that the empirical distribution of the eigenvalues of the matrix $\mT$ converges weakly to a probability distribution $\mathbb{P}_\kappa$. This implies that the following convergence in probability holds true
\begin{align}\label{T1}
\frac{1}{n}\text{Tr}[\mF \mOm^{-1} \mF^\top] \xrightarrow{p}  \mathbb{E}_\kappa\left[ { \kappa }/({\mu_1^2 \kappa + \mu_\star^2}) \right],
\end{align}
where the expectation is over the random variable $\kappa$ distributed according to the probability distribution $\mathbb{P}_\kappa$ defined in Assumption \ref{itm:ass_F}. This means that $T_{n,1}$ converges in probability as follows
\begin{align}\label{Tn1toT1}
&T_{n,1} \xrightarrow{p} T_1=\frac{ \mathbb{E}_\kappa\left[ { \kappa }/({\mu_1^2 \kappa + \mu_\star^2}) \right]}{ \mathbb{E}_\kappa\left[ { \mu_\star^2 }/({\mu_1^2 \kappa + \mu_\star^2}) \right]}.
\end{align}
Similarly, one can show that $T_{n,2}$ converges in probability as given in Section \ref{pre_ana_fm}. Note that the function $\widehat{f}_{n,3}$ is defined in the set 
\begin{align}
\mathcal{F}_{n,3}=\Big\lbrace q,\beta,t: \abs{q}\leq C_q,\beta\leq C_w,\beta \geq \abs{q}/\sqrt{T_{n,1}}, t >-\theta_n \Big\rbrace,\nonumber
\end{align}
which is a random set, where $\theta_n=\sigma_{\text{min}}(\mG)$, $\mG=(\mB_{\vv}^{\perp})^\top \mM^{-1} \mB_{\vv}^{\perp}$ and $\sigma_{\text{min}}(\mG)$ denotes the minimum eigenvalue of the matrix $\mG$. Additionally, the function $f_\vartheta$ is defined in the following set
\begin{align}
\mathcal{F}=\Big\lbrace q,\beta,t: \abs{q}\leq C_q,\beta\leq C_w,\beta \geq \abs{q}/\sqrt{T_{1}}, t >-\theta \Big\rbrace,\nonumber
\end{align}
where $\theta={1}/({\mu_\star^2+\mu_1^2 \kappa_{\text{max}}})$,
and $\kappa_{\text{max}}$ is defined in Assumption \ref{itm:ass_F}.
To work in the same set, we can perform a change of variable $\bar{q}=\abs{q}/\sqrt{T_{1}}$, $\bar{q}_n=\abs{q}/\sqrt{T_{n,1}}$, $\bar{t}=t+\theta$ and $\bar{t}_n=t+\theta_n$. Based on \eqref{Tn1toT1}, there exists $C_1>0$ such that $T_{n,1} \leq C_1$. This means that $T_{n,1}$ can be included in the constant $C_q$ in the sets $\mathcal{F}_{n,3}$ and $\mathcal{F}$ by Lemma \ref{prl_comp}. Then, both functions are defined in the set $\mathcal{S}$ given by
\begin{align}
\mathcal{S}=\Big\lbrace q,\beta: \abs{q}\leq C_q, \beta \leq C_w, \beta \geq \abs{q}, t > 0 \Big\rbrace.
\end{align}
First, assume that $q\in\mathbb{R}$ and $\beta\geq 0$ are fixed. Note that $\vh\in\mathbb{R}^m$ and $\vs\in\mathbb{R}^m$ are standard Gaussian random vectors and $\vy=\varphi(\rho \vs)+\Delta \vepsilon$, where $\vepsilon$ is a standard Gaussian random vector. Next, we study the assumption properties of 
\begin{align}\label{ME_emp}
I_{n}=\frac{1}{m} \sum_{i=1}^{m} \mathcal{M}_{\widehat{\ell}}\Big( \beta h_i + \mu_1 q s_i+\mu_0 \vartheta-y_i  ; x \Big),
\end{align}
for fixed $x\in\mathbb{R}$.
Given that the loss function is proper, continuous, and convex in $\mathbb{R}$, there exists $C>0$ such that
\begin{align}
&\abs{\mathcal{M}_{\widehat{\ell}}\Big( \beta h_i + \mu_1 q s_i+\mu_0 \vartheta-y_i  ; x \Big)} \leq C \nonumber\\
&~~~~~~~~~~~~~~~+ \frac{1}{2x} (\beta h_i + \mu_1 q s_i+\mu_0 \vartheta-y_i)^2.
\end{align}
This means that the the Moreau envelope is square integrable for fixed $x$, $q$, and $\beta$.
Then, using the WLLN, the following convergence in probability holds
\begin{align}\label{conv_Me}
I_{n} \xrightarrow{p}  \mathbb{E}\left[  \mathcal{M}_{\widehat{\ell}}\Big( \beta H + \mu_1 q S+\mu_0 \vartheta-Y  ; x \Big) \right],
\end{align}
for any fixed $q\in\mathbb{R}$ and $\beta\geq 0$, where the expectation is over the random variables $H$, $S$ and $Y$, and where $S$ and $H$ are independent standard Gaussian random variables and $Y=\varphi(\rho S)+\Delta \epsilon$, where $\epsilon$ is a standard Gaussian random variable independent of $S$ and $H$. \\ 
Next, we study the asymptotic behaviour of $T_{n,3}$, $T_{n,4}$ and $T_{n,5}$. For fixed $t>0$, define the function $\widehat{T}_{n,4}$ as follows
\begin{align}
\widehat{T}_{n,4}(t)&={T}_{n,4}(t-\theta_n)\nonumber\\
&= \frac{1}{m} \vg^\top (\mB_{\vv}^{\perp})^\top \left[ \mG-\theta_n \mI_{k-1}+ t \mI_{k-1} \right]^{-1} \mB_{\vv}^{\perp} \vg.\nonumber
\end{align}
Based on the analysis in \cite[Thereom 3.4]{couillet_debbah_2011}, the following convergence in probability holds
\begin{align}
\widehat{T}_{n,4}(t)- \frac{\eta}{k}~\text{Tr}[ \left( \mG -\theta_n \mI_{k-1} + t \mI_{k-1} \right)^{-1}  ] \xrightarrow{~p~} 0.\nonumber
\end{align}
Note that the matrix $\mG$ is given by $\mG=(\mB_{\vv}^{\perp})^\top \mM^{-1} \mB_{\vv}^{\perp}$, where $\mB_{\vv}^{\perp} \in\mathbb{R}^{k\times (k-1)}$ is formed by an orthonormal basis orthogonal to the vector $\vv\in\mathbb{R}^k$. Furthermore, observe that $\sigma_{\text{min}}(\mM^{-1})$ can be expressed as follows
\begin{align}
\sigma_{\text{min}}(\mM^{-1})=\frac{1}{\mu_\star^2+\mu_1^2 \sigma_{\text{max}}(\mF^\top \mF-\mF^\top \bar{\vxi} \bar{\vxi}^\top \mF) },\nonumber
\end{align}
where $\sigma_{\text{min}}(.)$ and $\sigma_{\text{min}}(.)$ denotes the minimum and the maximum eigenvalues and the matrix $\mM$ is given by $\mM=\mu_1^2 \mF^\top \mP^{\perp}_{\vxi} \mF+\mu^2_\star \mI_k$.
It is easy to see that the following asymptotic property holds true
\begin{align}
\sigma_{\text{max}}(\mF^\top \mF-\mF^\top \bar{\vxi} \bar{\vxi}^\top \mF)-\sigma_{\text{max}}(\mF^\top \mF) \xrightarrow{~p~} 0.
\end{align}
Combining this with the Cauchy's Interlace Theorem and Assumption \ref{itm:ass_F}, we obtain the asymptotic result
\begin{align}
\sigma_{\text{min}}(\mG)-\sigma_{\text{min}}(\mM^{-1}) \xrightarrow{~p~} 0.
\end{align}
Now, based on Assumption \ref{itm:ass_F}, the random quantity $\theta_n=\sigma_{\text{min}}(\mG)$ converges in probability as follows
\begin{align}
\theta_n \xrightarrow{~p~} \frac{1}{\mu_\star^2+\mu_1^2 \kappa_{\text{max}}},
\end{align}
where $\kappa_{\text{max}}$ is defined in Assumption \ref{itm:ass_F}.
The implies that the following asymptotic property holds
\begin{align}
\norm{ \left( \widehat{\mG}  + t \mI_{k-1} \right) - (\mB_{\vv}^{\perp})^\top ( \widehat{\mM} + t \mI_{k} ) \mB_{\vv}^{\perp}} \xrightarrow{~p~} 0,\nonumber
\end{align}
where $\widehat{\mG}=\mG-\theta_n \mI_{k-1}$ and $\widehat{\mM}={\mM}^{-1} - \theta^\prime_n \mI_{k}$, where $\theta^\prime_n=\sigma_{\text{min}}(\mM^{-1})$. 
Combining this with the convergence analysis before the result in \eqref{Tn1toT1}, we obtain the following asymptotic property
\begin{align}
\frac{1}{k}\text{Tr}\Big[ \Big( \widehat{\mG}+ t \mI_{k-1} \Big)^{-1}  \Big]- \frac{1}{k}\text{Tr}\Big[ ( \widehat{\mM}+ t \mI_{k} )^{-1}  \Big] \xrightarrow{p} 0.\nonumber
\end{align}
Based on \eqref{mC} and \eqref{conv_haar}, we obtain the following asymptotic property
\begin{align}
\frac{1}{k}\text{Tr}\Big[ ( \widehat{\mM} + t \mI_{k} )^{-1}  \Big] - \frac{1}{k}\text{Tr}\Big[ ( \widehat{\mOm} + t \mI_{k} )^{-1}  \Big] \xrightarrow{p} 0,\nonumber
\end{align}
where $\widehat{\mOm}=\mOm^{-1}-\theta_n^\prime \mI_{k}$. Let's first assume that $\theta_n^\prime$ is a not a random variable and study the following function
\begin{align}\label{gn_conv}
g_n(\zeta)=\frac{1}{k} \text{Tr}[ ( \mOm^{-1}-\zeta \mI_{k}+ t \mI_{k} )^{-1}  ],
\end{align}
defined for $0\leq \zeta \leq \theta+\epsilon_t$, where $\epsilon_t>0$ is selected such that \eqref{gn_conv} is well-defined for fixed $t>0$ and where $\kappa_{\text{max}}$ is defined in Appendix \ref{itm:ass_F}. For fixed $\zeta$ and $t$, note that the following convergence in probability holds true
\begin{align}
g_n(\zeta) \xrightarrow{p} \frac{1}{\delta} \mathbb{E}_\kappa\left[ \frac{ \mu_\star^2 + \mu_1^2 \kappa }{1+(t-\zeta)(\mu_1^2 \kappa + \mu_\star^2) } \right]+ \frac{(1-{1}/{\delta})\mu_\star^2}{1+(t-\zeta) \mu_\star^2}, \nonumber
\end{align} 
where the expectation is over the random variable $\kappa$ distributed according to the probability distribution $\mathbb{P}_\kappa$ defined in Assumption \ref{itm:ass_F}. Now, using the bounded convergence theorem, we obtain the following convergence property
\begin{align}
g_n(\theta_n) \xrightarrow{p} \frac{1}{\delta} \mathbb{E}_\kappa\left[ \frac{ \mu_\star^2 + \mu_1^2 \kappa }{1+(t-\theta)(\mu_1^2 \kappa + \mu_\star^2) } \right]+ \frac{(1-{1}/{\delta})\mu_\star^2}{1+(t-\zeta) \mu_\star^2}, \nonumber
\end{align} 
where $\theta=1/(\mu_1^2 \kappa_{\text{max}}+\mu_\star^2)$.
Therefore, the function $T_{n,4}$ defined in the set $\lbrace t: t>-\theta_n \rbrace$ converges in probability to the function $T_4$ defined as 
\begin{align}
T_4(t)=\frac{\eta}{\delta} \mathbb{E}_\kappa\left[ \frac{ \mu_\star^2 + \mu_1^2 \kappa }{1+t(\mu_1^2 \kappa + \mu_\star^2) } \right] +\eta(1-\frac{1}{\delta}) \frac{\mu_\star^2}{1+t \mu_\star^2},\nonumber
\end{align}
in the set $\lbrace t: t>-\theta \rbrace$.
Similarly, one can show that the function $T_{n,3}$ converges in probability to the function $T_3$ defined in Section \ref{pre_ana_fm} and the function $T_{n,5}$ converges in probability to zero. Based on the analysis in Appendix \ref{pf_convx_prop}, the function defined as follows
\begin{align}
f(x) = \mathbb{E}\left[  \mathcal{M}_{\widehat{\ell}}\Big( \beta H + \mu_1 q S+\mu_0 \vartheta-Y  ; x \Big) \right],
\end{align}
and the function defined in \eqref{ME_emp} are both jointly convex in $q\in\mathbb{R}$ and $x>0$. Moreover, \cite[Theorem II.1]{andersen1982} states that pointwise convergence of convex functions in compact sets implies uniform convergence. Then, the convergence in \eqref{conv_Me} is uniform in compact sets of the variables $q$ and $x$. Based on \cite[Theorem 2.26]{var_ana}, the Moreau envelope inside the expectation in \eqref{conv_Me} is jointly convex and differentiable with respect to $q$ and $x>0$. Combining this with \cite[Theorem 7.46]{stoch_opt}, the asymptotic function in \eqref{conv_Me} is continuous in $q$ and $x>0$. Given that $t > 0$ is fixed and Assumption \ref{itm:ass_F}, the fixed values $T_1$ and $T_{4}(t)$ are strictly positive and bounded. Then, using the uniform convergence and the continuity property, we conclude that
\begin{align}
&\frac{1}{m} \sum_{i=1}^{m} \mathcal{M}_{\widehat{\ell}}\Big( \beta h_i + \mu_1 \sqrt{T_{n,1}} q s_i+\mu_0 \vartheta-y_i  ; \frac{T_{n,4}(t)}{\lambda} \Big) \xrightarrow{p} \nonumber\\
&~~~~~~\mathbb{E}\left[  \mathcal{M}_{\widehat{\ell}}\Big( \beta H + \mu_1 \sqrt{T_{1}} q S+\mu_0 \vartheta-Y  ; \frac{T_{4}(t)}{\lambda} \Big) \right],
\end{align}
where the expectation is over the random variables $H$, $S$ and $Y$.
This shows the desired pointwise convergence which completes the proof of Lemma \ref{sop_pt_conv}.
\subsection{Proof of Lemma \ref{convx_prop}: Convexity Property}\label{pf_convx_prop}
Consider the change of variable $t_{\text{new}}=t+\theta$. Then, the cost function $f_\vartheta$ can be expressed as follows
\begin{align}
&\hspace{-2mm}f_\vartheta(q,\beta,t)=\frac{\lambda q^2}{2 T_1}  \Big(\mu_1^2 \theta T_1 - \mu_1^2 t T_1 +\frac{T_1/e}{ \mathbb{E}_\kappa\left[ { \gamma_1(\kappa)}/{(\gamma_2(\kappa)+t)} \right]}\Big)  \nonumber\\
&\hspace{-2mm}-\frac{t \lambda}{2} \beta^2  +\frac{\theta \lambda}{2} \beta^2+\mathbb{E} \Big[ \mathcal{M}_{\widehat{\ell}}\Big( V(\vartheta,q,\beta);\frac{T_{4}(t-\theta) Z}{\lambda} \Big) \Big],
\end{align}
where $\gamma_1(\kappa)=\kappa/(\mu_1^2 \kappa + \mu_\star^2)$ and $\gamma_2(\kappa)=1/(\mu_1^2 \kappa + \mu_\star^2)-\theta$, the random variable $\kappa$ is distributed according to the distribution $\mathbb{P}_\kappa$ defined in Assumption \ref{itm:ass_F}, and where the function $V$ is linear in the variables $\vartheta$, $q$ and $\beta$ and is given in Section \ref{pre_ana_fm}.  
Note that the function $f_\vartheta$ is now define in the set
\begin{align}
\mathcal{S}=\lbrace \vartheta,q,\beta,t: \abs{\vartheta}\leq C_\vartheta,  \abs{q}\leq C_q,\beta\leq C_w, \beta \geq \frac{\abs{q}}{\sqrt{T_1}}, t >0 \rbrace.\nonumber
\end{align}
Based on \cite[Theorem 7.46]{stoch_opt}, all functions we introduce later are twice differentiable.\\
{\bf Property 1}: Given Assumption \ref{itm:ass_F}, the terms $q\to\mu_1^2 \theta \lambda q^2/2$ and $\beta \to {\theta \lambda \beta^2}/{2}$ are strongly convex in the variable $q$ and $\beta$, respectively.\\
Next, the objective is to show that the function $f_\vartheta$ is strictly concave in the variable $t$. To this end, fix $\vartheta$, $q$ and $\beta$ in the feasibility set and define the function $g$ as follows
\begin{align}
g(t)&= - \mu_1^2 t T_1 +\frac{T_1}{e \mathbb{E}_\kappa\left[ { \gamma_1(\kappa)}/{(\gamma_2(\kappa)+t)} \right]},
\end{align}
 in the set $\mathbb{R}^+$.
Given that the statements in Assumption \ref{itm:ass_F} are valid, the functions $\gamma_1$ and $\gamma_2$ are strictly positive and bounded functions. The function $g$ is differentiable where the first derivative is given by
\begin{align}
g^\prime(t)=-\mu_1^2 T_1 +\frac{T_1}{e} \frac{\mathbb{E}_\kappa\left[ { \gamma_1(\kappa)}/{(\gamma_2(\kappa)+t)^2} \right]}{\mathbb{E}_\kappa\left[ { \gamma_1(\kappa)}/{(\gamma_2(\kappa)+t)} \right]^2}.
\end{align}
Based on the Cauchy--Schwarz inequality, we obtain the following inequality
\begin{align}
\frac{\mathbb{E}_\kappa\left[ { \gamma_1(\kappa)}/{(\gamma_2(\kappa)+t)^2} \right]}{\mathbb{E}_\kappa\left[ { \gamma_1(\kappa)}/{(\gamma_2(\kappa)+t)} \right]^2} \geq \mathbb{E}_\kappa\left[  \gamma_1(\kappa) \right]^{-1} > \mu_1^2.
\end{align}
Combining this with the fact that $e \leq 1$ shows that the first derivative of the function $g$ is strictly positive which means that the function $g$ is a strictly increasing function. Now, define the function $\widetilde{g}$ as follows
\begin{align}
\widetilde{g}(t)=\mathbb{E}_\kappa\left[ { \gamma_1(\kappa)}/{(\gamma_2(\kappa)+t)} \right].
\end{align}
The function $\widetilde{g}$ is strictly positive and twice differentiable. Note that the first and second derivatives of this function can be expressed as follows
\begin{align}
\begin{cases}
\widetilde{g}^\prime(t)=-\mathbb{E}_\kappa\left[ { \gamma_1(\kappa)}/{(\gamma_2(\kappa)+t)^2} \right] \nonumber\\
\widetilde{g}^{\prime\prime}(t)=\mathbb{E}_\kappa\left[ { 2 \gamma_1(\kappa)}/{(\gamma_2(\kappa)+t)^3} \right].
\end{cases}
\end{align}
Using the Cauchy--Schwarz inequality, the following result is always true
\begin{align}
2\widetilde{g}^\prime(t)^2-\widetilde{g}^{\prime\prime}(t)\widetilde{g}(t) < 0.
\end{align}
This implies that the function $1/\widetilde{g}$ is strictly concave in the variable $t$. Therefore, we conclude that the function $g$ is strictly concave in the variable $t$. Using similar analysis, it can be shown that the function $h:t\to \lambda/(ZT_4(t-\theta))$ is concave in its argument. \\
{\bf Property 2}: Based on the above analysis and the properties stated in \cite[Section 3.2]{convex_opt}, the function $f_\vartheta$ is strictly concave in the variable $t$ for fixed feasible $\vartheta$, $q$ and $\beta$. \\
Next, the objective is to study the convexity properties of the function  ${f}_\vartheta$ in the variable $\vartheta$, $q$ and $\beta$ for fixed feasible $t$. To this end, fix $t>0$ and define the function $\widehat{f}_\vartheta$ as follows
\begin{align}\label{ref_f}
\widehat{f}_\vartheta(q,\beta)&=\sup_{t>0}~\frac{\lambda q^2}{2 T_1}  g(t) -\frac{t \lambda}{2} \beta^2 + \mathbb{E} \Big[ \mathcal{M}_{\widehat{\ell}}\Big( V(\vartheta,q,\beta);\frac{1}{h(t)} \Big) \Big] \nonumber\\
&= \sup_{t>0} \min_{0\leq \tau \leq C_{\tau}}~\frac{\lambda q^2}{2 T_1}  g(t) + \mathbb{E} \Big[ \mathcal{M}_{\widehat{\ell}}\Big( V(\vartheta,q,\beta);\frac{1}{h(t)} \Big) \Big]\nonumber\\
&~~~~~~~~~~~~+\frac{\tau^2 t}{2}-\sqrt{\lambda} \beta t \tau,
\end{align}
where $C_{\tau}$ is a sufficiently large constant. The reformulation in \eqref{ref_f} is valid given that the optimal $\tau$ is $\tau^\star=\sqrt{\lambda} \beta$ and $\beta$ is bounded. Observe that the cost function of the optimization problem given in \eqref{ref_f} is concave in $t$ and convex in $\tau$ for fixed feasible $\vartheta$, $q$ and $\beta$. Then, based on \cite[Corollary 3.3]{sion1958}, the function $\widehat{f}_\vartheta$ can be expressed as follows
\begin{align}
\widehat{f}_\vartheta(q,\beta)&= \inf_{0<\tau \leq C_{\tau}} \sup_{t>0}~\frac{\lambda q^2}{2 T_1}  g(t) + \mathbb{E} \Big[ \mathcal{M}_{\widehat{\ell}}\Big( V(\vartheta,q,\beta);\frac{1}{h(t)} \Big) \Big]\nonumber\\
&~~~~~~~~~~~~~~~~~+\frac{\tau^2 t}{2}-\sqrt{\lambda} \beta t \tau.
\end{align}
Now, we perform the change of variable $t=t_{\text{new}}/\tau$ which leads to the following equivalent formulation
\begin{align}\label{f_ref_f}
\widehat{f}_\vartheta(q,\beta)&= \inf_{0<\tau \leq C_{\tau}}  \sup_{t>0}~\frac{\lambda q^2}{2 T_1}  g(t/\tau) +\frac{\tau t}{2}-\sqrt{\lambda} \beta t \nonumber\\
&~~~~~~~~~~+ \mathbb{E} \Big[ \mathcal{M}_{\widehat{\ell}}\Big( V(\vartheta,q,\beta);\frac{1}{h(t/\tau)} \Big) \Big].
\end{align}
To show that the function $\widehat{f}_\vartheta$ is jointly convex in $\vartheta$, $q$ and $\beta$, it suffices to show that the cost function of the optimization problem given in \eqref{f_ref_f} is jointly convex in the variables $\vartheta$, $q$, $\beta$ and $\tau$ for any fixed feasible $t$. Based on the properties stated in \cite[Section 3.2]{convex_opt} and to show the joint convexity of the function
\begin{align}
(\vartheta,q,\beta,\tau) \to \mathbb{E} \Big[ \mathcal{M}_{\widehat{\ell}}\Big( V(\vartheta,q,\beta);\frac{1}{h(t/\tau)} \Big) \Big],
\end{align}
it suffices to show that the function $b:(x,\tau) \to x^2/(2\widehat{h}(\tau))$ is jointly convex in its arguments, where the function $\widehat{h}$ is given by $\tau \to ZT_4(t/\tau-\theta)/\lambda$. Note that the Hessian of the function $b$ is given by
\begin{align}
\mH_b(x,\tau)=\begin{bmatrix}
\frac{1}{\widehat{h}(\tau)} & -\frac{x \widehat{h}^\prime(\tau)}{\widehat{h}(\tau)^2} \\
-\frac{x \widehat{h}^\prime(\tau)}{\widehat{h}(\tau)^2} & \frac{x^2 (2\widehat{h}^\prime(\tau)^2-\widehat{h}^{\prime\prime}(\tau)\widehat{h}(\tau))}{2 \widehat{h}(\tau)^3}
\end{bmatrix}.
\end{align}
Next, we prove that the Hessian matrix $\mH_b$ is positive semi-definite for any $x$ and $\tau$. First, the function $h$ is strictly positive which means that ${1}/{\widehat{h}(\tau)}$ is strictly positive. Then, based on the Schur complement condition for positive semi-definiteness, the positive semi-definiteness of the Hessian matrix is guaranteed when the following quantify 
\begin{align}
I_{x,\tau}&=\frac{x^2 (2\widehat{h}^\prime(\tau)^2-\widehat{h}^{\prime\prime}(\tau)\widehat{h}(\tau))}{2 \widehat{h}(\tau)^3} - \frac{x^2 \widehat{h}^\prime(\tau)^2}{\widehat{h}(\tau)^3}\nonumber\\
&=-\frac{x^2 \widehat{h}^{\prime\prime}(\tau)}{2 \widehat{h}(\tau)^2},
\end{align}
is nonnegative for any $x$ and $\tau$. Then, it suffices to show that $\widehat{h}^{\prime\prime}(\tau) \leq 0$ for any $\tau$. Note that the second derivative of the function $\widehat{h}$ is given by
\begin{align}
\widehat{h}^{\prime\prime}(\tau)=-\frac{2\eta Z}{d \lambda} \mathbb{E} \left[ \frac{{\gamma}_2(\kappa)t}{({\gamma}_2(\kappa)\tau+t)^3} \right]-\frac{2 \eta Z}{\lambda}(1-\frac{1}{d}) \frac{\widehat{\gamma}_1 t}{ (\widehat{\gamma}_1 \tau +t)^3 },\nonumber
\end{align}
which is clearly non-positive for any $\tau>0$, where $\widehat{\gamma}_1=1/\mu_\star^2-\theta > 0$.\\
{\bf Property 3}: The above analysis shows that the function defined as follows
\begin{align}
(\vartheta,q,\beta,\tau) \to \mathbb{E} \Big[ \mathcal{M}_{\widehat{\ell}}\Big( V(\vartheta,q,\beta);\frac{1}{h(t/\tau)} \Big) \Big],
\end{align}
is jointly convex in its arguments.\\
Next, the objective is to show that the function $(q,\tau)\to q^2 g(t/\tau)$ is jointly strictly convex in its arguments. Note that it suffices to show that the function given by 
\begin{align}\label{g_hat_def}
\widehat{g}(q, \tau)=\frac{q^2}{2} \bar{g}(\tau),~\text{where}~\bar{g}(\tau)=\frac{1}{ \mathbb{E} \left[ \frac{ \widetilde{\gamma}_1(\kappa)\tau}{{\gamma}_2(\kappa)\tau+t} \right]} - \frac{t}{\tau},
\end{align}
is jointly strictly convex in its argument, where $\widetilde{\gamma}_1(\kappa)=\mu_1^2 \kappa/(\mu_\star^2+\mu_1^2 \kappa)$ is in the open set $(0,1)$. The function $\widehat{g}$ is twice differentiable where its Hessian matrix is given by
\begin{align}
\mH_{\widehat{g}}(q,\tau)=\begin{bmatrix}
\bar{g}(\tau)& q \bar{g}^\prime(\tau) \\
q \bar{g}^\prime(\tau)  & \frac{q^2}{2} \bar{g}^{\prime\prime}(\tau) 
\end{bmatrix}.
\end{align}
Next, we prove that the Hessian matrix $\mH_{\widehat{g}}$ is positive definite. Given that the function $\bar{g}$ is strictly positive, it suffices to show that the following quantity
\begin{align}
I_{q,\tau}=\bar{g}^{\prime\prime}(\tau)\bar{g}(\tau)-2\bar{g}^\prime(\tau)^2,
\end{align}
is strictly positive for any feasible $q$ and $\tau$. Note that $I_{q,\tau}$ is always strictly positive if the function $1/\bar{g}$ is strictly concave. It can be easily check that the first derivative of the function $1/\bar{g}$ is given by
\begin{align}
(1/\bar{g})^\prime(\tau)=t \frac{ \mathbb{E} \left[ \frac{ \widetilde{\gamma}_1(\kappa)}{({\gamma}_2(\kappa)\tau+t)^2} \right] -\mathbb{E} \left[ \frac{ \widetilde{\gamma}_1(\kappa)}{{\gamma}_2(\kappa)\tau+t} \right]^2 }{ \left( 1-t  \mathbb{E} \left[ { \widetilde{\gamma}_1(\kappa)}/{({\gamma}_2(\kappa)\tau+t)} \right] \right)^2 }.
\end{align}
From the Cauchy--Schwarz inequality, one can show that the function defined as follows
\begin{align}
\tau \to \mathbb{E} \left[ \frac{ \widetilde{\gamma}_1(\kappa)}{({\gamma}_2(\kappa)\tau+t)^2} \right] -\mathbb{E} \left[ \frac{ \widetilde{\gamma}_1(\kappa)}{{\gamma}_2(\kappa)\tau+t} \right]^2,
\end{align}
is a positive and strictly decreasing function in $\tau > 0$. This means that the first derivative of the function $(1/\bar{g})$ is strictly decreasing. \\
{\bf Property 4}: The above analysis shows that the function $\widehat{g}$ defined in \eqref{g_hat_def} is jointly strictly convex in $(q,\tau)$. \\
Property 3 and 4 show that the function $\widehat{f}_\vartheta$ is jointly convex in $(\vartheta,q,\beta)$. Combining this with the result in Property 1 prove that the function $f_\vartheta$ is jointly strongly convex in $(q,\beta)$ with a strong convexity parameter given by $\min(\mu_1^2,1)\theta \lambda$. Based on the properties stated in \cite[Section 3.2]{convex_opt} and the above four properties, the function defined as follows
\begin{align}\label{phi_def}
\phi(\vartheta)=\min_{(q,\beta) \in {\mathcal{F}}_{q,\beta}} \sup_{ t > 0} f_\vartheta (q,\beta,t),
\end{align}
is convex in the variable $\vartheta$ in the set $\lbrace \vartheta: \abs{\vartheta} \leq C_\vartheta \rbrace$. It remains to show that the function $\phi$ has a unique minimizer. Based on the above properties and \cite[Corollary 3.3]{sion1958}, the minimizer of the function $\phi$ solves the following optimization problem
\begin{align}\label{phi_st}
\phi^\star&=\min_{\substack{(q,\beta) \in {\mathcal{F}}_{q,\beta}\\0<\tau \leq C_{\tau}}} \sup_{ t > 0 } \frac{\lambda q^2}{2 T_1}  g(t/\tau) +\frac{\tau t}{2}-\sqrt{\lambda} \beta t+\mu_1^2 \theta \lambda q^2/2 \nonumber\\
&+ \min_{ \abs{\vartheta}\leq C_\vartheta}  \mathbb{E} \Big[ \mathcal{M}_{\widehat{\ell}}\Big( V(\vartheta,q,\beta);\frac{1}{h(t/\tau)} \Big) \Big]+{\theta \lambda \beta^2}/{2}.
\end{align}
Now, Property 1, 2, 3 and 4 show that the optimization problem given in \eqref{phi_st} has a unique optimal $q$, $\beta$, $\tau$, and $t$. Then, it suffices to show that the following optimization problem
\begin{align}\label{varth_opt_unq}
\vartheta^\star(q,\beta,\tau,t)=\argmin_{ \abs{\vartheta}\leq C_\vartheta}  \mathbb{E} \Big[ \mathcal{M}_{\widehat{\ell}}\Big( V(\vartheta,q,\beta);\frac{1}{h(t/\tau)} \Big) \Big],
\end{align}
has a unique solution and the optimal solution $\vartheta^\star$ is a continuous function in the variables $(q,\beta,\tau,t)$. Next, we use the definitions in \cite{ess_conv}. Note that the loss function $\widehat{\ell}$ is closed, proper and strictly convex. Based on \cite[Theorem 26.3]{ess_conv}, the conjugate of the loss function defined as $\widehat{\ell}^\star$ is essentially smooth. Based on \cite[Example 11.26]{var_ana}, the conjugate of the Moreau envelope $\mathcal{M}_{\widehat{\ell}}$ is given by
\begin{align}
\mathcal{M}_{\widehat{\ell}}^\star(.,h(t/\tau))=\widehat{\ell}^\star+\frac{h(t/\tau)}{2} \abs{.}^2.
\end{align}
Given that $\widehat{\ell}^\star$ is essentially smooth, the conjugate of the Moreau envelope $\mathcal{M}_{\widehat{\ell}}$  is essentially smooth. Then, based on \cite[Theorem 26.3]{ess_conv}, the Moreau envelope $\mathcal{M}_{\widehat{\ell}}$ is essentially strictly convex. Given that the feasibility set of $\vartheta$ is convex, the Moreau envelope $\mathcal{M}_{\widehat{\ell}}$ is strictly convex in $\vartheta$ in the feasibility set. Therefore, the optimization problem \eqref{varth_opt_unq} has a unique solution. Based on the discussion after \cite[Theorem 7.43]{stoch_opt}, the cost function of the optimization problem \eqref{varth_opt_unq} is continuous. Note that it is also strictly convex in $\vartheta$ and the feasibility set is convex and compact. Then, using \cite[Theorem 9.17]{sundaram_1996}, the function $\vartheta^\star$ is continuous in its arguments. This means that the minimizer of the function $\phi$ defined in \eqref{phi_def} is unique.\\
{\bf Property 5}: The above analysis shows that $\phi(\vartheta)$ is convex in the variable $\vartheta$ in the set $\lbrace \vartheta: \abs{\vartheta} \leq C_\vartheta \rbrace$ and it has a unique minimizer.\\
This completes the proof of Lemma \ref{convx_prop}.
\subsection{Proof of Proposition \ref{cons_sop}: Consistency of the SOP}\label{pf_cons_sop}
Note that the domain of the cost functions of the formulations given in \eqref{V2_form} and \eqref{detprob} are not the same. To work in the same set, we can perform the change of variable $\bar{q}_{\text{new}}=\abs{q}/\sqrt{T_{1}}$, $\bar{q}_{n,\text{new}}=\abs{q}/\sqrt{T_{n,1}}$, $\bar{t}_{\text{new}}=t+\theta$ and $\bar{t}_{n,\text{new}}=t+\theta_n$. Moreover, one can extend the analysis in Appendix \ref{pf_convx_prop} to show that the cost function  $\widehat{f}_{n,3}$ of the optimization problem $\widehat{\mathfrak{V}}_{n,3}$ given in \eqref{SOP_v1} is concave in the variable $t$. Moreover, the cost function  $\widehat{f}_{n,3}$ converges pointwisely as given in Lemma \ref{sop_pt_conv} where its asymptotic function satisfies the following asymptotic property
\begin{align}
\lim_{t\to+\infty}f_\vartheta(q,\beta,t)=-\infty,
\end{align}
for any fixed feasible $(q,\beta) \in {\mathcal{F}}_{q,\beta}$ and $q^2\neq \beta^2$. Assuming that $q^2\neq \beta^2$ and using the result in \cite[Lemma B.1]{chris:151}, we obtain the following asymptotic property
\begin{align}\label{conv_fn3_fv}
\sup_{t>-\theta_n} \widehat{f}_{n,3}(q,\beta,t) \xrightarrow{p} \sup_{t > -\theta} f_\vartheta(q,\beta,t).
\end{align} 
If $q^2 = \beta^2$, the supremum in \eqref{conv_fn3_fv} occurs at $+\infty$ and the closed-form expression can be found. Moreover, using the analysis in Appendix \ref{pf_sop_pt_conv}, we can show that the result in \eqref{conv_fn3_fv} is still valid in this case.
Now, define $\widehat{f}_{n,2}$ as the cost function of the minimization problem given in \eqref{V2_form}. Using the property stated in Lemma \ref{par_unf_con}, the convergence in \eqref{conv_fn3_fv} implies the following asymptotic result
\begin{align}\label{conv_fn2_fv}
\widehat{f}_{n,2}(q,\beta) \xrightarrow{p} f_\vartheta(q,\beta) =\sup_{t > -\theta} f_\vartheta(q,\beta,t),
\end{align} 
valid for any fixed feasible $(q,\beta)$. Based on Lemma \ref{convx_prop}, the function $f_\vartheta$ is jointly strongly convex in $q$ and $\beta$ where a strong convexity parameter is $S=\theta \lambda\min(\mu_1^2,1)>0$. Next, the objective is to show that the function $\widehat{f}_{n,2}$ is jointly convex in $q$ and $\beta$ with probability going to $1$ as $n$ goes to $+\infty$. To this end, assume that $\zeta\in(0,1)$ and $q_1$, $q_2$, $\beta_1$ and $\beta_2$ are feasible such that $\beta_1 \neq \beta_2$ or $q_1 \neq q_2$. Note that 
\begin{align}\label{sconv_def}
&f_\vartheta(\zeta q_1+ (1-\zeta) q_2,\zeta \beta_1+ (1-\zeta) \beta_2) \leq\zeta f_\vartheta(q_1,\beta_1)\nonumber\\
& +(1-\zeta) f_\vartheta(q_2,\beta_2)-\frac{S}{2} \zeta(1-\zeta) (q_1-q_2)^2\nonumber\\
&-\frac{S}{2} \zeta(1-\zeta) (\beta_1-\beta_2)^2.
\end{align}
Based on the asymptotic property in \eqref{conv_fn2_fv}, for any fixed feasible $q$ and $\beta$ and $\epsilon>0$, the following inequality 
\begin{align}\label{pw_conv_fne}
\abs{\widehat{f}_{n,2}(q,\beta)-f_\vartheta(q,\beta)} < \epsilon,
\end{align}
holds with probability going to $1$ as $n$ goes to $+\infty$. Combining \eqref{sconv_def} and \eqref{pw_conv_fne}, the following inequality 
\begin{align}
&\widehat{f}_{n,2}(\zeta q_1+ (1-\zeta) q_2,\zeta \beta_1+ (1-\zeta) \beta_2) \leq\zeta \widehat{f}_{n,2}(q_1,\beta_1)\nonumber\\
& +(1-\zeta) \widehat{f}_{n,2}(q_2,\beta_2)+\frac{S}{2} \zeta(1-\zeta) (q_1-q_2)^2\nonumber\\
&+\frac{S}{2} \zeta(1-\zeta) (\beta_1-\beta_2)^2+3\epsilon,
\end{align}
holds with probability going to $1$ as $n$ goes to $+\infty$. Given that $q_1\neq q_2$ or $\beta_1\neq\beta_2$, $\zeta\in(0,1)$ and $S>0$, we conclude that the following inequality
\begin{align}
&\widehat{f}_{n,2}(\zeta q_1+ (1-\zeta) q_2,\zeta \beta_1+ (1-\zeta) \beta_2) < \zeta \widehat{f}_{n,2}(q_1,\beta_1)\nonumber\\
& +(1-\zeta) \widehat{f}_{n,2}(q_2,\beta_2),
\end{align}
holds with probability going to $1$ as $n$ goes to $+\infty$. This means that the function $\widehat{f}_{n,2}$ is strictly convex in $q$ and $\beta$ on events with probability going to $1$ as $n$ goes to $+\infty$. Then, based on \cite[Theorem II.1]{andersen1982}, the convergence in \eqref{conv_fn2_fv} is uniform in compact sets of the variables $q$ and $\beta$. Moreover, based on the discussion after \cite[Theorem 7.43]{stoch_opt}, the cost function of the minimization problem \eqref{detprob} is continuous in $q$ and $\beta$. Then, based on \cite[Theorem 2.1]{NEWEY19942111}, we obtain the following asymptotic results
\begin{equation}
\widehat{O}^{\star}_{n,2}  \xrightarrow{p} \widehat{O}^{\star}~\mathrm{and}~\mathbb{D}( \widehat{\mathcal{S}}^{\star}_{n,2},\widehat{\mathcal{S}}^{\star} )  \xrightarrow{p} 0,
\end{equation}
where $\mathbb{D}( \mathcal{A},\mathcal{B} )$ denotes the deviation between the sets $\mathcal{A}$ and $\mathcal{B}$ and is defined as $\mathbb{D}( \mathcal{A},\mathcal{B} )=\sup_{\vx_1\in\mathcal{A}} \inf_{\vx_2\in\mathcal{B}} \norm{\vx_1-\vx_2}$. This completes the proof of Proposition \ref{cons_sop}.
\subsection{Proof of Proposition \ref{ff_perf}: Feature Formulation Performance}\label{pf_ff_perf}
Assume that $q^\star_\vartheta$ and $\beta^\star_\vartheta$ defined in Proposition \ref{swi_norm} are the unique optimal solutions of the optimization problem \eqref{detprob}. Let $\phi_n(\vartheta)$ and $\phi^c_{n}(\vartheta)$ be the optimal cost values of the formulation in $\widehat{\mathfrak{V}}_{n,2}$ with feasibility sets ${\mathcal{P}}_{q,\beta}$ and $\mathcal{S}^c_{n,\vartheta,\epsilon}$ defined in Proposition \ref{swi_norm}, respectively. Moreover, define $\phi(\vartheta)$ and $\phi^c(\vartheta)$ as the optimal cost values of the deterministic optimization problem \eqref{detprob}, with feasibility sets ${\mathcal{F}}_{q,\beta}$ and $\mathcal{S}^c_{\vartheta,\epsilon}={\mathcal{F}}_{q,\beta} \setminus \lbrace (q,\beta): \abs{q-q^\star_\vartheta} < \epsilon, \abs{\beta-\beta^\star_\vartheta} < \epsilon\rbrace$, respectively. Based on the analysis in Appendix \ref{pf_cons_sop}, the optimal cost $\phi_n(\vartheta)$ converges in probability to $\phi(\vartheta)$ as $n$ goes to $\infty$ and the optimal cost $\phi^c_{n}(\vartheta)$ converges in probability to $\phi^c(\vartheta)$ as $n$ goes to $\infty$, for any fixed $\epsilon>0$. Given that the optimization problem given in \eqref{detprob} is jointly strongly convex in the variables $q$ and $\beta$, the third assumption in Proposition \ref{swi_norm} is also satisfied. Then, we obtain the following convergence result
\begin{equation}
\abs{ \Phi_n(\vartheta) -\phi_n(\vartheta) } \overset{p}{\longrightarrow} 0,~\text{and}~\mathbb{P}( \widehat{\vw}_{n,\vartheta} \in \mathcal{S}_{n,\vartheta,\epsilon} )  \overset{n\to\infty}{\longrightarrow} 1,\nonumber
\end{equation}
for any fixed $\epsilon>0$, where $\Phi_n(\vartheta)$ and $\widehat{\vw}_{n,\vartheta}$ are the optimal cost and the optimal solution of the primary problem \eqref{PO4}. Note that the above analysis is performed for fixed $\vartheta$ in the set $\lbrace \vartheta: \abs{\vartheta} \leq C_\vartheta \rbrace$. To finalize our proof, we show that the assumptions in Proposition \ref{in_vartheta} are all satisfied. Note that the above analysis shows that the assumptions (1) and (2) in Proposition \ref{in_vartheta} are satisfied. Based on Lemma \ref{convx_prop}, the function $\vartheta \to \phi(\vartheta)$ is convex in $\vartheta$ and has a unique minimizer $\vartheta^\star$. Therefore, the following convergence in probability holds
\begin{align}
\abs{ \Phi_n -\phi(\vartheta^\star) }& \overset{p}{\longrightarrow} 0,~\text{and}~\mathbb{P}( \widehat{\vw}_n \in cl(\mathcal{S}_{n,\vartheta^\star,\epsilon}) )  \overset{n\to \infty}{\longrightarrow} 0,\nonumber\\
&~~~~~\text{and}~\widehat{\vartheta}_n^\star \overset{p}{\longrightarrow} \vartheta^\star,
\end{align}
for any $\epsilon>0$, where $\Phi_n$ and $(\widehat{\vw}_n,\widehat{\vartheta}_n^\star)$ are the optimal cost and the optimal solution of the problem \eqref{cons_po}, and where $cl(\mathcal{S}_{n,\vartheta^\star,\epsilon} )$ denotes the closure of the set $\mathcal{S}_{n,\vartheta^\star,\epsilon}$. Based on \cite[Theorem 1.6]{stoch_opt}, we have the property $cl(\mathcal{S}_{n,\vartheta^\star,\epsilon})=\mathcal{S}_{n,\vartheta^\star,\epsilon}$. This completes the proof of Proposition \ref{ff_perf}.

\bibliographystyle{IEEEtran}
\bibliography{reference,refs}

\end{document}